\documentclass{aa}  

\usepackage{natbib}
\usepackage{graphicx}
\usepackage{txfonts}
\usepackage[colorlinks=true,urlcolor=blue,citecolor=blue]{hyperref}
\newcommand{\kms}{km $s^{-1}$}
\newcommand{\ha}{H$\alpha$}
\newcommand{\hb}{H$\beta$}
\newcommand{\siidoublet}{[\ion{S}{ii}]$\lambda6716,31$}

\newcommand{\niiauroral}{[\ion{N}{ii}]$\lambda5755$}
\newcommand{\oiidoublet}{[\ion{O}{ii}]$\lambda7320,30$}
\newcommand{\nii}{[\ion{N}{ii}]$\lambda6583$}
\newcommand{\niileft}{[\ion{N}{ii}]$\lambda6548$}
\newcommand{\sii}{[\ion{S}{ii}]$\lambda6716$}
\newcommand{\siisum}{[\ion{S}{ii}]$\lambda6716+31$}
\newcommand{\siirat}{[\ion{S}{ii}]$\lambda6731/16$}
\newcommand{\oiii}{[\ion{O}{iii}]$\lambda5007$}
\newcommand{\siii}{[\ion{S}{iii}]$\lambda9068$}
\newcommand{\hii}{\ion{H}{ii}}

\newcommand{\te}{$T_{\rm e}$}
\newcommand{\den}{$N_{\rm e}$}
\newcommand{\fielda}{\textsf{A}}
\newcommand{\fieldb}{\textsf{B}}
\newcommand{\fieldc}{\textsf{C}}
\newcommand{\fieldd}{\textsf{D}}
\newcommand{\fielde}{\textsf{E}}
\newcommand{\fieldi}{\textsf{I}}
\newcommand{\fieldj}{\textsf{J}}
\newcommand{\fieldp}{\textsf{P}}
\newcommand{\fieldq}{\textsf{Q}}
\newcommand{\metal}{$12 + \log_{10}{\rm {O/H}}$}

\begin{document}

   \title{MUSE crowded field 3D spectroscopy in NGC 300}

   \subtitle{III.\ Characterizing extremely faint HII regions and diffuse ionized gas}

   \author{Genoveva Micheva\inst{1}\thanks{\email{gmicheva@aip.de}}
          \and
          Martin M. Roth\inst{1}
          \and
          Peter M.\ Weilbacher\inst{1}
          \and
          Christophe Morisset\inst{2}
          \and
          N.\ Castro\inst{1}
          \and
          A. Monreal Ibero\inst{3}
          \and 
          Azlizan A. Soemitro\inst{1}
          \and
          Michael V. Maseda\inst{4}
          \and
          Matthias Steinmetz\inst{1}
          \and
          Jarle Brinchmann\inst{3,5}
          }

   \institute{
    Leibniz-Institute for Astrophysics Potsdam (AIP), An der Sternwarte 16, 14482 Potsdam, Germany\\
        \and
    Universidad Nacional Aut\'{o}noma de M\'{e}xico, Instituto de Astronom\'{\i}a, AP 106,  Ensenada 22800, BC, M\'{e}xico
        \and
           Leiden Observatory, Niels Bohrweg 2, 2333 CA Leiden, The Netherlands
           \and
           Department of Astronomy, University of Wisconsin-Madison, 475 N. Charter St., Madison, WI 53706 USA
           \and
           Instituto de Astrof\'{\i}sica e Ci\^{e}ncias do Espa\c{c}o, Universidade do Porto, CAUP, rua das Estrelas, 4150-762, Porto, Portugal
             }
    
   \date{Submitted May 14 2022; Accepted October 3 2022}

  \abstract
  % context heading (optional)
   {There are known differences between the physical properties of \hii~and diffuse ionized gas (DIG). However, most of the studied regions in the literature are relatively bright, with $\log_{10}L(H\alpha) [erg/s]\gtrsim37$.}  
   {We compiled an extremely faint sample of $390$ \hii~regions with a median \ha~luminosity of $34.7$ in the flocculent spiral galaxy NGC 300, derived their physical properties in terms of metallicity, density, extinction, and kinematics, and performed a comparative analysis of the properties of the DIG.}
  % aims heading (mandatory)
   {We used MUSE data of nine fields in NGC 300, covering a galactocentric distance of zero to $\sim450$ arcsec ($\sim4$ projected kpc), including spiral arm and inter-arm regions. We binned the data in dendrogram leaves and extracted all strong nebular emission lines. We identified \hii~and DIG regions and compared their electron densities, metallicity, extinction, and kinematic properties. We also tested the effectiveness of unsupervised machine-learning algorithms in distinguishing between the \hii~and DIG regions.}
  % methods heading (mandatory)
   {The gas density in the \hii~and DIG regions is close to the low-density limit in all fields. The average velocity dispersion in the DIG is higher than in the \hii~regions, which can be explained by the DIG being 1.8 kK hotter than \hii~gas. The DIG manifests a lower ionization parameter than \hii~gas, and the DIG fractions vary between $15\mbox{-}77\%$, with strong evidence of a contribution by hot low-mass evolved stars and shocks to the DIG ionization. Most of the DIG is consistent with no extinction and an oxygen metallicity that is indistinguishable from that of the \hii~gas. We observe a flat metallicity profile in the central region of NGC 300, without a sign of a gradient.}
  % results heading (mandatory)
   {The differences between extremely faint \hii~and DIG regions follow the same trends and correlations as their much brighter cousins. Both types of objects are so heterogeneous, however, that the differences within each class are larger than the differences between the two classes.}
  % conclusions heading (optional), leave it empty if necessary 
   {}


   \maketitle
%
%-------------------------------------------------------------------

\section{Introduction}
\hii~regions are tracers of recent star formation in the interstellar medium (ISM) and typically display well-detected nebular emission lines such as the Balmer lines, [\ion{O}{iii}]$\lambda\lambda4959,5007$,  [\ion{S}{ii}]$\lambda\lambda6716,6731$, and [\ion{N}{ii}]$\lambda\lambda6548,6583$ \citep[e.g.,][]{Shields1990}. The diffuse ionized gas (DIG) is another component of the ISM, contributing a significant part ($\sim50\%$) of the total \ha~luminosity in late-type spirals \citep[e.g.,][]{Haffner2009}. This component is often also referred to as the warm ionized medium (WIM) in the literature \cite[e.g.,][]{Haffner2009, Weilbacher2018}. 
The properties of the DIG have been shown to be significantly different not only from those of classical \hii~regions \citep{Madsen2006}, but also from one location to the next within the same galaxy, with a range of measured electron densities and temperatures, and a great variation in the [\ion{O}{iii}]$\lambda\lambda4959,5007$/\ha,  [\ion{S}{ii}]$\lambda\lambda6716,6731$/\ha, and [\ion{N}{ii}]$\lambda\lambda6548,6583$/\ha~emission line ratios \citep[e.g.,][]{Mathis2000, Haffner2009}. This led to the conclusion that there is no single emission line value or threshold in the equivalent width of the lines that allows for a separation of DIG-dominated spaxels from those with non-DIG emission \citep[e.g.,][]{Tomicic2021}. The DIG has been shown to be ionized by Lyman continuum radiation escaping from the nearby \hii~regions at least partially \citep[e.g.][]{Reynolds1984, Ferguson1996, Oey2007, Chevance2020, denBrok2020, DellaBruna2020, Belfiore2022}. This is supported by the fact that the DIG is usually observed in the vicinity of \hii~regions \citep[e.g.,][]{Ferguson1996, Zurita2000,Zurita2002,Madsen2006,Howard2018}. In order to study DIG-dominated regions, therefore, a selection criterion based on physical proximity to nearby \hii~regions is often used \citep[e.g.,][]{Weilbacher2018, DellaBruna2020,denBrok2020}. There are also alternative scenarios for the DIG ionization in which the DIG emission is attributed to turbulence \citep{Binette2009}, shocks \citep{Collins2001}, and contributions by hot evolved low-mass stars (HOLMES) as defined by \citet{FloresFajardo2011}, and supported by \citet[][]{Belfiore2022}. None of these, however, offer a robust and systematic solution to determining the location of the DIG-dominated regions. 
Understanding the characteristics of \hii~and DIG regions is fundamentally important for understanding the star formation process in galaxies. Most of the \hii~regions studied in the literature are rarely fainter than $10^{36}$ erg/s in \ha~luminosity \citep[e.g.,][]{Youngblood1999, Bradley2006, Hakobyan2007, Weilbacher2018, DellaBruna2020}, with the DIG being similarly bright. To study fainter samples of both \hii~and DIG regions, we must therefore look to nearby galaxies.

NGC 300 is a flocculent spiral galaxy in the Sculptor Group and one of the closest galaxies to the Local Group \citep{Sersic1966,Gieren2005}. Due to its proximity and therefore the higher spatial resolution it offers, it is the subject of numerous studies on its own merit and as a case study \citep[e.g.,][]{Sersic1966, Deharveng1988, Galvin2012,Stasinska2013,Faesi2014,Faesi2016,Gazak2015,Niederhofer2016,Hillis2016,Rodriguez2016,Kang2016,Riener2018,Roth2018,Mondal2019,Gross2019,McLeod2019}.  
The MUSE observations of NGC 300 are presented at length in the pilot paper of \citet{Roth2018}, which showed the use of MUSE for crowded field 3D spectroscopy. The observed fields in NGC 300 are purposefully chosen to cover much of the central regions of this galaxy while avoiding the brightest \hii~regions. Our work is a continuation of this project, but concentrates exclusively on the study of the \hii~regions and the DIG. 
In this paper we study extremely faint \hii~regions and even fainter DIG. We aim to construct a sample of the \hii~regions and localize the DIG. We make use of the high spatial resolution that MUSE affords in NGC 300 to characterize both samples in terms of their physical properties, study their spatial distribution, and highlight any detected differences between these ISM components.

Similar to \citet{Roth2018}, the first paper in this series, the adopted distance to NGC 300 is $1.88$ Mpc \citep{Gieren2005,Bresolin2005}. The sample of \hii~and DIG regions we use in our analysis is available online\footnote{The catalog is available in electronic form at the CDS via anonymous ftp to via anonymous ftp to cdsarc.u-strasbg.fr (130.79.128.5) or via \url{http://cdsweb.u-strasbg.fr/cgi-bin/qcat?J/A+A/?/?}.}.

The paper is organized as follows: In Section \ref{sec:data} we briefly present the data and give details of the line extraction. The classification of emission line regions is explained in Section \ref{sec:bpt}. In Section \ref{sec:prop} and its subsections, we derive physical properties for the \hii~and DIG regions such as metallicity, extinction, and electron density, as well as kinematic properties such as velocity dispersion and shear. To increase the signal-to-noise ratio of the data, we stack the spectra per field in Section \ref{sec:stacks}. We discuss our findings in Section \ref{sec:discuss} and summarize our conclusions in Section \ref{sec:conclude}.

%----------------------------------------------------------------- 
   \begin{figure}
   \centering
   \includegraphics[width=8.0cm]{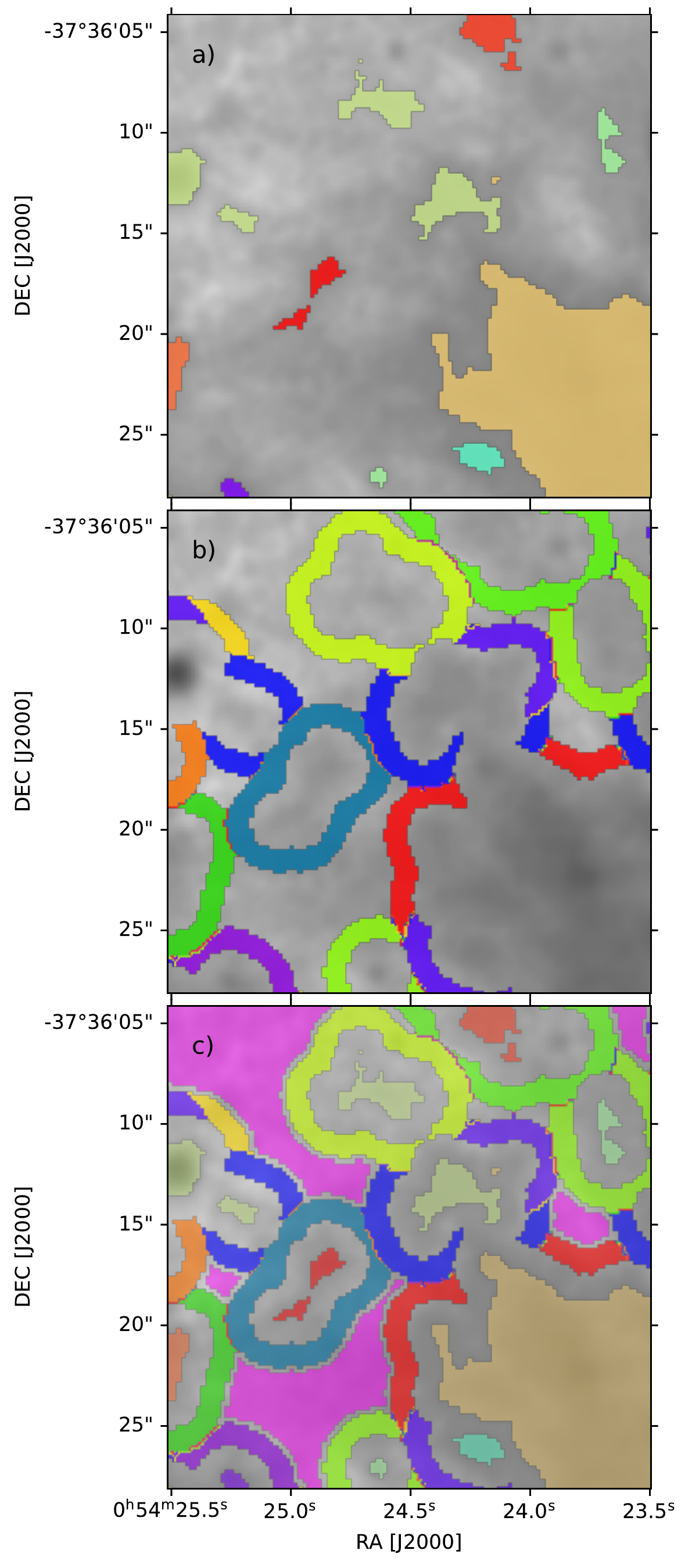}
      \caption{Region definitions for an example region in field P. Overplotted are a) dendrogram leaves, b) DIG ring regions, which also serve as sky background regions, and c) dendrogram leaves and DIG rings both, as well as the manually added DIG regions, shaded in pink.}
         \label{fig:showcase}
   \end{figure}
%----------------------------------------------------------------- 

\section{Data}\protect\label{sec:data}
Several fields of NGC\,300 were observed as part of the Guaranteed Time
Observations with the MUSE instrument \citep{Bacon2010} in 2014 to
2016\footnote{ESO program IDs 094.D-0116, 095.D-0173, 097.D-0348, and
  0102.B-0317}.
All observations used the wide-field mode, which covers $1\arcmin\times1\arcmin$
, and individual exposures of 900\,s were made, with 90\degr\ rotations in
between exposures and with an offset sky exposure taken after two
science exposures. The initial observations including all exposures taken in
2014 and 2015 are described by \citet{Roth2018}. They included the fields \fielda,
\fieldb, \fieldc, \fieldd, \fielde, \fieldj, and \fieldi. All of these observations were carried out without adaptive optics (AO) support (extended Wide Field Mode with natural seeing, WFM-NOAO-E), with a contiguous wavelength coverage of
4600--9350\,\AA. While all of the fields were targeted to be 1.5\,h deep
($6\times900$\,s), fields \fieldb, \fielde, and \fieldj{} had only partial coverage. After
combination, fields \fieldb{} and \fieldc{} had an effective image quality of 1\farcs0 or lower.
% 2016 data
In August and September 2016, observations of fields \fieldd{} (two new exposures) and \fielde{} (four new exposures) were completed using the same mode. Two new fields (\fieldp{} and \fieldq) at
larger radii from the galaxy center were added, both receiving the full
coverage $6\times900$\,s in good atmospheric conditions.
% 2018 AO data
In October and November 2018, the innermost three fields (\fielda, \fieldb, and \fieldc) were
reobserved, this time with AO support \citep{Stroebele2012} in good
conditions with an average external seeing of 0\farcs75. They were again
observed in extended mode (Wide Field Mode with Adaptive Optics - Extended, WFM-AO-E, 4600--9350\,\AA), except for two exposures
of field \fielda,{} which by mistake were taken in nominal mode (Wide Field Mode with Adaptive Optics - Nominal, WFM-AO-N,
4700--9350\,\AA). In AO-supported observations, a wide gap around the Na\,D line
blocks out the laser light (about 5750--6010\,\AA).
Again, 120\,s offset sky observations followed each on-target exposure.
Illumination flat-fields were taken before or after the observing blocks, and
a standard star was observed in the same instrument mode in the same night.

\subsection{Reductions}
The data were reduced with the standard MUSE pipeline \citep{Weilbacher2020}, but we used the version integrated into the MUSE-Wise \citep{Vriend2015} environment.
MUSE-Wise automatically associates the appropriate calibrations for each science
exposure. Due to the long time span of the observations, the data were processed with
different software versions. The basic CCD-level processing of the data of 2014 and 2015
is the same as used by \citet{Roth2018} and was performed with v1.0 and v1.2; the 2016 data
use v1.6. A stack of all exposures was created with v2.4. The AO data of 2018
then used a development version (v2.5.4) that soon after was publicly released as
v2.6\footnote{\url{https://data.aip.de/projects/musepipeline.html}}.
As usual for most MUSE data, all science exposures were corrected for the CCD bias, but did not correct for the negligible dark current. Flat-fields and arc exposures taken in the morning after
the science data were used to correct for pixel-to-pixel response, to locate the illuminated areas
on the CCDs, and to assign wavelengths to each pixel. Spatial positions in the field of view
were taken from the geometric calibration specific to the corresponding observing run.
Further flat-fielding was applied to minimize spatial slice-to-slice variations using
the corresponding illumination-flat exposure and to correct global gradients across the
field with sky-flats taken in the same mode within a few days of the science exposure.

The data were then corrected for atmospheric refraction and were flux calibrated. Because the
observations were obtained during clear or photometric nights, a single standard star in the
same night and instrument mode was used to determine the response curve and also to apply
a correction of the telluric absorption features. The sky subtraction first employed
the creation of a sky continuum spectrum from the offset sky field taken closest in time.
To do this, the sky spectrum was averaged over 80\% of the sky field, and the emission lines
were fitted and subtracted. The fitted sky emission line fluxes were then taken as
first-guess inputs to the fit of the sky lines over a 5--10\% fraction of the science
field, from which the sky continuum was directly subtracted. The telluric \ha{}
emission is only imperfectly removed by this procedure; all other lines typically leave
1--2\% residuals in the science data. The data were further corrected for distortions
using an astrometric calibration of the corresponding observing run, and were shifted to
barycentric velocity.
One AO exposure of field \fieldb{} (with header DATE-OBS=2018-11-10T00:16:43) was
contaminated by a satellite. After finding the trail in visual inspection of the
cube, the pixels corresponding to $\pm$5 pixels from the peak were masked and the
data were processed again.
Relative offsets were corrected before combining all exposures of each field. 
In case of the AO fields, all offsets were computed relative to the Gaia
DR2 \citep{Lindegren2018} sources. For the non-AO fields, offsets were computed
relative to the first exposure, and the world coordinates of the final cubes were
manually adjusted to have absolute astrometry in agreement with the Gaia DR2 positions.
The absolute positional accuracy is on the order of one spatial MUSE pixel or better,
that is, $<0\farcs2$.
%
%                                                One column figure
%----------------------------------------------------------------- 
   \begin{figure}
   \centering
   \includegraphics[width=\hsize]{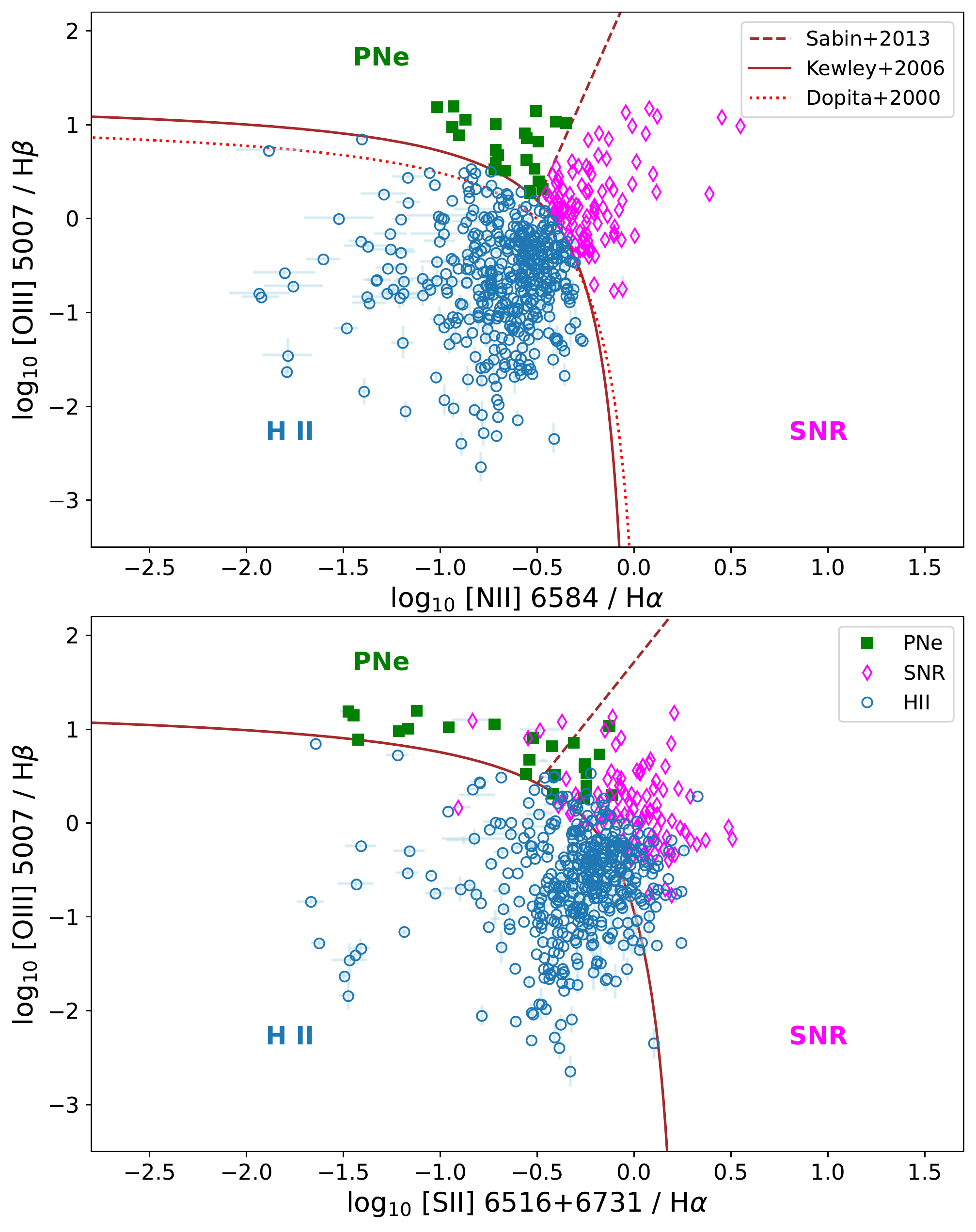}
      \caption{BPT diagram in [\ion{N}{ii}]$\lambda6584$/\ha~ (top) and \siidoublet/\ha~(bottom). The objects are classified in the [\ion{N}{ii}]$\lambda6584$/\ha~diagram via the \citet{Kewley2006} (solid line) star formation limit and the PNe-SNR (dashed line) separation lines. The \citet{Dopita2000} (dotted line) is also shown for reference. The color-coding in the \siidoublet/\ha~diagram is based on the [\ion{N}{ii}]$\lambda6584$/\ha~classification.}
         \label{fig:bpt}
   \end{figure}
%----------------------------------------------------------------- 

  \begin{figure*}
   \centering
   \includegraphics[width=17.5cm]{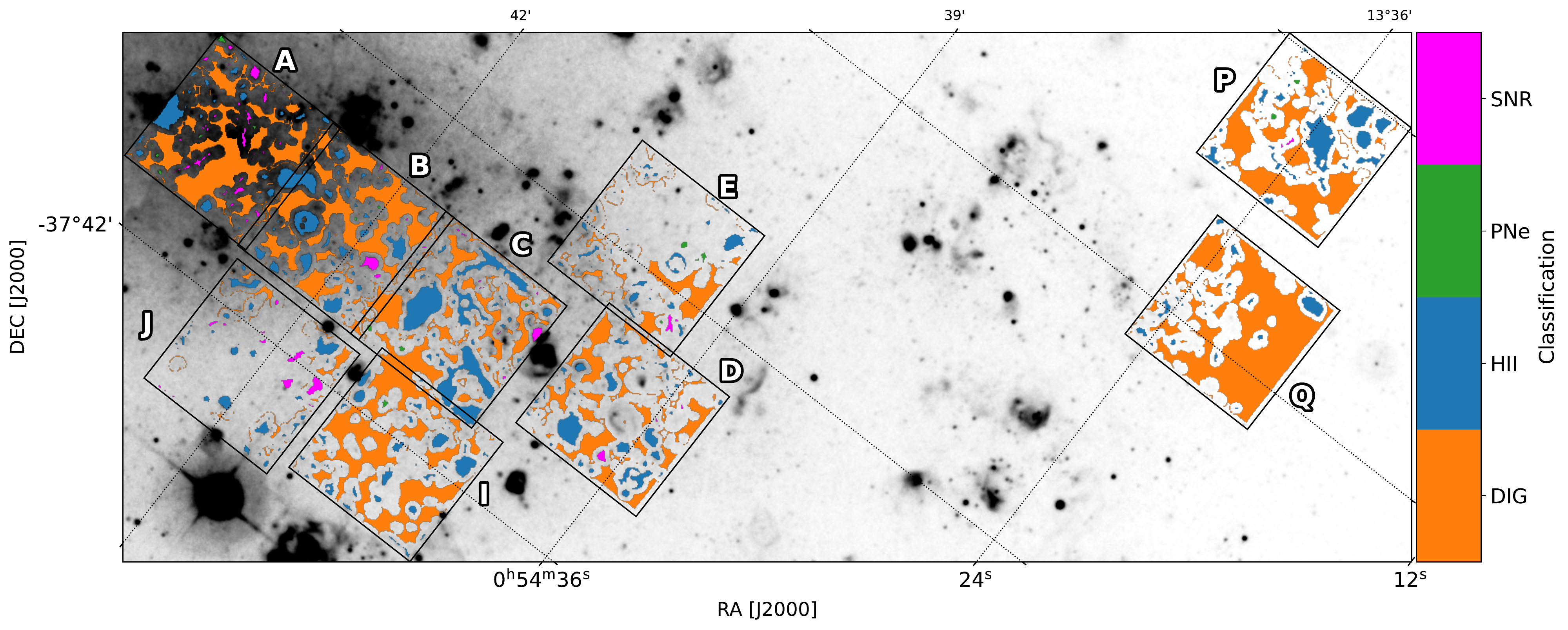}
      \caption{2D map of the BPT classification. }
         \label{fig:2Dclass}
   \end{figure*}
 
All cubes were constructed with the default $0\farcs2\times0\farcs2$ spatial sampling
and have approximately $60\arcsec\times60\arcsec$ spatial coverage. The wavelength
coverage is about 4600--9350\,\AA{} and sampled at 1.25\,\AA{} per pixel. The exact
wavelength value of the first pixel depends on the velocity correction applied to
the data.

\subsection{Definition of regions}\label{sec:defreg}
Before the emission line flux could be extracted, we prepared the data in the following way. A first-order approximation of the \ha~spatial distribution was obtained for all fields for the AO and non-AO data using the \ha~narrowband filter range defined in \citet{Roth2018}. To reduce noise, the \ha~maps were smoothed with a 2D Gaussian kernel with a standard deviation equal to $1.25$, using \textsc{astropy.convolve}. Unfortunately, the signal-to-noise ratio (S/N) in individual spaxels is low, and hence reliable spectral information on the scales of individual pixels often cannot be obtained. Therefore, we binned the preliminary \ha~maps in dendrograms, using the \textsc{astrodendro}\footnote{\footnotesize https://dendrograms.readthedocs.io/en/stable/} package \citep{Robitaille2019}. Through trial and error, we determined that the best parameters for dendrogram extraction, resulting in a reduced number of spurious detections and yet capturing all main structures in the fields, are  \verb+min_delta+$=0.005$ and \verb+min_npix+$= 10$. For the purpose of illustration, in Figure \ref{fig:showcase}a we show an example region in field P and some typical dendrogram leaf masks. We manually examined the dendrogram tree, and found that more often than not, only the compact peaks of otherwise large filaments and bubbles were captured by the automatic algorithm. We merged down and trimmed leaves and branches until the resulting shapes traced the entirety of the visible filaments and bubbles. This procedure is subjective in the sense that there are cases in which the morphological appearance suggested to merge individual leaves that were not reported as connected by the automatic procedure. The spectra in each dendrogram leaf were summed.

Next, as an adaptation of classical aperture photometry techniques \citep[e.g.,][]{Stetson1987}, the diffuse background contribution was estimated locally around each dendrogram leaf in a ring with a width of 7 pixels, with a gap between the dendrogram and the background region of 5 pixels. The ring shapes were determined via binary dilation of the leaf masks. These ring regions are illustrated in Figure \ref{fig:showcase}b. Nearby dendrogram leaves overlapping the rings in crowded areas were excluded from the ring masks. For each background ring, all spectra were ordered in increasing continuum levels, as measured in the wavelength range of $6320\AA \le \lambda \le6480\AA$, where no strong lines could be seen. This ordering was used to identify the bottom 25th$^{}$ percentile of the background spectra. The median of these was taken as the final representation of the local background and was subtracted from the corresponding dendrogram leaf. The underlying assumption here is that the diffuse background in each ring is representative of the background inside its corresponding dendrogram leaf, and that the convolution of the true image of a nebula with the point spread function (PSF) has become negligible at the location of the ring. 

The ring masks serve the additional purpose of representing the DIG regions. In a separate process, the spectra in each ring mask were summed to represent the local DIG. This method resulted in large gaps with sizes on the order of hundreds of spaxels between leaves and DIG rings, which remain unassigned to any of the binned regions and therefore outside of our analysis. The diffuse gas signal there is very low, and we binned them into bins that were as large as possible, resulting in an additional $113$ DIG regions with average equivalent radii of $27.8$ pc. These regions were manually added to the list of DIG regions. An example is shown in Figure \ref{fig:showcase}c, where these regions are shaded pink and fill the available space between ring regions.    

\begin{table}
\caption{Emission lines that were input to \textsc{pPXF}}\label{tab:ppxflines}
\begin{tabular}{ll}
\hline\hline
line & wavelength \\
     & \ \ \ [\AA] \\
\hline
H$\beta$ & 4861.320 \\
H$\alpha$ & 6562.791 \\
Pa20 & 8392.397 \\
Pa19 & 8413.318 \\
Pa18 & 8437.956 \\
Pa17 & 8467.254 \\
Pa16 & 8502.483 \\
Pa15 & 8545.383 \\
Pa14 & 8598.392 \\
Pa13 & 8665.019 \\
Pa12 & 8750.472 \\
Pa11 & 8862.782 \\
Pa10 & 9014.909 \\
Pa9 & 9229.014 \\
\ion{He}{i}\,5016 & 5015.678 \\
\ion{He}{i}\,5876$^*$ & 5875.624 \\
\ion{He}{i}\,6678 & 6678.152 \\
\ion{He}{i}\,7065 & 7065.231 \\
\ion{He}{i}\,7281 & 7281.351 \\
\ion{He}{ii}\,4686 & 4685.676 \\{}
[\ion{S}{ii}]6716 & 6716.44 \\{}
[\ion{S}{ii}]6731 & 6730.816 \\{}
[\ion{O}{ii}]7320 & 7319.99 \\{}
[\ion{O}{ii}]7330 & 7330.73 \\{}
[\ion{S}{iii}]6312 & 6312.06 \\{}
[\ion{N}{i}]5200 & 5199.837 \\{}
[\ion{Fe}{iii}]5270 & 5270.4 \\{}
[\ion{Ar}{iii}]7136 & 7135.79 \\{}
\ion{[Ar}{iii}]7751 & 7751.11 \\{}
[\ion{O}{iii}]4959 & 4958.911 \\{}
[\ion{O}{iii}]5007 & 5006.843 \\{}
[\ion{N}{ii}]6548 & 6548.05 \\{}
[\ion{N}{ii}]6584 & 6583.45 \\{}
[\ion{N}{ii}]5755$^*$& 5754.59 \\{}
\end{tabular}\\
$^*$: only for non-AO data
\end{table}
\subsection{Line extraction}\label{sec:stellarbkg}
We measured emission line fluxes of all sources using the \textsc{pPXF} Python program
\citep[v6.7.17][]{Cappellari2004,Cappellari2017}. The advantage of this program is that it can model
any residual stellar contribution to the extracted spectra, and fit emission lines in NGC\,300
as well as the residual telluric \ha{} at the same time. After testing various spectral
libraries of individual stars and simple stellar populations (SSPs), we settled on the
GALEV \citep{Kotulla2009} SSPs computed with the \citet{Munari2005} spectral library as created
by \citet{Weilbacher2018}.
As a second component, we input 32 typical lines of hydrogen and helium as well as significant
forbidden lines as listed\ in Table~\ref{tab:ppxflines} in the wavelength range of the MUSE data. We masked the wavelength ranges around
strong sky emission lines, and for AO, we also masked the region in the Na\,D gap. We excluded the [\ion{O}{i}]
lines from the fit because they are still in the masked area at the redshift of NGC\,300. All
object emission lines were fit as one component with consistent velocity, and hence all
contribute to the fit of the ionized gas kinematics simultaneously.
As a third component to the fit, we included an \ha{} line at near-zero redshift, which effectively
models the strong residual of the telluric \ha{} emission.
Because our main aim was to extract the emission line fluxes, we included a multiplicative polynomial
of degree 15 in the fit and iteratively fit each spectrum. The first iteration determines the
best fit and allows us to compute the residual noise. We used the residual noise in the wavelength
range 6050--6250\,\AA{} to scale the variance of each spectrum. This is useful because binning the
MUSE cubes causes the actual variance to be underestimated because of correlation across
spaxels (see \citealt{Weilbacher2020} and, e.g., \citealt{Herenz2020}). Improving the variance allowed us to obtain realistic
error estimates for the kinematics in a second iteration of the fit. We also used the scaled
residual noise to build a 6-pixel sliding standard deviation spectrum, which served as input to
100 Monte Carlo iterations of the emission line fit. We therefore expect to obtain realistic flux error estimates for all fitted emission lines.
As output of \textsc{pPXF,} we then obtained kinematics ($V$, $\sigma$) for both stars and ionized gas,
error estimates for them, and emission line fluxes and corresponding errors for every
spectrum.

\subsection{Sample sizes}\label{sec:sample_sizes}
The total number of dendrogram leaves is $661$. We discarded $68$ leaves whose spectra looked like those from emission line stars (compact regions, broad \ha, and none or extremely weak other emission lines), and an additional $45$ leaves that were completely noise dominated after the background subtraction. In Section \ref{sec:bpt} we describe the further cleaning of these leaves, which decreased the total number to $499$ dendrogram leaves.

The total number of DIG regions are $774$. Of these, $661$ were created by binary dilation of the leave masks, and $113$ were the manually added, large, and extremely faint DIG regions described in Section \ref{sec:defreg}. The further cleaning of the DIG regions, described in the next section, resulted in a total of $696$ DIG regions.

\section{Classification of emission line objects via a BPT diagram}\protect\label{sec:bpt}
To distinguish between \hii~regions, supernova remnants (SNR), and planetary nebulae (PNe), the traditionally used diagnostics are based on the relative strengths of the \ha, \nii, and the sum of [\ion{S}{ii}]$\lambda6716+$[\ion{S}{ii}]$\lambda6731$ (hereafter \siisum) emission lines. These diagnostics are defined in greater detail in \citet{Fesen1985}, \citet{Frew2010}, and \citet{Sabin2013}, for example, and were used for MUSE datacubes to identify PNe in \citep{Roth2021}.

%----------------------------------------------------------------- 
   \begin{figure*}
   \centering
   \includegraphics[width=17.0cm]{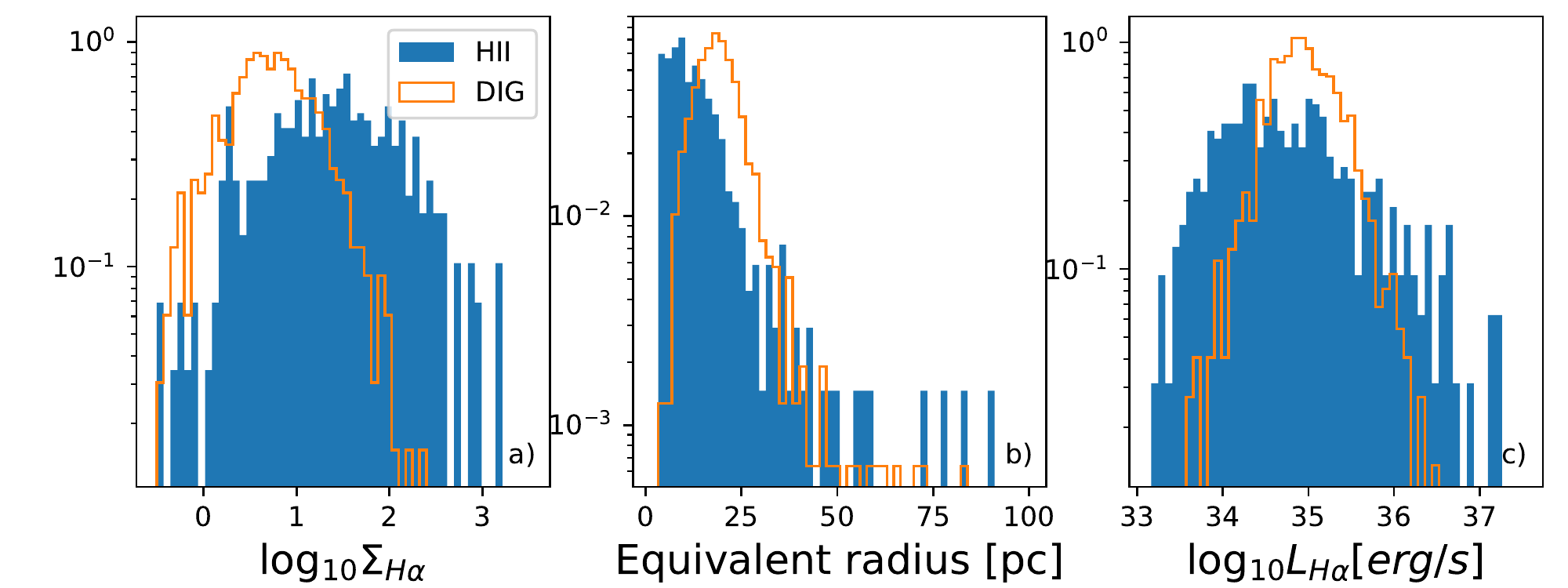}
      \caption{Normalized histograms of (a) the \ha~surface brightness $\Sigma_{H\alpha}$ [$10^{-20} $erg/s/cm$^2$/arcsec$^2$] of the \hii~and DIG dendrogram regions, (b) the equivalent radius [pc], and c) the \ha~luminosity [erg/s].}
         \label{fig:hist1}
   \end{figure*}
%----------------------------------------------------------------- 
%----------------------------------------------------------------- 
   \begin{figure}
   \centering
   \includegraphics[width=9.0cm]{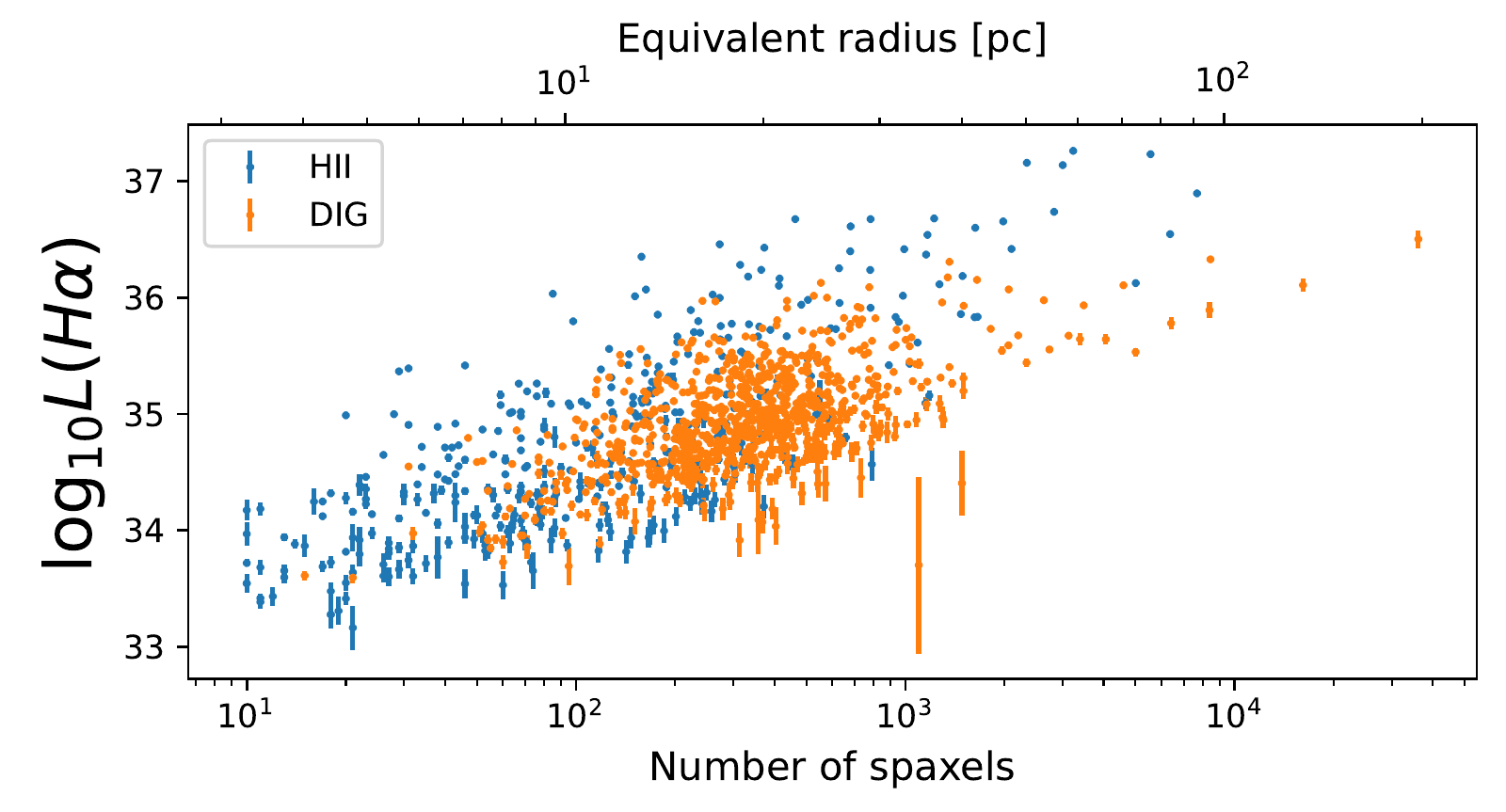}
      \caption{\ha~luminosity [erg/s] as a function of region size in spaxels (bottom x-axis) or equivalent radius in pc (top x-axis).}
         \label{fig:Lum}
   \end{figure}
%----------------------------------------------------------------- 
In Figure \ref{fig:bpt} we show the \citet{Baldwin1981} diagram (BPT) with respect to \siisum~ and \nii and the theoretical limit of star-forming regions of \citet{Kewley2006} and the PNe/SNR separation line \citep{Sabin2013}. These separation lines are typically used in the literature to distinguish between \hii, PNe, and SNR \citep[e.g.,][]{Frew2010,Sabin2013}. The \citet{Kewley2006} line is the revised extreme starburst line of \citet{Kauffmann2003}, obtained via a semi-empirical fit to the outer bound of galaxy spectra. To alleviate the concern that \hii~regions may not share the same properties as star-forming galaxies, and hence that the Kewley line is not an appropriate separator, we also plot the line from \citet{Dopita2000}, which is based on instantaneous burst modeling of \hii~regions. The figure shows that the Kewley and Dopita lines are very similar in the \nii/\ha~diagram. We therefore performed the formal separation of the objects into \hii, PNe, and SNR in the \nii-BPT diagram, but we kept in mind that their location with respect to \siisum/\ha~may vary. This is shown in Figure \ref{fig:bpt}, where objects classified based on \nii/\ha~(top panel) cross over the separation lines in the \siisum/\ha\text{ } diagram (bottom panel), highlighting the continuous nature of the sample properties. While our goal is to obtain a sample of \hii~regions, we are aware that the emission line properties comprise a continuous distribution, and hence we cannot claim that the blue data points in Figure \ref{fig:bpt} constitute a pure \hii-region sample.  

Without cleaning the sample, we obtain $429$ \hii-regions, $96$ SNR, and $23$ PNe over all nine fields with the \nii/\ha~classification.  With the \siisum/\ha~classification, the numbers instead are $334$, $203$, and $17$, respectively.  The numbers do not add up to the same total because some dendrogram leaves have a low S/N in \nii and some in \siisum, and so the number of points in the top and bottom panels of Figure \ref{fig:bpt} is different. Next, we adjusted the labels for five extended bubble and filament-like structures in fields C, D, and P, which otherwise fall in the PNe region in the \nii-based BPT diagram. Because PNe are compact objects, we assigned either an \hii~ or SNR label to these structures, depending on their label in the \siisum-based BPT diagram. The assigned SNR labels are consistent with the $log_{10}$(\siisum/\ha)$>-0.5$ criterion used to separate PNe from SNR in \citet{Roth2021}. We compared our resulting labels to the list of identified PNe in \citet{Roth2018} and find that while most of the $22$ objects in common have matching PNe labels, $6$ of the objects are not identified as PNe by the \nii-based BPT diagram. We visually examined these dendrogram leaves to confirm that they are small, compact, and somewhat symmetrical regions, and manually relabeled these regions as PNe. The number of PNe detections is consistent with the more targeted search for such objects in Paper IV of this series, which is concerned with the planetary nebula luminosity function in NGC 300 (Soemitro et al., in prep.).  We  filtered the remaining sample to discard all spectra with an $S/N<5$ in \ha~and the [\ion{S}{ii}]$\lambda\lambda6716,6731$ (hereafter, \siidoublet) doublet. Because the MUSE pointings purposefully avoid bright \hii~regions \citep{Roth2018}, the population of \hii~regions that we observe is likely older, fainter, and not very strongly ionized. We therefore did not add a selection criterion based on \oiii~emission. Finally, we visually examined the spectrum of each region and the pPXF fit to the emission lines and discarded any remaining clearly failed fits. The final numbers are $390$ \hii-regions, $82$ SNR, and $27$ PNe over all nine fields, based on the \nii-BPT diagram. 

The DIG regions were cleaned separately by removing spectra that were too noisy, resulting in either a failed pPXF fit, or in large uncertainties consistent with non-detections in the brightest forbidden lines. This manual inspection removed $78$ regions, leaving a total of $696$ DIG regions. 

The spatial distribution of objects is shown in Figure \ref{fig:2Dclass}. The \hii~regions can be found in every pointing, with a noticeable dearth of regions in the inter-arm field \fieldj. The DIG is also ubiquitous, but again, field \fieldj~is the exception. The SNR are mostly found in the central field \fielda, with a possible second number density peak in the inter-arm field \fieldj, trailing the spiral arm that goes through fields \fieldc, \fieldd, and \fieldi. We note that field \fielde~only partially covers the spiral arm, and partially covers another inter-arm region, which likely accounts for the lack of detected regions in the upper part of \fielde, which overlaps with the inter-arm region.   

In Appendix \ref{sec:ml} we attempt to classify the emission line objects into \hii, SNR, and PNe via machine learning. We conclude, however, that an unsupervised learning algorithm such as UMAP and a labeling algorithm such as HDBscan are unable to separate the objects into the desired groups when the feature sets listed in Table \ref{tab:mlfeatures} are used.  
%-------------------------------------------------------------
%                 A figure as large as the width of the column
%-------------------------------------------------------------
   \begin{figure}
   \centering
   \includegraphics[width=\hsize]{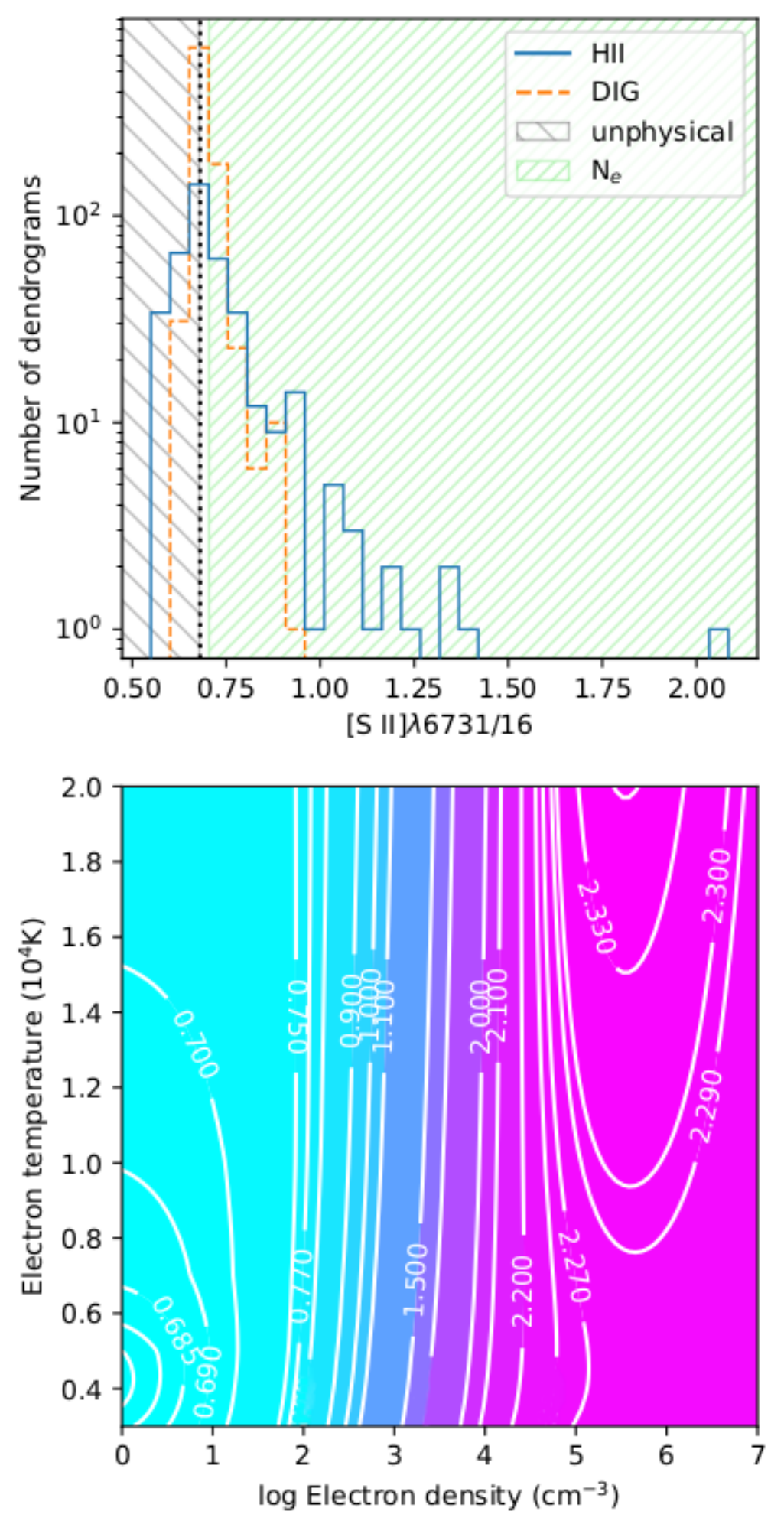}
      \caption{Sample electron densities. (Top) Histogram of the \siirat~ratio for \hii~(solid blue lines) and DIG (dashed orange lines). The area hatched with backslashes represents nonphysical values, and the forward slash covers the range of \den~that is fairly invariant with \te. (Bottom) \siirat~emissivity grid for $3$kK $\le \rm T_{\rm e} \le 20$ kK and $1\le \log_{10}$ \den [cm$^{-3}$] $\le7$. Contours of \siirat~are overplotted in white, and the corresponding value is shown inline. 
              }
         \label{fig:temden}
   \end{figure}
   
\section{Physical properties}\protect\label{sec:prop}
\subsection{Brightness and size}
The \ha~surface brightness $\Sigma_\mathrm{H\alpha}$ of the \hii~dendrograms ranges between $3\times10^{-21}$ and  $1.6\times10^{-17}$ erg s$^{-1}$ cm$^{-2}$ arcsec$^{-2}$, with a mean and median of $8.0\times10^{-19}$ and  $2.5\times10^{-19}$ erg s$^{-1}$ cm$^{-2}$ arcsec$^{-2}$, respectively, as illustrated by the histograms in Figure \ref{fig:hist1}a. We obtained their equivalent radius via $r_\mathrm{equiv} = \sqrt{N_\mathrm{pix}/\pi}$, shown in Figure \ref{fig:hist1}b. The radius ranges between $r_\mathrm{equiv}\in[3.3\rm{pc},91\rm{pc}]$, with a mean and median of $14.3$, and $11.7$ pc, respectively. The diffuse regions are about eight times fainter on average, with an \ha~surface brightness $\in[1.4\times10^{-22}, 2.2\times10^{-18}]$ erg s$^{-1}$ cm$^{-2}$ arcsec$^{-2}$, with a mean and median of $1.0\times10^{-19}$ and $5.2\times10^{-19}$ erg s$^{-1}$ cm$^{-2}$ arcsec$^{-2}$, respectively. They are also larger on average by design because they are constructed via dilated masks, $r_\mathrm{equiv}\in[4\rm{pc},197\rm{pc}]$, with a mean and median $20$ and $18.8$ pc, respectively.  

In terms of \ha~luminosity, the \hii~regions range between $\log_{10}L_\mathrm{H\alpha} [\text{erg}/\text{s}]\in(33.2, 37.3)$, and have a mean and median of $34.8$ and $34.7$, respectively. This makes our \hii~sample one of the faintest in the literature to date. The DIG, represented by the ring masks in Section \ref{sec:stellarbkg}, is similarly faint, with $\log_{10}L_\mathrm{H\alpha} [\text{erg}/\text{s}]\in(33.6, 36.5)$, and a mean and median of $35.0$ and $34.9$, respectively. This is visualized in Figure \ref{fig:hist1}c. In Figure \ref{fig:Lum} we display the luminosity-size relation. We note that the median luminosity of $34.9$ corresponds to regions that are $\gtrsim100$ spaxels in size ($\gtrsim 10$ pc equivalent radius), and hence we conclude that the extreme faintness of the sample cannot be the result of random peaks in the noise and must instead be due to the physically small sizes.  

Although we do not analyze SNR and PNe regions in this paper, we mention the range of their \ha~luminosities for completeness. For PNe, $\log_{10}L_\mathrm{H\alpha} [\text{erg}/\text{s}]\in(33.1, 36.1)$, with a mean equal to the median of $34.6$. For SNR, $\log_{10}L_\mathrm{H\alpha} [\text{erg}/\text{s}]\in(33.1, 36.7)$, with a mean and median $34.5$ and $34.3$, respectively.

%
%                                                One column figure
%----------------------------------------------------------------- 
   \begin{figure*}[ht!]
   \centering
   \includegraphics[width=17.5cm]{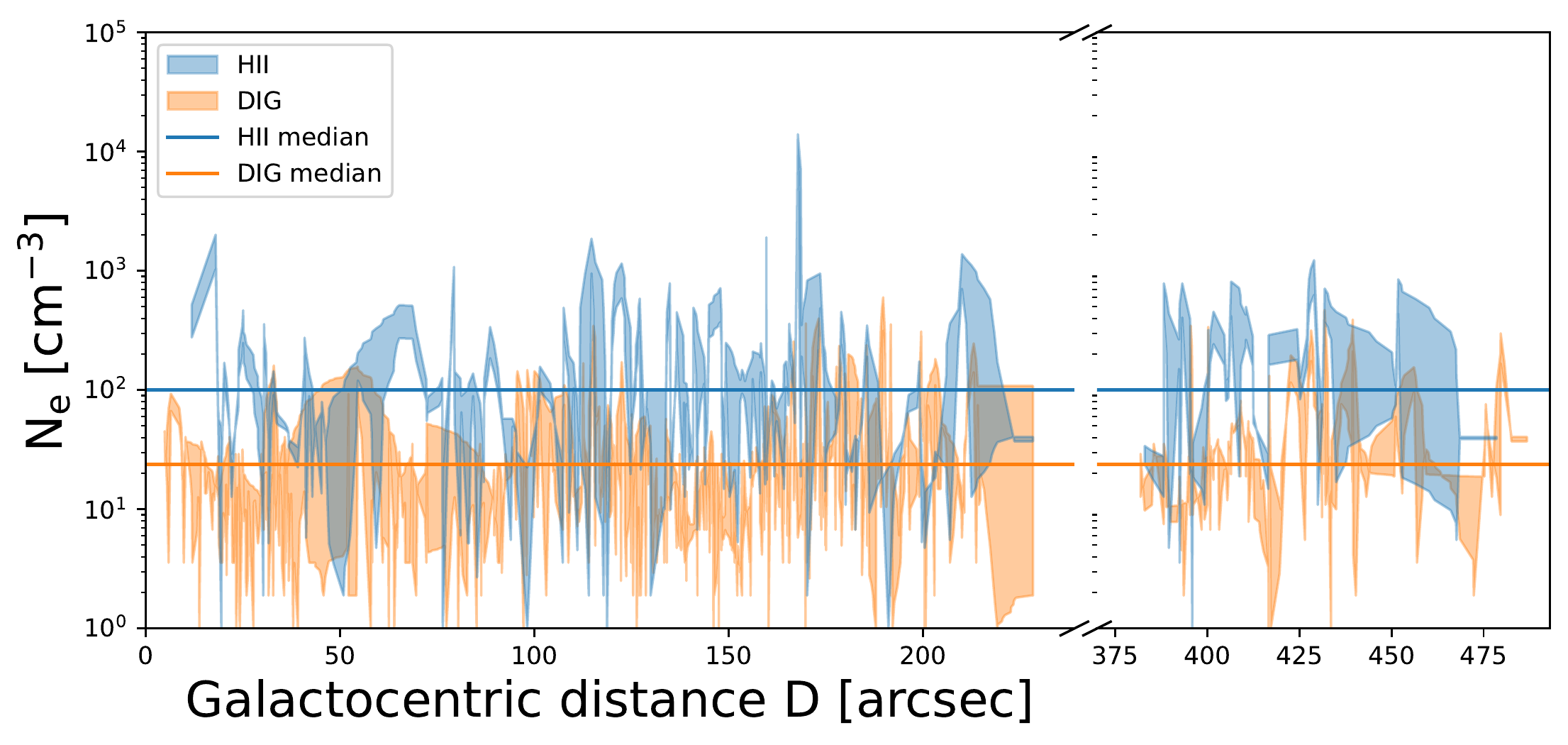}
      \caption{Electron density \den~as a function of galactocentric distance D for \hii~(blue) and DIG regions (orange). At each D, the width of the \den~range is shown, obtained from \den(\te$=20$ kK) and \den(\te$=4$kK). The medians for \hii~and DIG are also shown with solid lines.  }
         \label{fig:ne}
   \end{figure*}
%----------------------------------------------------------------- 
%
%                                                One column figure
%----------------------------------------------------------------- 
   \begin{figure*}[h!]
   \centering
   \includegraphics[width=17.5cm]{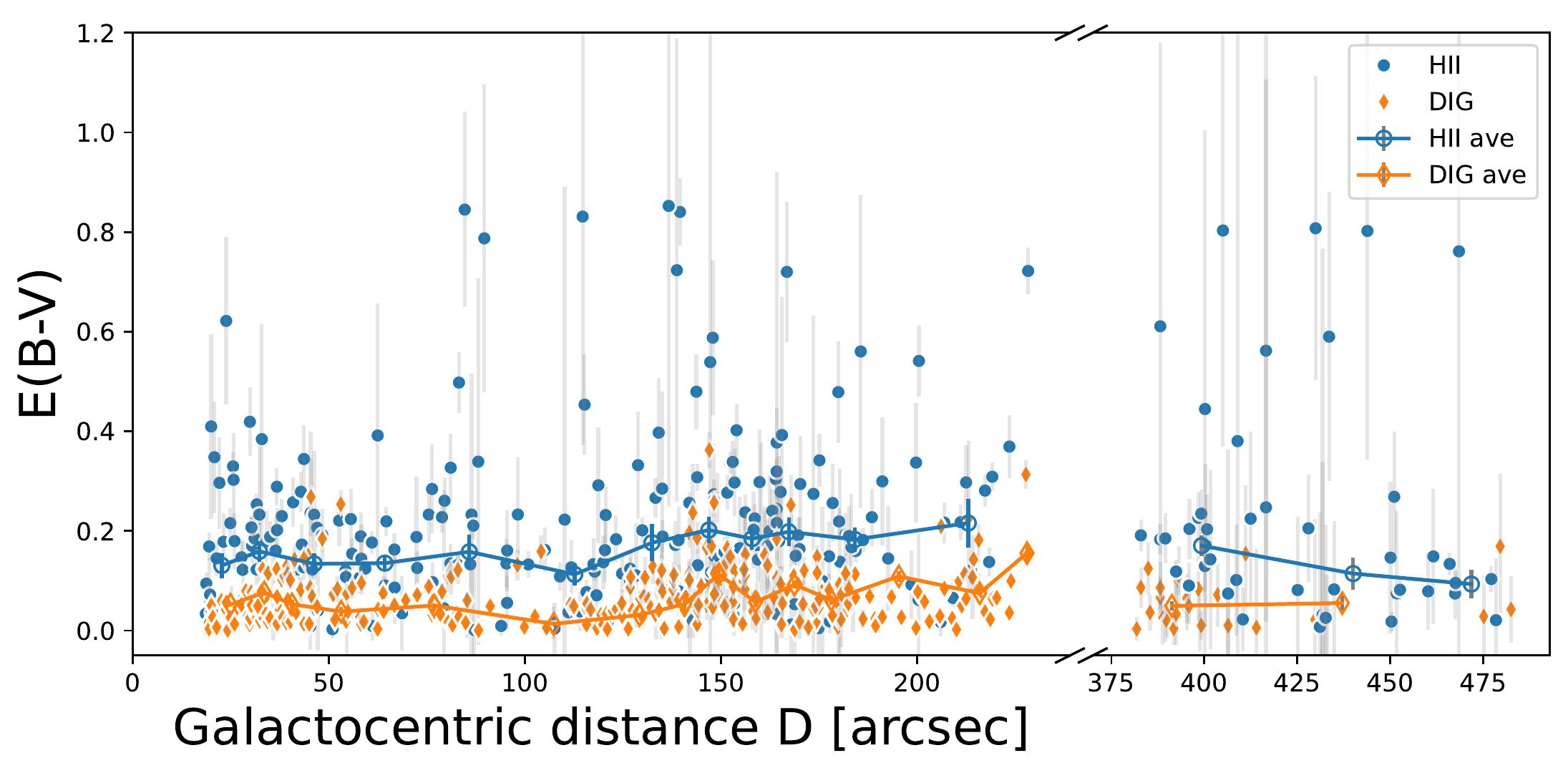}
   \includegraphics[width=17.5cm]{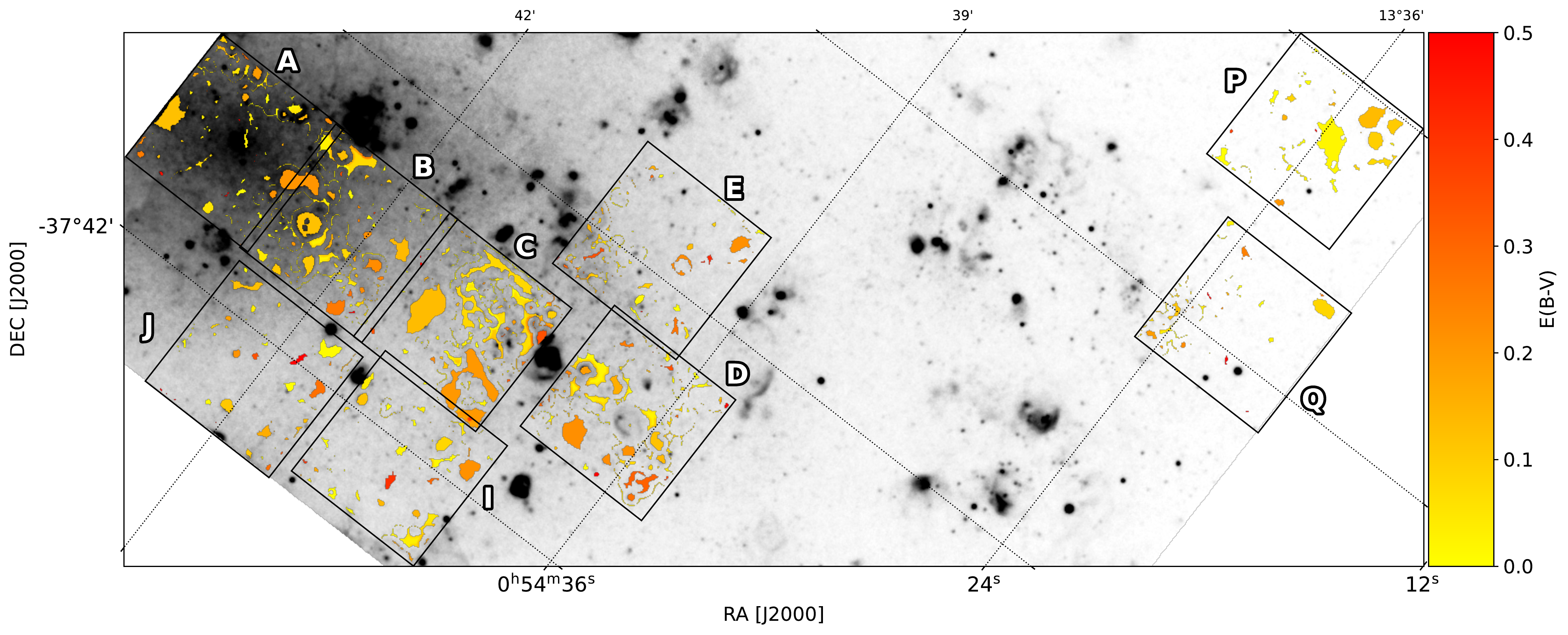}
      \caption{Sample nebular extinction. (Top) $E(B-V)$ as a function of galactocentric distance D for individual \hii~(black dots) and DIG regions (blue diamonds). Uncertainty-weighted averages in bins of $20$ objects are also shown in red circles for \hii~and in orange diamonds for DIG. The propagated error bars are often smaller than the markers. The x-axis position of each uncertainty-weighted average is the average galactocentric distance for the dendrograms inside each bin. (Bottom) 2D map of $E(B-V)$. }
         \label{fig:ebv}
   \end{figure*}
%----------------------------------------------------------------- 
\subsection{Electron temperature and density}\protect\label{sec:temden}
Unfortunately, the auroral [\ion{S}{iii}] $\lambda6312$ and [\ion{N}{ii}] $\lambda5755$ lines in our data are so weak as to be completely in the noise of individual regions, and hence we cannot directly map an [\ion{S}{iii}]- or [\ion{N}{ii}]-based estimate of the electron temperature \te. This is likely due to the near-solar metallicity of NGC\,300 \citep[e.g.,][]{Stasinska2013}. The [\ion{O}{iii}] $\lambda4363$ line is outside of the wavelength coverage, and in addition the [\ion{O}{ii}] doublet at $\lambda\lambda 7320,7330$ is also extremely noisy. The ratio \siirat~is an electron density (\den) tracer and is well detected in the data.  

For calculations of \den~, we used the PyNeb \citep{Luridiana2015,Morisset2020} software package, with sulfur atomic data from \citet{Podobedova2009} and collision strengths from \citet{Tayal2010}. In Figure \ref{fig:temden} we show the \siirat~distribution of the \hii~and DIG regions, as well as the theoretical \siirat~emissivity grid as a function of \te~and \den. The theoretical limits of the \siirat~ratio are $0.6806\lesssim$ \siirat $\lesssim2.351$. Because the observed distribution peaks at the theoretical minimum, implying very low densities for most of the regions (Figure \ref{fig:temden}, upper panel), we would expect some regions to fall left of this minimum due to statistical scatter. In Appendix \ref{sec:S2S2} we motivate our choice to include \siirat~down to 0.548 for \hii~regions and 0.638 for the DIG.

The theoretical model in the lower panel of Figure \ref{fig:temden} is degenerate in \te~for \siirat~between approximately $0.706$ to $2.27$, but the \siirat~contours are fairly straight, and hence we can give an acceptably narrow range of possible \den~values from the \siirat~ratio alone. The peak of the \siirat~histogram falls in the small range of $0.6806\lesssim$\siirat$\lesssim0.70$, where the contours curve with \te. This range is highly sensitive to the accuracy of the atomic data, and hence here we can only say that \den$\lesssim30$ cm$^{-3}$, but are unable to make a statement about \te.  

Figure \ref{fig:ne} shows the estimated range in \den~as a function of galactocentric distance for \hii~and DIG regions. \citet{Toribio2016} reported electron temperatures in the range of $7\mbox{-}10$ kK for a handful of \hii~regions in NGC 300, but they are much larger and brighter than our sample. We therefore did not constrain the temperature range based on their results and instead assumed that all of our \hii~regions lie within a temperature range of \te$=4\mbox{-}20$ kK, which is hardly constraining. This resulted in a range of possible \den, represented by a blue-shaded area between the low and high \den~limits for each data point. The same was done for the DIG regions, but shaded orange in the figure. We also computed the average \den~between the extremes for each object and from these values the medians of the full samples, with \den$\approx 100$ cm$^{-3}$ for \hii~(solid blue line) and \den$\approx 23$ cm$^{-3}$ for DIG (solid orange line).

\begin{table}
    \caption{Strong-line method comparison of $12+\log_{10}{O/H}$.}
    \label{tab:z}
    \centering
    \begin{tabular}{ccc}
    \hline\hline
    Field& O3N2$^{\dagger}$ & S$^{\ddagger}$\\
    \hline
    \multicolumn{3}{c}{\hii}\\
    A & $8.593\pm 0.021$& $8.358\pm 0.014$\\
    B & $8.581\pm 0.012$& $8.345\pm 0.010$\\
    C & $8.530\pm 0.011$& $8.286\pm 0.013$\\
    D & $8.539\pm 0.018$& $8.284\pm 0.022$\\
    E & $8.528\pm 0.011$& $8.261\pm 0.009$\\
    I & $8.499\pm 0.014$& $8.288\pm 0.010$\\
    J & $8.495\pm 0.016$& $8.302\pm 0.012$\\
    P & $8.379\pm 0.018$& $8.237\pm 0.018$\\
    Q & $8.501\pm 0.021$& $8.106\pm 0.025$\\
    All &$8.530\pm 0.006$& $8.295\pm 0.006$\\
    \multicolumn{3}{c}{DIG}\\
    A &$8.469\pm0.003$& $8.521\pm0.014$\\
    B &$8.519\pm0.002$& $8.405\pm0.007$\\
    C &$8.495\pm0.004$& $8.365\pm0.007$\\
    D &$8.478\pm0.004$& $8.355\pm0.009$\\
    E &$8.521\pm0.005$& $8.293\pm0.006$\\
    I &$8.463\pm0.003$& $8.373\pm0.006$\\
    J &$8.512\pm0.010$& $8.453\pm0.025$\\
    P &$8.378\pm0.006$& $8.171\pm0.006$\\
    Q &$8.469\pm0.004$& $8.026\pm0.005$\\
    All&$8.482\pm 0.002$& $8.379\pm 0.005$\\
    \hline
    \end{tabular}
        {\\$\dagger$ - \citet{Marino2013}, $\ddagger$ - \citet{Pilyugin2019} \par}
\end{table}

%----------------------------------------------------------------- 
%
%                                                One column figure
%----------------------------------------------------------------- 
   \begin{figure*}
   \centering
   \includegraphics[width=17.5cm]{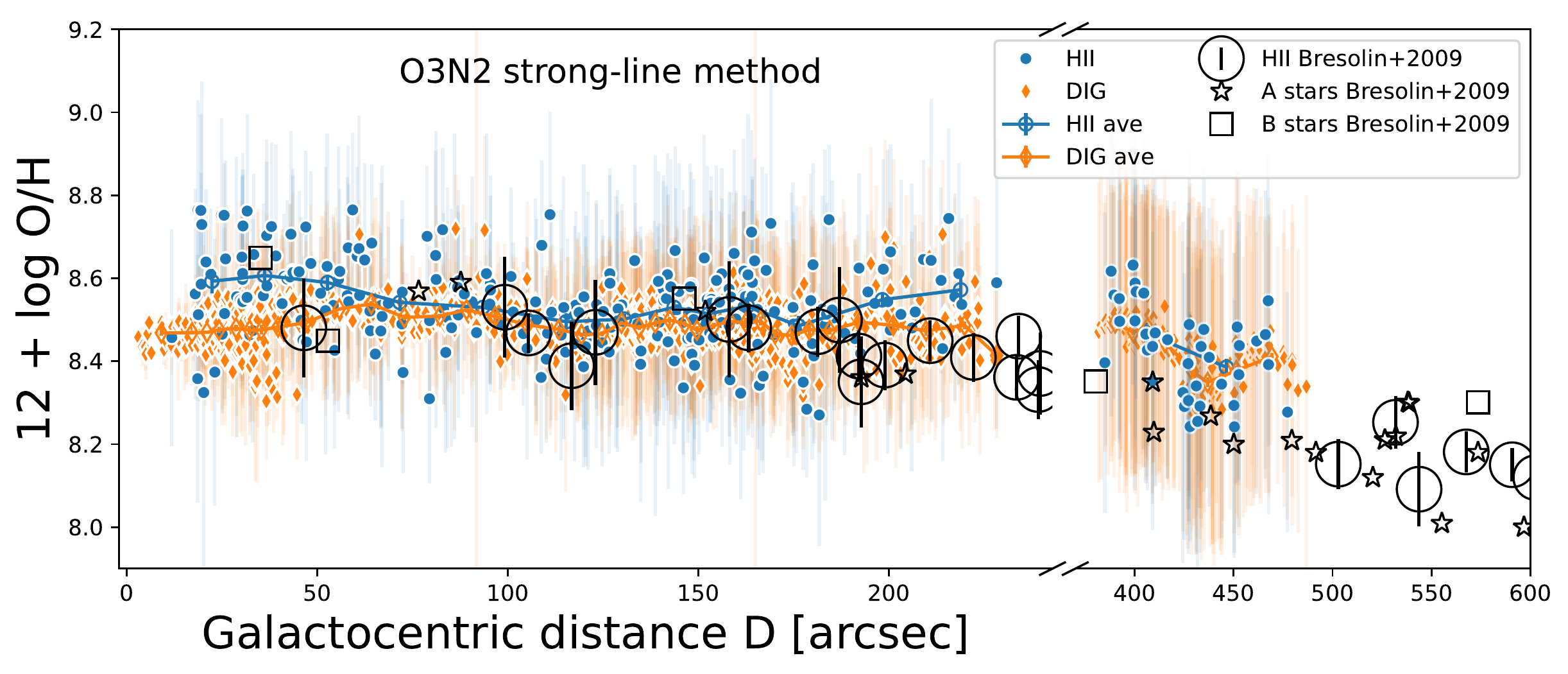}
   \includegraphics[width=17.5cm]{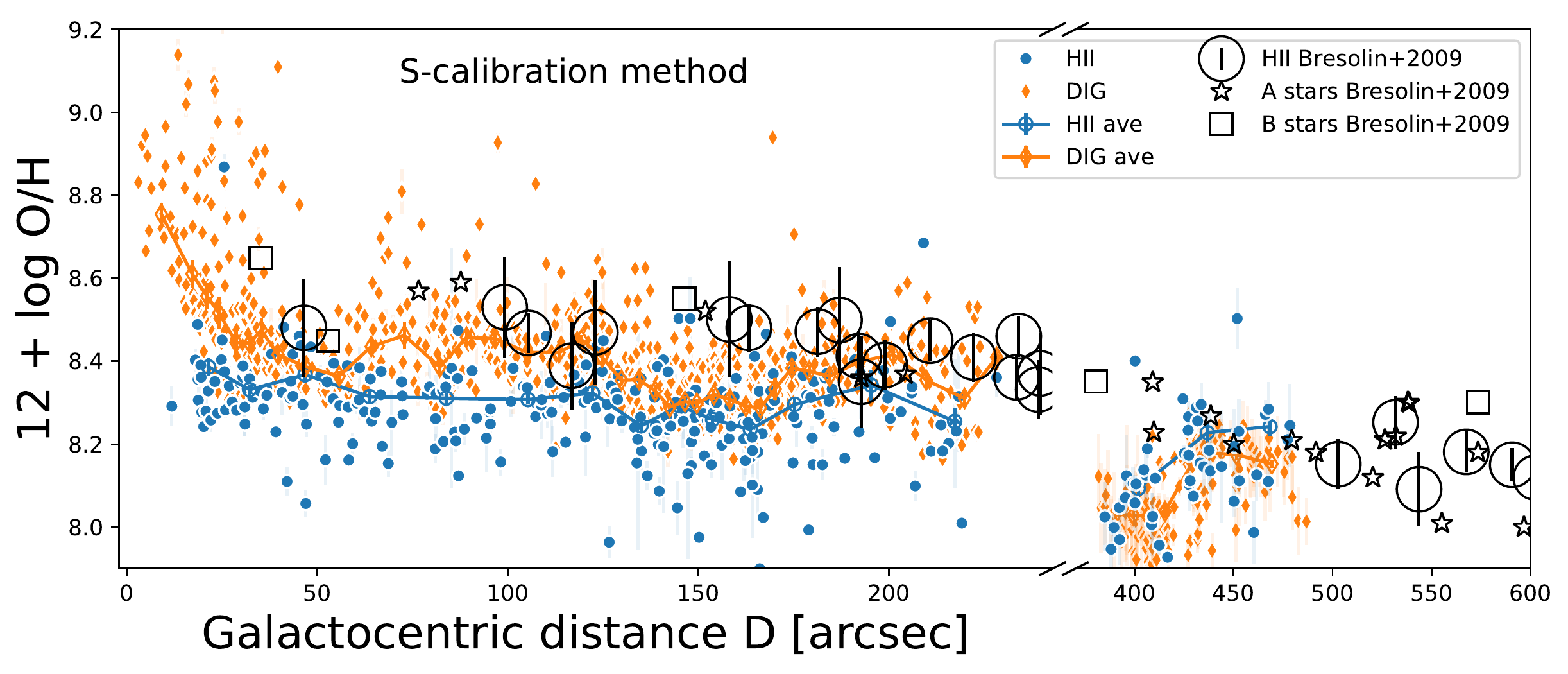}
      \caption{Sample metallicity. (Top) \metal~as a function of galactocentric distance D, using the \citet{Marino2013} O3N2 calibration. Same legend as in Figure \ref{fig:ebv}, except for the data by \citet{Bresolin2009} (open gray circles), which were obtained via the direct-\te\  method for NGC\,300. (Bottom) Same as top, but using the \citet{Pilyugin2016} S-calibration.}
         \label{fig:z}
   \end{figure*}
%----------------------------------------------------------------- 
%----------------------------------------------------------------- 
%
%                                                One column figure
%----------------------------------------------------------------- 
   \begin{figure*}
   \centering
   \includegraphics[width=17.5cm]{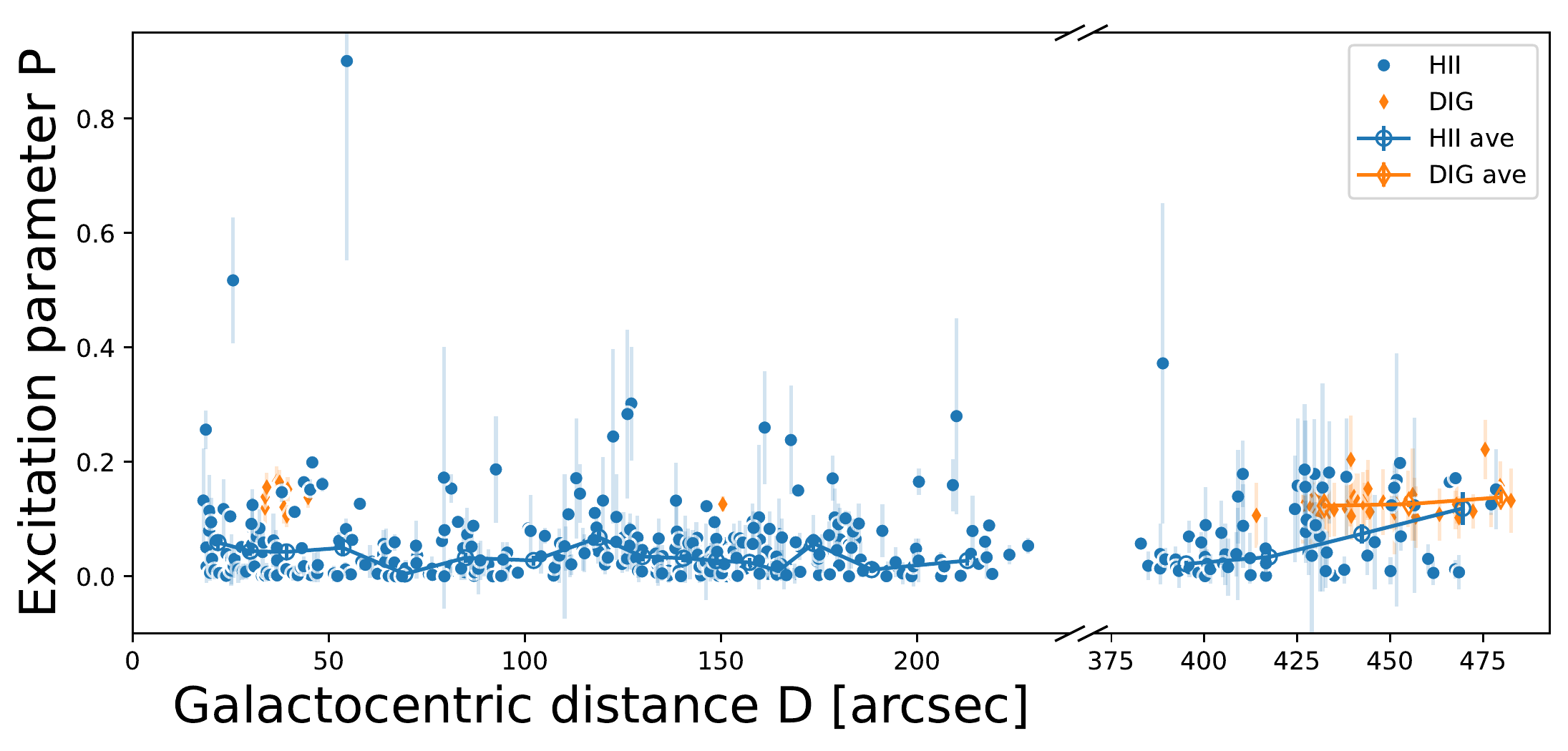}
   \caption{Excitation parameter P$\equiv S_3/(S_2 + S_3)$ as a function of galactocentric distance D. }
         \label{fig:P}
   \end{figure*}
%----------------------------------------------------------------- 
\subsection{Extinction}\protect\label{sec:ebv}
The wavelength range covers both \ha~and \hb, allowing us to estimate the dust attenuation via the Balmer decrement. \citet{Roussel2005} investigated extinction variations in NGC 300 and reported that compact and diffuse \hii~regions may be associated with very different extinction laws. However, $80\%$ of the \hii~regions in their sample are larger than $90$ pc in diameter, and $100\%$ are larger than $55$ pc, while $93\%$ of our sample are smaller than $30$ pc in projected diameter, and $98\%$ are smaller than $55$ pc. It is therefore not immediately obvious whether we can apply their findings in the choice of extinction law.  Furthermore, \citet{Tomicic2017} found that a weakly or nonattenuated DIG affects the global attenuation. Global attenuation is not included in most dust-gas geometry models. We therefore chose the \citet{Cardelli1989} extinction law, which is applicable to the dense and to the diffuse ISM, with the default value of $R_V=3.1$.  

Figure \ref{fig:ebv} shows $E(B-V)$ as a function of galactocentric distance D for individual \hii~and DIG regions. By propagating the uncertainties of the objects, we also constructed the uncertainty-weighted average $E(B-V)$ in bins containing 20 objects each for both DIG and \hii,  which show a similar fluctuation in the data, but with a narrower range. While individual regions show highly variable $E(B-V)$, the general trend is for a flat radial profile for both \hii~and DIG, with the former being $\sim0.1$ mag redder than the latter.

Because the 1D plot with galactocentric distance erases any possible azimuthal trends in the fields, we also show in the bottom panel of Figure \ref{fig:ebv} the 2D map across all pointings and for all objects with derivable $E(B-V)$, including PNe and SNR. We see very little measurable variation across all pointings, but individual \hii~or SNR regions occasionally reach higher extinction values. The Balmer decrements of most of the DIG are consistent with the theoretical value, and hence an $E(B-V)=0$. These values are omitted. Apparently, only DIG regions in close proximity to \hii~regions have $E(B-V)>0$. The predominantly low DIG extinction values support the typical assumption in the literature that the DIG suffers no extinction \citep[e.g.,][]{Belfiore2022}.

On a final note, the range of extinction values for our \hii~regions is consistent with the value of \citet{Roussel2005}, who obtained an A(\ha)$=0.6$ mag average extinction with RMS$=0.64$ mag for $23$ \hii~regions in NGC 300, which compares well to our A(\ha)$=0.53$ mag average with RMS$=0.67$.

  \begin{figure*}
   \centering
   \includegraphics[width=17.5cm]{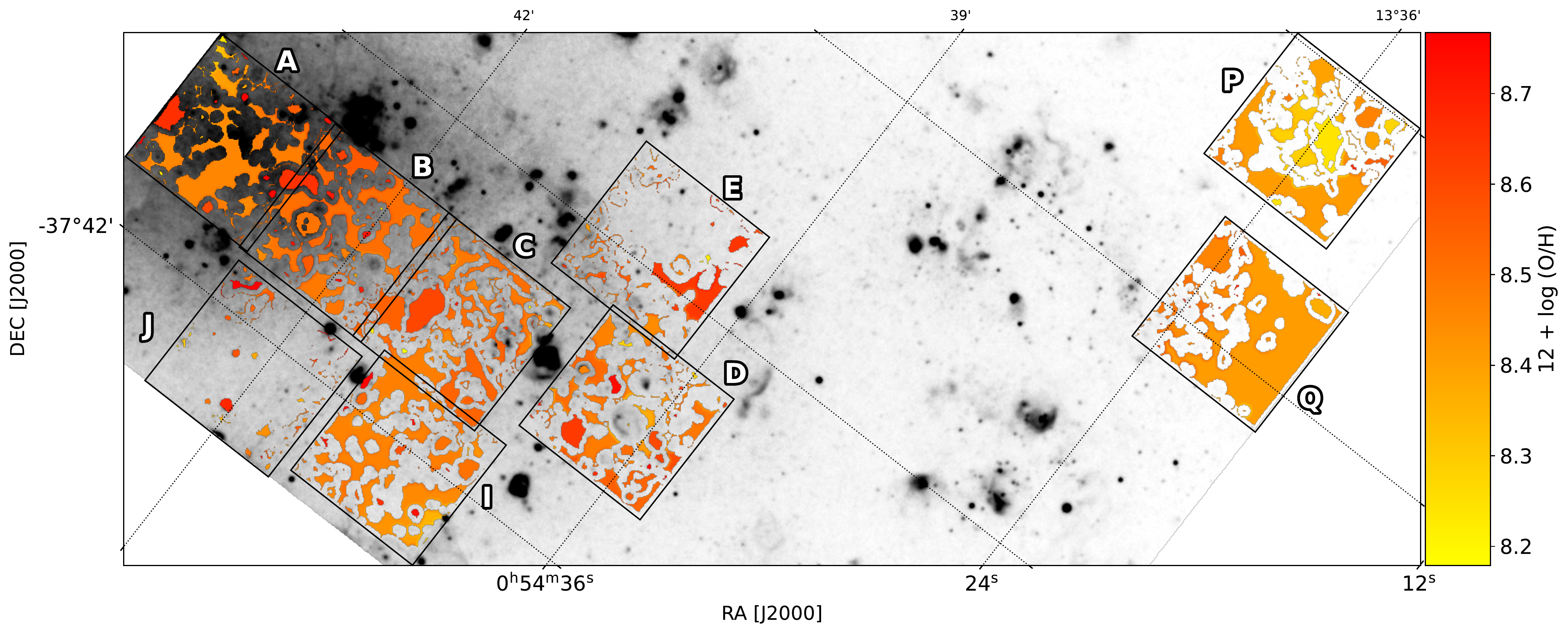}
      \caption{2D map of $12+\log_{10}{O/H}$ from the O3N2 calibration for \hii~and DIG regions. }
         \label{fig:2Dz}
   \end{figure*}

\subsection{Metallicity}\protect\label{sec:z}
The lack of detections in the auroral lines of [\ion{S}{iii}]$\lambda6312$ and [\ion{N}{ii}]$\lambda5755$ implies that we cannot use the so-called direct method, which is based on the electron temperature \te, to estimate the metallicity across our \hii~and DIG samples. Instead, we have to resort to the strong-line method based on [\ion{O}{iii}] and [\ion{N}{ii}] that was first introduced by \citet{Alloin1979}, which was later recalibrated by \citet{Pettini2004}. We used the more recent recalibration by \citet{Marino2013}. The O3N2 index is defined as O3N2$\equiv \log_{10}{ \{ ([\ion{O}{iii}]\lambda5007 / \rm H\beta  ) / ([\ion{N}{ii}]\lambda6583 / \rm H \alpha)\} }$, and is valid for the range $-1< \rm{O3N2}<1.9$. This is a suitable index for our data because NGC\,300 is fairly metal rich in the low electron density regime, and all of these lines are well detected in our \hii~and DIG samples.

Figure \ref{fig:z} shows the resulting \metal~as a function of galactocentric distance D for individual \hii~and DIG regions, as well as the uncertainty-weighted averages in bins containing 20 objects each. The O3N2 calibration by \citet{Marino2013} was obtained for \hii~regions and hence may not automatically apply for the DIG. However, by showing that chemically homogeneous \hii-DIG pairs show minimum offset ($0.01\mbox{-}0.04$ dex) and little dispersion in the metallicity differences ($0.05$ dex) when using the O3N2 diagnostic, \citet{Kumari2019} concluded that the O3N2 metallicity calibration for \hii~regions can be used for DIG regions as well. A comparison of the \hii~and DIG region metallicity is therefore still meaningful, and we overplot the DIG results in Figure \ref{fig:z}.

We observe a flat metallicity gradient for $D\lesssim245$ arcsec $\approx2.2$ kpc. We have propagated the individual uncertainties of each line to the O3N2 index and then to the metallicity. \citet{Bresolin2009} suggested that strong-line diagnostics such as O3N2 (and N2) consistently underestimate the slope of abundance gradients because they ignore the effects of the ionization parameter, as extensively demonstrated by \citet{Pilyugin2016}. Therefore, as a sanity check, we overplot the NGC\,300 \metal~measurements by \citet{Bresolin2009} in the figure. These authors used auroral lines to obtain the \hii~electron temperature and then the metallicity via the so-called direct method. We note that our data show a flat gradient only in the central region, while the observed metallicity beyond  $D\gtrsim400$ arcsec $\approx3.6$ kpc seems consistent with the gradient suggested by the Bresolin data points at $D\gtrsim500$ arcsec. Via single linear fit to their data, these authors found a y-intercept of \metal$=8.57\pm0.02$. A linear regression to our data gives a similar y-intercept of \metal$=8.58\pm0.01$, which is consistent with the Bresolin value. Furthermore, we note that within the uncertainties shown in Figure \ref{fig:z}, our measurements are fully consistent with those of Bresolin at all galactocentric distances, not just in the y-intercept or at large distances. When it is superimposed on our data, the \citet{Bresolin2009} direct method metallicity measurements themselves appear to be also consistent with a flat inner disk gradient. Similarly, the results in \citet{ToribioSanCipriano2016} are consistent with those of \citet{Bresolin2009} in terms of oxygen abundance measurements of giant \hii~regions in NGC 300. We saw no indication of a steepening of the metallicity gradient in the inner disk, as suggested by \citet{VilaCostas1992}. In both \citet{VilaCostas1992} and \citet{Bresolin2009}, the number of data points near the galactic center ($D<1$ kpc $\approx110$ arcsec) is much lower than in our data, which could explain the different conclusions reached by these authors.  A flatter metallicity gradient for NGC 300 was also reported by \citet{Stasinska2013}, who used the direct $T_e$ method and $37$ giant and compact \hii~regions, most of which are at comparable galactocentric distances as our sample ($\leq 0.5 R_{25}$ in their figure 8 corresponds to $\leq292.5$ arcsec in our Figure \ref{fig:z}). 

However, we must also point out that the intrinsic uncertainty in the O3N2 index is on the order of $0.02$ dex \citep{Marino2013}. Such a large uncertainty range would easily be able to accommodate a metallicity gradient. The inherent insensitivity of the O3N2 index to variations in what \citet{Pilyugin2016} called the "excitation parameter P" is perhaps reason enough to dismiss the observed flat metallicity gradient in Figure \ref{fig:z} as an artifact of the methodology. We refer to this parameter as P$_O$, where the subscript indicates that it is based on oxygen. It is possible that O3N2 consistently underestimates the slope only in the central disk regions ($D\lesssim245$ arcsec $\approx2.2$ kpc) where P$_O$ might vary rapidly, and not at larger distances where it is fairly constant. We therefore also examined the resulting metallicity when the so-called S-calibration \citep{Pilyugin2016} was used, based on R3$\equiv  (\rm I_{ [\ion{O}{iii}](\lambda4959 + \lambda5007)} / \rm I_{H\beta}  )$, S2$\equiv  (\rm I_{ [\ion{S}{ii}](\lambda6717 + \lambda6731)} / \rm I_{H\beta}  )$, and N2$\equiv  (\rm I_{ [\ion{N}{ii}](\lambda6548 + \lambda6584)} / \rm I_{H\beta}  )$. This is shown in the lower panel of Figure \ref{fig:z}. Consistent with our previous findings, the \hii~regions continue to display a flat metallicity gradient in the central fields, albeit at $\sim0.2$ dex lower metallicity. When the direct-\te~metallicity of \citet{Bresolin2009} is considered as the true metallicity, then the S-calibration systematically underestimates the true metallicity for \hii~regions. For the DIG, we now instead observe a steepening of the gradient at very small D$\lesssim25$ arcsec. In the outer fields at D$\gtrsim400$ arcsec, the results from the S-calibration method suggest that the metallicity for both \hii~and DIG regions has dipped lower than for the inner regions, and steadily increases until it reaches the \citet{Bresolin2009} direct-\te~values for \hii-regions. The A and B star metallicity measurements of \cite{Bresolin2009} suggest no such rapid increase in the outer fields, however. We summarize the obtained metallicity values per field in Table \ref{tab:z}, where we list the formal uncertainties obtained through error propagation.

\begin{figure}[h!]
   \centering
   \includegraphics[width=8cm]{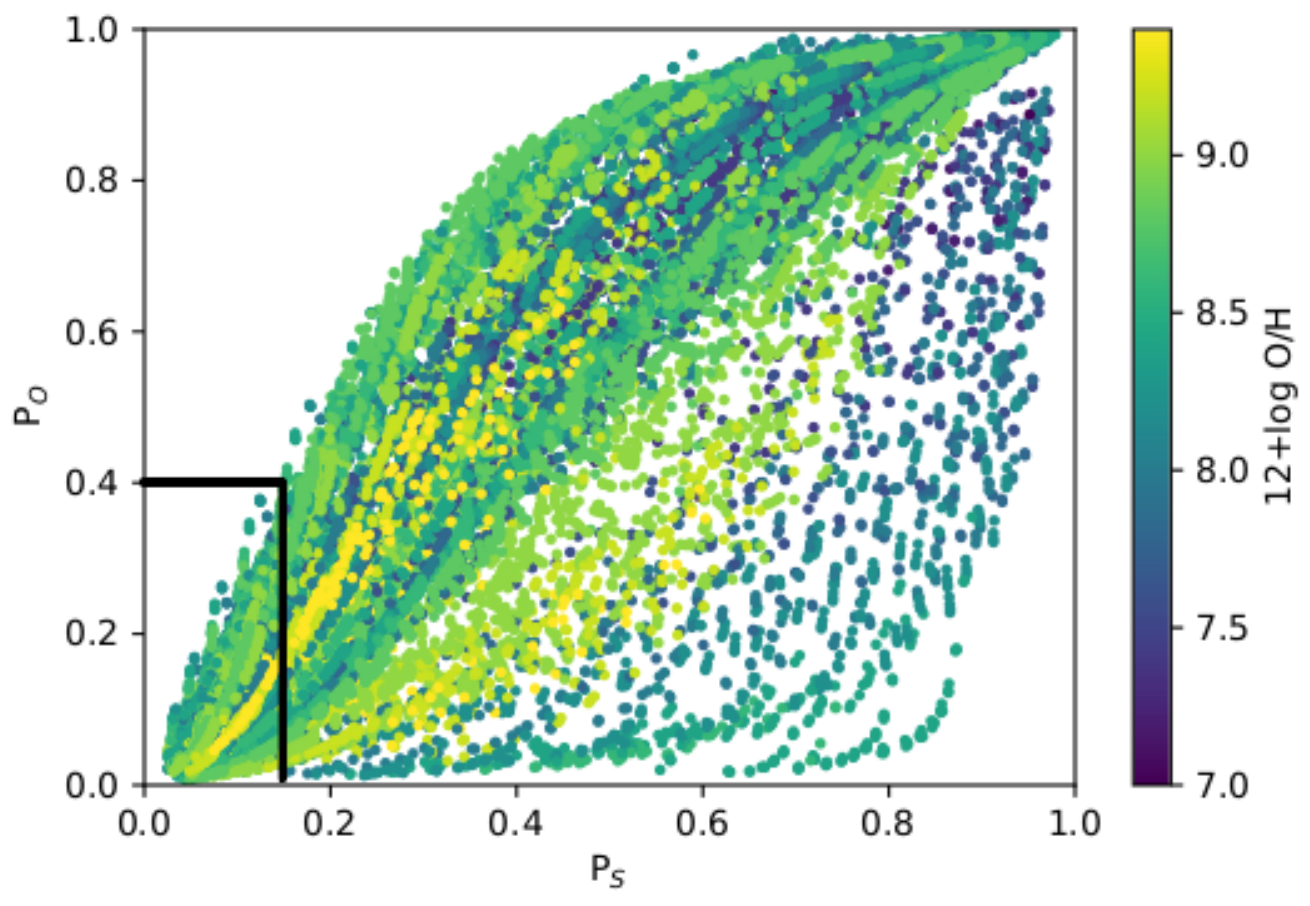}
      \caption{3MdB models showing the relation between the P$_O$ parameter and our P$_S$ substitute. The limit of P$_S\leq0.15$ and the corresponding P$_O\lesssim0.4$ are indicated with solid lines. See Section \ref{sec:z} for more details. }
         \label{fig:3mdb}
   \end{figure}

As another sanity check and in an attempt to understand why the \hii~metallicities from both methods show a similarly flat behavior although one method accounts for variations in P$_O$ (the S-calibration) and the other does not (the O3N2 calibration), we constructed and examined the parameter P$_O$ itself. \citet{Pilyugin2016} defined it as P$_O\equiv R_3/(R_2 + R_3)$, where 
R2$\equiv  (\rm I_{ [\ion{O}{II}](\lambda3727 + \lambda3729)} / \rm I_{H\beta}  )$, and R3 as above. The MUSE data cover the \oiidoublet~doublet, but the detections are insignificant and the propagated uncertainties on P are too large, with relative errors $\gtrsim100\%$. The definition of P$_O$ as given above traces the ionization more than the excitation, because the difference between ionization potentials\footnote{Ionization potentials for [\ion{O}{III}], [\ion{O}{II}], [\ion{S}{III}], and [\ion{S}{II}] are $35.1,13.6,23.3,\text{and }10.4$ eV, respectively \citep{BIEMONT1999117, Martin1990, Martin1993}.} of R2 and R3 is $\Delta E_\mathrm{ion}=21.5$ eV, while the difference between collisional excitation potentials\footnote{Collisional excitation potentials for \oiii, [\ion{O}{ii}]$\lambda3727,29$, \siii, and \siidoublet~are $2.48,3.32,1.37, \text{and }1.84$ eV, respectively \citep{NIST_ASD}.} is only $\Delta E_\mathrm{exc} = 0.8$ eV. We can therefore attempt an approximation of P$_O$ by using other ionic species of fairly large ionization potential differences, but only marginally different excitation potentials. The only option with our data is the \siii~and \siidoublet~doublet, with $\Delta E_{ion}=13$ eV and $\Delta E_\mathrm{exc} = 0.5$ eV. Our modified parameter is then P$_S\equiv S_3/(S_2 + S_3)$, where
S3$\equiv  (\rm I_{ [\ion{S}{III}](\lambda9068)} / \rm I_{H\beta}  )$, and S2 as above, and the subscript indicates that it is based on sulfur, not oxygen. The result is shown in Figure \ref{fig:P}. The P$_S$ parameter for \hii~regions shows a flat monotone behavior with galactocentric distance D, which explains why the S calibration metallicity profile shows a flat gradient similar to that of the O3N2 method. The uncertainty-weighted average P$_S$ parameter for all \hii~regions from all fields is $\left<P_S\right>=0.028\pm0.002$. At this low level, there is little to no difference between the P$_O$ of \citet{Pilyugin2016} and our P$_S$ variant. However, $90\%$ of the data points in Figure \ref{fig:P} lie below a value of P$_S\leq0.15$. We therefore proceed with this higher limit on P$_S$. To translate P$_S$ values into P$_O$ values, we used the Mexican million models database \citep[3MdB;][]{Morisset2015} to obtain a grid of  Cloudy \citep[v17.02,][]{Ferland2017} models. We selected only a subset of the 3MdB\_17 model grid under the reference BOND\_2 and based on the selection criteria of \citet[][their equations 1, 2, and 3]{Amayo2021}, describing the relations between the ionization parameter U and the ratio N/O to the oxygen abundance O/H. For the remaining set of models, we defined P$_O\equiv $[\ion{O}{iii}] $\lambda5007 / ($ [\ion{O}{iii}] $\lambda5007 +$ [\ion{O}{ii}]$\lambda3726,29)$ and P$_S\equiv $[\ion{S}{iii}]$\lambda9068 /( $[\ion{S}{iii}]$\lambda9068 + $ [\ion{S}{ii}]$\lambda6716,31) $ and plot the resulting models in figure \ref{fig:3mdb}. The figure shows that for P$_S\leq0.15,$ the value of P$_O$ can vary up to P$_O\lesssim0.4$. With this value, we can finally check the consistency of our metallicity results. A P${}_O \lesssim 0.4$ is indeed consistent with metallicities of \metal$\sim8.4\mbox{-}8.7$ \citep[][their figure 2]{Pilyugin2016}, which comfortably brackets our results in Figure \ref{fig:z}. 

An enlightening comparison could have been made for the DIG regions, but most have insignificantly detected \siii. The few available DIG data points at galactocentric distances $D>425$ arcsec (orange diamonds in Figure \ref{fig:P}) are not enough to hazard a guess about radial variations in the DIG physical conditions.

In conclusion, we can note that a linear regression on the \hii~data has y-intercepts  \metal$=8.58\pm0.19$ and $8.33\pm0.06$ with the O3N2 and S calibrations, respectively, where we have estimated the error via bootstrapping. Similarly, the DIG data has y-intercepts \metal$=8.51\pm0.02$ and $8.57\pm0.15$ with the O3N2 and S calibrations, respectively. A final comparison of our \hii~metallicity to the literature shows that the O3N2 calibration results are consistent with \citet{Stasinska2013}, who obtained a y-intercept of $12+\log_{10}{O/H}=8.48\pm0.03$ for $37$ compact and giant \hii~regions in NGC 300, based on the direct $T_e$ method. Because two independent results from the literature are consistent with the O3N2 results in our sample, we consider only the O3N2-based metallicity for
the remainder of this paper.

For completeness, in Figure \ref{fig:2Dz} we also show the 2D map of the O3N2-based metallicity. Taken at face value, the apparent decrease in metallicity throughout the outer fields \fieldp~and \fieldq\  is consistent with expectations of lower metal abundances at larger radii in spiral galaxies \citep[e.g.,][]{Holwerda2005}. 

   \begin{table}
      \caption{Uncertainty-weighted averages of velocity V, velocity dispersion $\sigma$, and shear velocity $V_\mathrm{shear}$ in \kms~ for the stellar continuum, \hii~regions, and DIG. \protect\label{tab:kin}}
         \label{tab:kinematics}
         \centering
         \begin{tabular}{ccccc}
           \hline\hline
           & \multicolumn{4}{c}{Stars}\\
           & $\left<V\right>$ & $\left<\sigma\right>$ & $V_\mathrm{shear}$ &$V_\mathrm{shear}/\left<\sigma\right>$ \\
            A & 144.3 $\pm$ 0.4 & 33.0 $\pm$ 0.2 & 9.4 $\pm$ 5.6 & 0.3 $\pm$ 0.3 \\
            B & 156.6 $\pm$ 0.3 & 34.3 $\pm$ 0.3 & 5.9 $\pm$ 0.8 & 0.2 $\pm$ 0.0 \\
            C & 163.5 $\pm$ 0.5 & 37.7 $\pm$ 0.6 & 11.4 $\pm$ 6.5 & 0.3 $\pm$ 0.3 \\
            D & 160.8 $\pm$ 0.9 & 59.0 $\pm$ 1.3 & 23.1 $\pm$ 6.4 & 0.4 $\pm$ 0.3 \\
            E & 179.7 $\pm$ 0.4 & 27.0 $\pm$ 0.5 & 9.1 $\pm$ 4.6 & 0.3 $\pm$ 0.2 \\
            I & 160.3 $\pm$ 0.4 & 27.5 $\pm$ 0.5 & 10.3 $\pm$ 2.3 & 0.4 $\pm$ 0.1 \\
            J & 151.3 $\pm$ 0.7 & 20.0 $\pm$ 0.9 & 10.9 $\pm$ 1.3 & 0.5 $\pm$ 0.1 \\
            P & 199.9 $\pm$ 1.4 & 55.0 $\pm$ 3.3 & 30.6 $\pm$ 9.7 & 0.6 $\pm$ 0.4 \\
            Q & 206.2 $\pm$ 1.2 & 84.2 $\pm$ 3.5 & 27.5 $\pm$ 15.1 & 0.3 $\pm$ 0.6 \\
            \hline
            & \multicolumn{4}{c}{\hii~regions}\\
            A & 152.6 $\pm$ 1.3 & 20.7 $\pm$ 0.9 & 21.8 $\pm$ 9.2 & 1.1 $\pm$ 0.4 \\
            B & 158.2 $\pm$ 1.1 & 21.0 $\pm$ 1.1 & 21.6 $\pm$ 16.4 & 1.0 $\pm$ 0.8 \\
            C & 172.1 $\pm$ 1.5 & 22.0 $\pm$ 1.2 & 17.8 $\pm$ 3.2 & 0.8 $\pm$ 0.2 \\
            D & 187.5 $\pm$ 1.0 & 20.6 $\pm$ 1.3 & 16.0 $\pm$ 2.8 & 0.8 $\pm$ 0.1 \\
            E & 190.2 $\pm$ 0.8 & 20.4 $\pm$ 1.5 & 13.5 $\pm$ 9.2 & 0.7 $\pm$ 0.5 \\
            I & 176.3 $\pm$ 0.9 & 22.2 $\pm$ 1.3 & 14.6 $\pm$ 6.6 & 0.7 $\pm$ 0.3 \\
            J & 160.4 $\pm$ 1.4 & 21.2 $\pm$ 1.5 & 21.9 $\pm$ 0.4 & 1.0 $\pm$ 0.1 \\
            P & 194.1 $\pm$ 1.9 & 25.2 $\pm$ 2.2 & 29.5 $\pm$ 12.6 & 1.2 $\pm$ 0.5 \\
            Q & 205.0 $\pm$ 1.8 & 26.1 $\pm$ 1.2 & 28.3 $\pm$ 5.0 & 1.1 $\pm$ 0.2 \\
            \hline
            & \multicolumn{4}{c}{DIG}\\
            A & 152.4 $\pm$ 0.5 & 30.3 $\pm$ 0.3 & 9.3 $\pm$ 3.4 & 0.3 $\pm$ 0.2 \\
            B & 162.0 $\pm$ 0.4 & 28.2 $\pm$ 0.5 & 10.2 $\pm$ 3.0 & 0.4 $\pm$ 0.1 \\
            C & 175.0 $\pm$ 0.6 & 27.8 $\pm$ 0.4 & 12.5 $\pm$ 3.3 & 0.5 $\pm$ 0.2 \\
            D & 190.9 $\pm$ 0.4 & 21.7 $\pm$ 0.4 & 7.2 $\pm$ 1.2 & 0.3 $\pm$ 0.1 \\
            E & 190.2 $\pm$ 0.4 & 19.2 $\pm$ 0.5 & 7.5 $\pm$ 1.0 & 0.4 $\pm$ 0.1 \\
            I & 177.6 $\pm$ 0.3 & 26.6 $\pm$ 0.4 & 7.8 $\pm$ 4.3 & 0.3 $\pm$ 0.2 \\
            J & 158.4 $\pm$ 2.0 & 23.2 $\pm$ 1.3 & 22.6 $\pm$ 0.5 & 1.0 $\pm$ 0.1 \\
            P & 207.9 $\pm$ 0.6 & 20.7 $\pm$ 0.5 & 10.5 $\pm$ 2.0 & 0.5 $\pm$ 0.1 \\
            Q & 212.7 $\pm$ 0.4 & 25.1 $\pm$ 1.0 & 8.2 $\pm$ 2.6 & 0.3 $\pm$ 0.1 \\
            \hline
         \end{tabular}
   \end{table}
   \begin{figure}
   \centering
   \includegraphics[width=8.5cm]{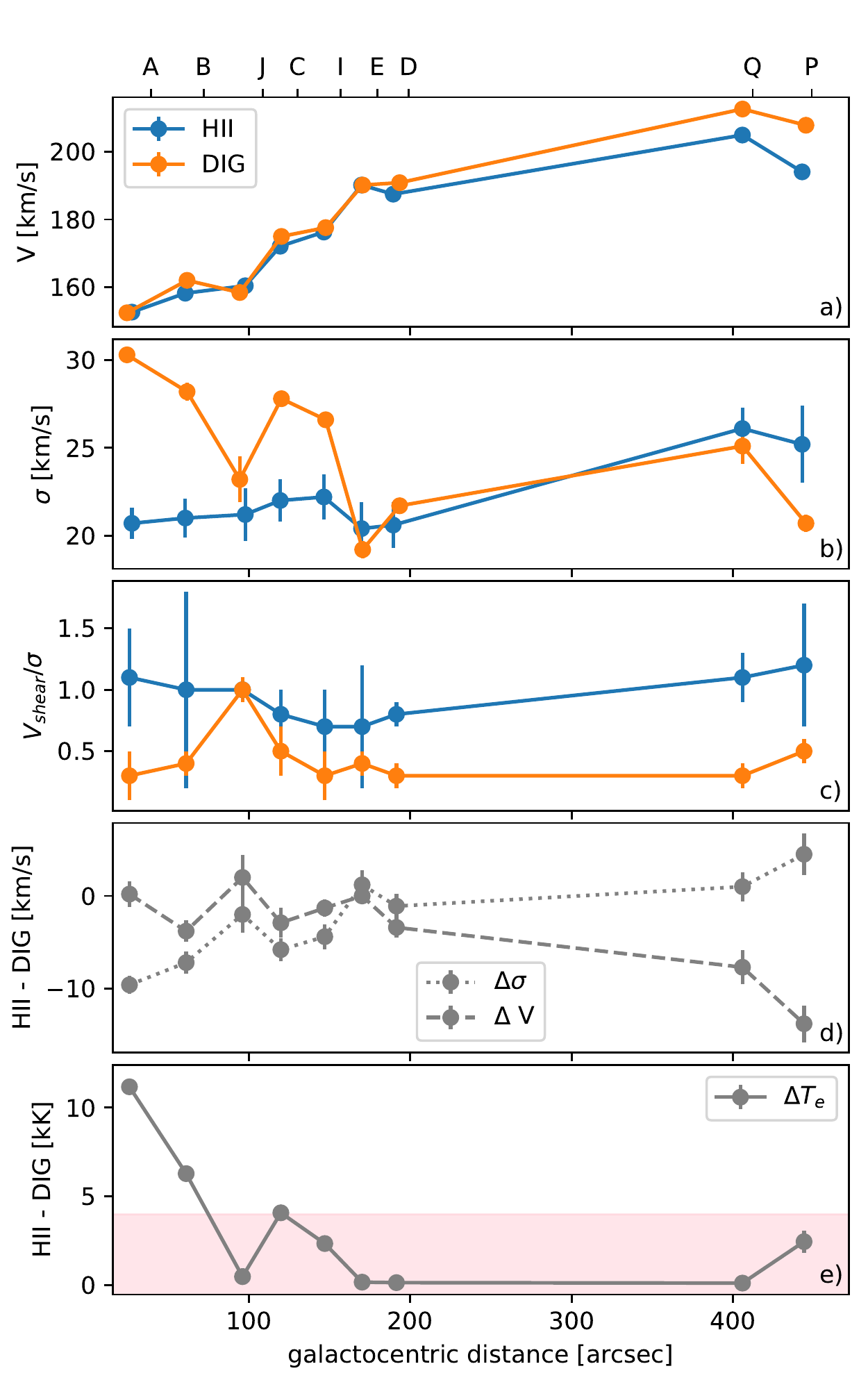}
      \caption{Sample kinematics. a) Velocity. b) Velocity dispersion. c) $V_\mathrm{shear}/\sigma$. d) $\Delta$V and $\Delta\sigma$ between \hii~and DIG. e) $\Delta T_e$ between \hii~and DIG, assuming $\Delta\sigma$ is solely due to the thermal component. The pink shaded region is the $\Delta T_e$ range found in the literature. Panels b) and c) have the same legend as in a). All panels share the galactocentric distance in arcsec as x-axis. The twin axis shows the corresponding field labels.  }
         \label{fig:1Dkine}
   \end{figure}

\subsection{Kinematics}\protect\label{sec:kine}
Table \ref{tab:kin} shows the kinematic properties of the fields for \hii~and DIG. For completeness, we also show the results of the stellar continuum. At the risk of stating the obvious, the local stellar background is best represented by itself and not by the residual of the local stellar background and the nearby stellar background, which is our method to obtain the \hii~region spectra before the pPXF line extraction. Therefore, we ran a separate pPXF extraction on spectra from \hii~region dendrogram leaves, but without subtraction of the diffuse background approximation. We stress that this pPXF run was solely used for the purpose of obtaining precise stellar continuum velocities $V_\mathrm{cont}$ and velocity dispersions $\sigma_\mathrm{cont}$. 

The focal point of our analysis, however, are the \hii~and DIG regions. We note that the values of $\sigma_{\rm HII}$ and $\sigma_{\rm DIG}$ fall below the instrumental resolution ($47$ km s$^{-1}$ at \ha), which means that it is unclear whether they are reliable. Similar values were obtained by \citet{denBrok2020} for 41 star-forming spirals in the MUSE Atlas of Discs, and these authors performed idealized simulations to examine the intrinsic compared to the measured dispersion of the gas. They concluded that although the absolute values of the dispersion are biased below $\sigma=25$ km s$^{-1}$, the relative values are robust. Additionally, it has been demonstrated that pPXF robustly measures dispersion below the instrumental dispersion \citep{Cappellari2017}. This enables us to perform a comparative analysis of the \hii~and DIG regions. 

%----------------------------------------------------------------- 
%
%                                                Two column figure
%----------------------------------------------------------------- 

   \begin{figure*}[h!]
   \centering
   \includegraphics[width=17.5cm]{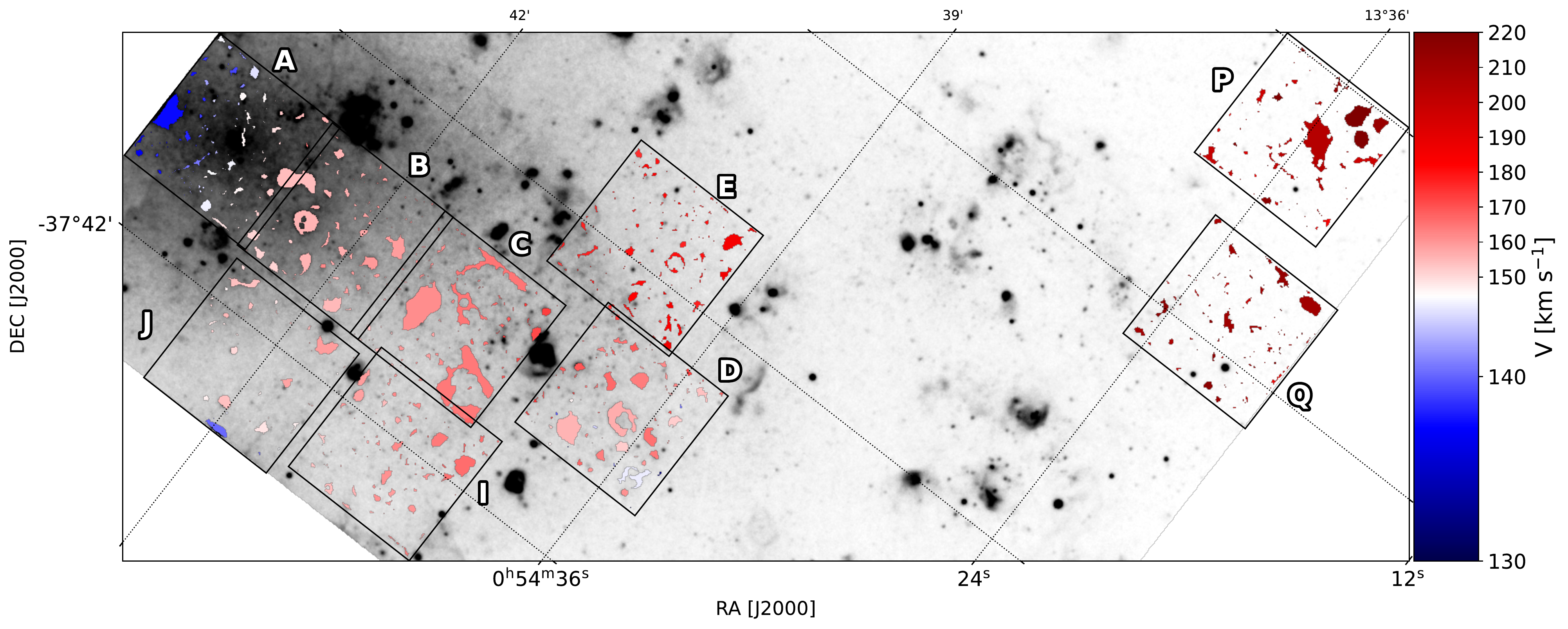}
   \includegraphics[width=17.5cm]{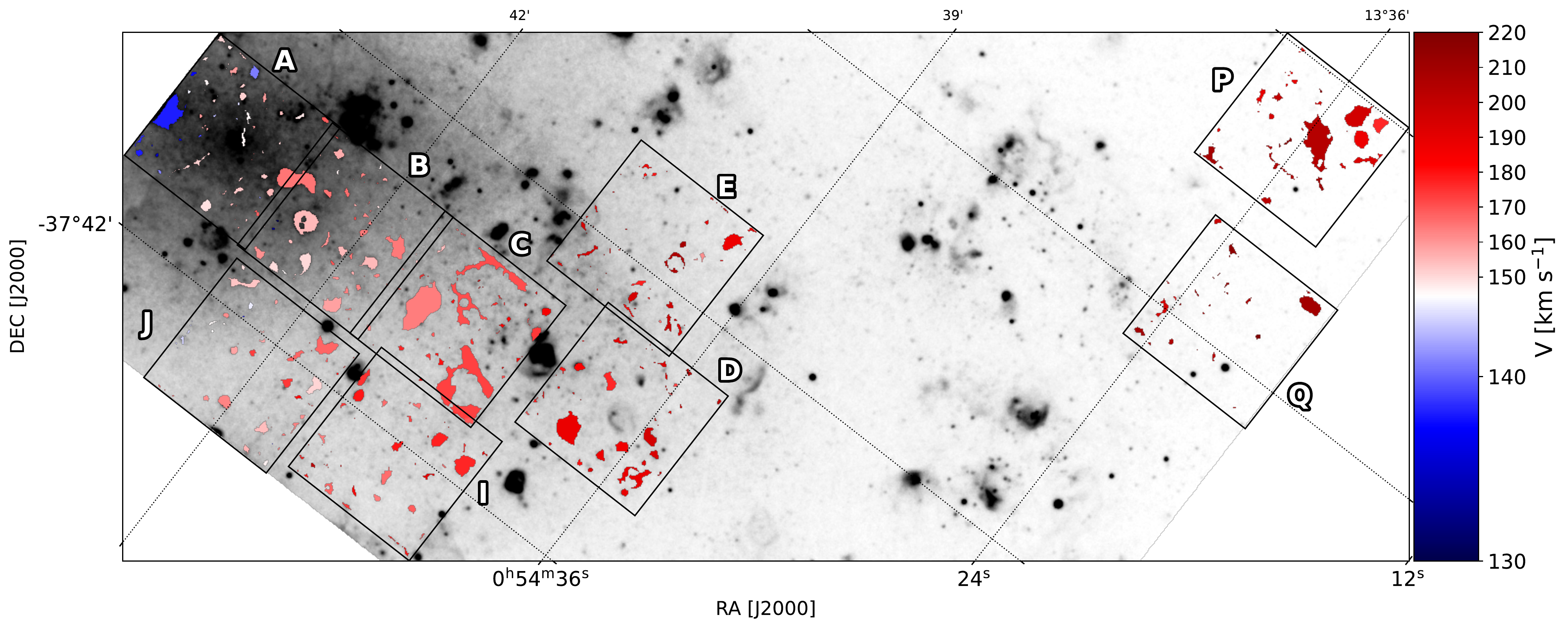}
   \includegraphics[width=17.5cm]{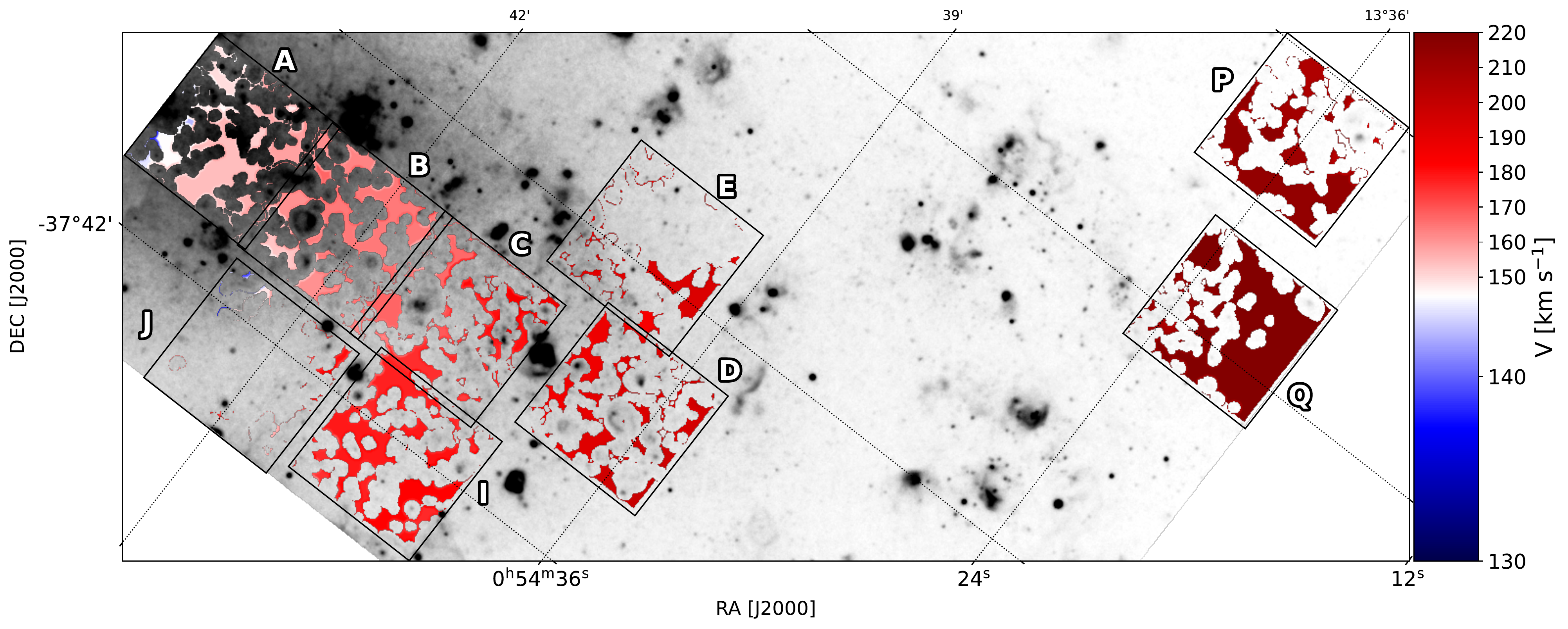}
      \caption{Velocity in km s$^{-1}$ for stars (top), \hii~regions (middle), and DIG (bottom). We note that the color map is a divergent two-slope color map, linear in both slopes around the centered midpoint (white color) at 144.34 km/s in all panels, which is the uncertainty-weighted average velocity of the stars in Field \fielda.}
         \label{fig:v}
   \end{figure*}

  \begin{figure*}[h!]
   \centering
   \includegraphics[width=17.5cm]{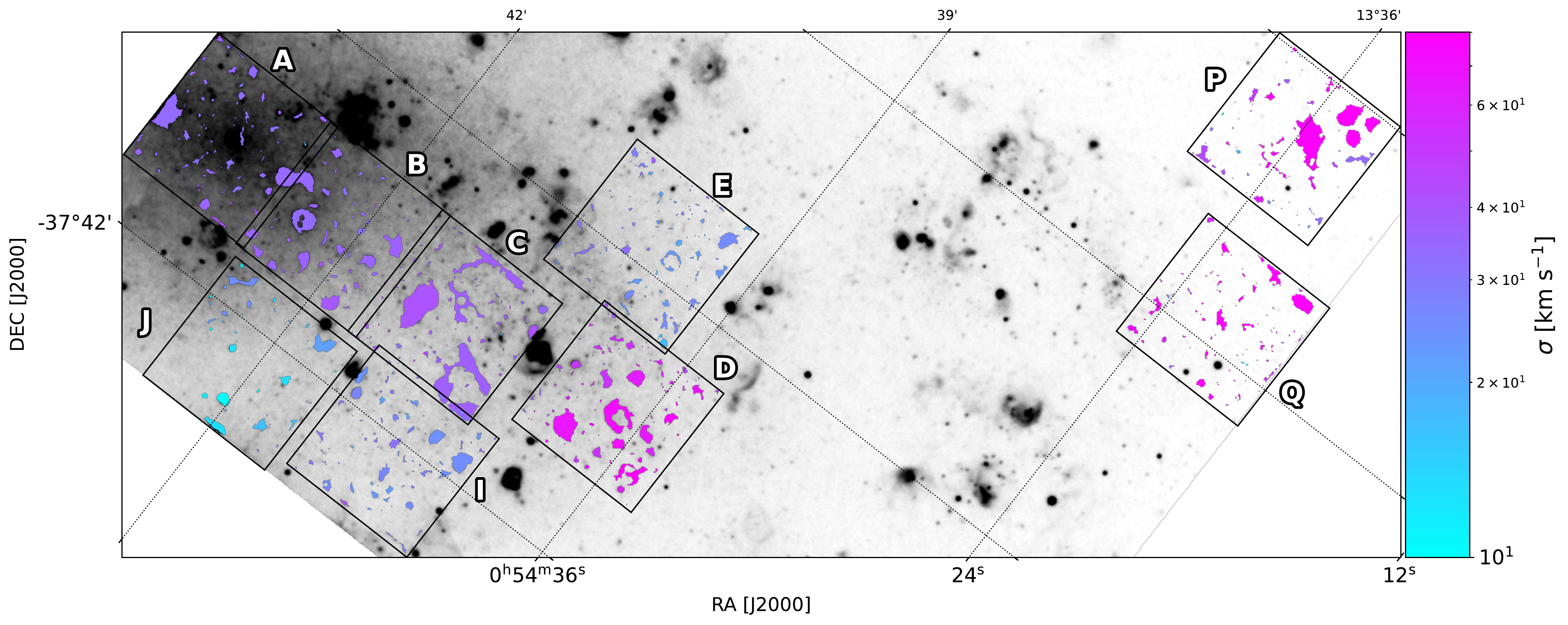}
   \includegraphics[width=17.5cm]{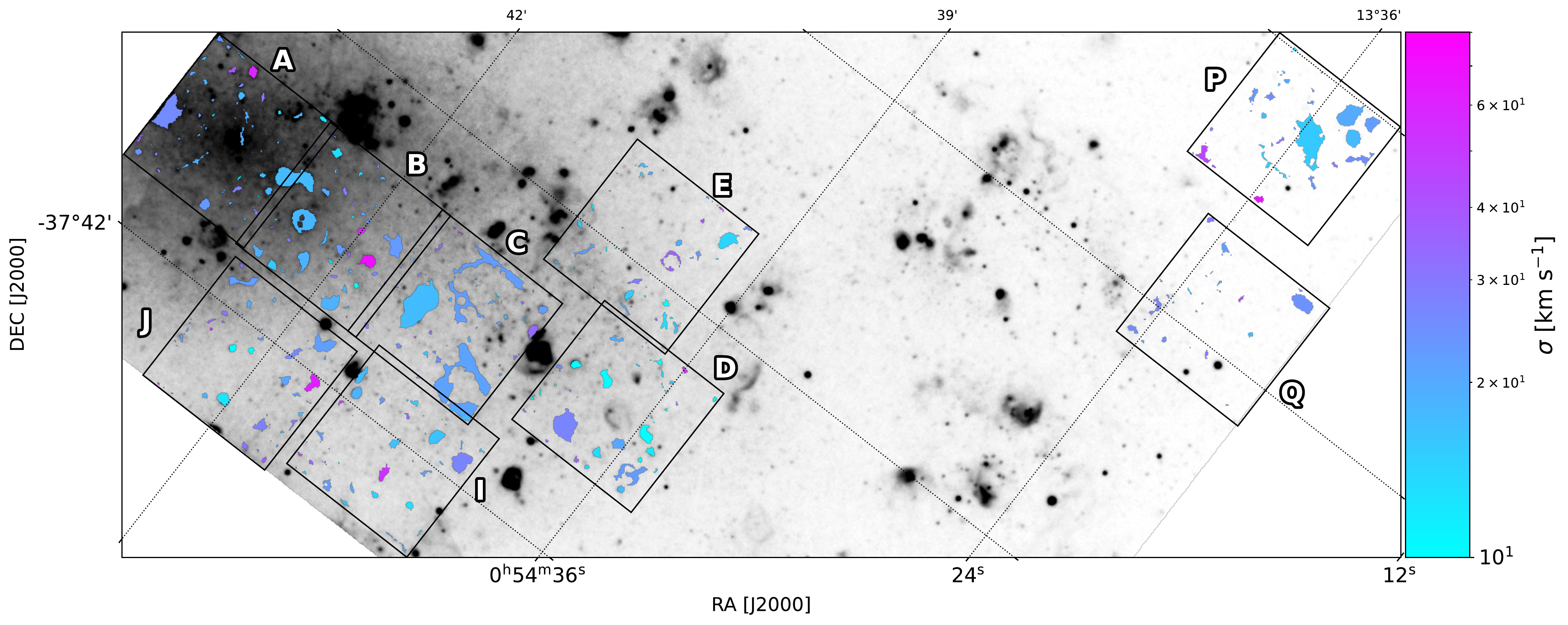}
   \includegraphics[width=17.5cm]{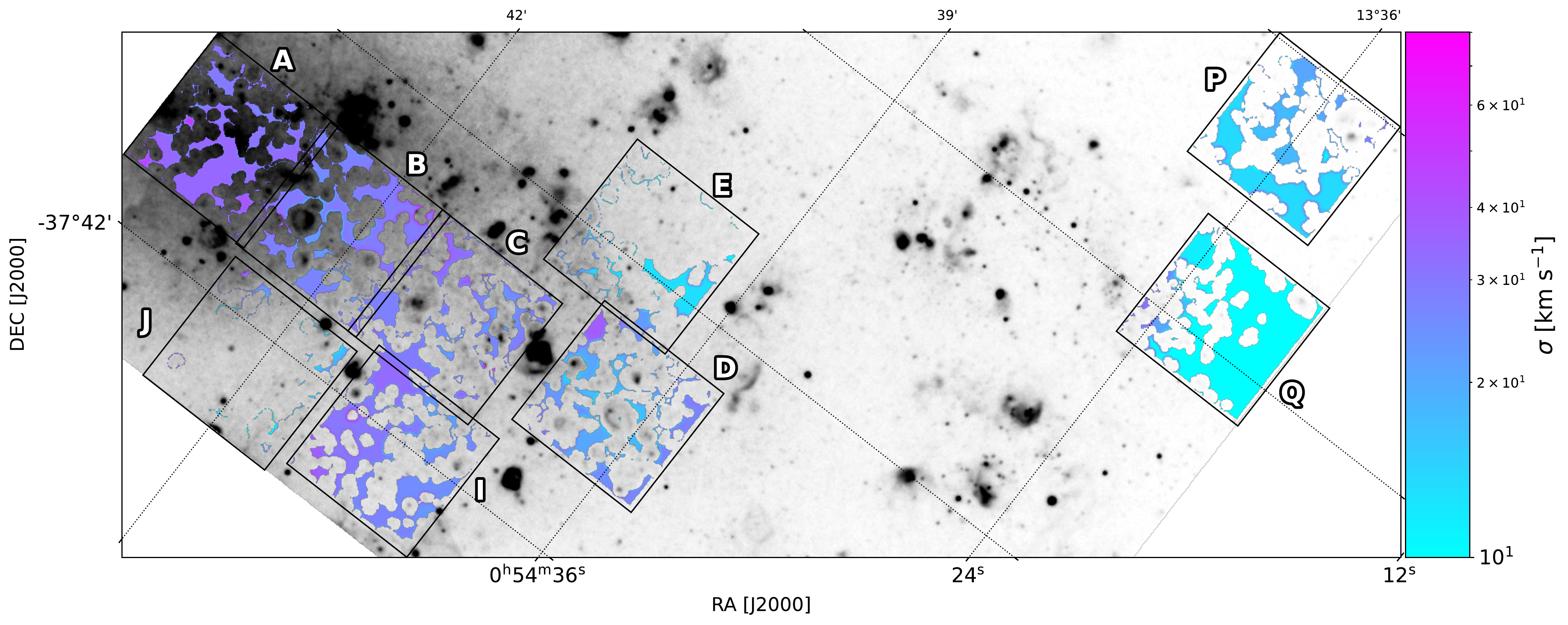}
      \caption{Velocity dispersion in km s$^{-1}$ for stars (top), \hii~regions (middle), and DIG (bottom). }
         \label{fig:vdisp}
   \end{figure*}

If the velocities $V$ and velocity dispersions $\sigma$ were weighted by the \ha~flux, then the handful of bright regions in our sample would dominate the results. This is not representative of the sample, and therefore we take the uncertainty-weighted averages instead in Table \ref{tab:kin}. The lowest uncertainty points have the highest weights. The shear velocity is a measure of the large-scale motion of the gas, and we measured it as $V_\mathrm{shear}=(V_\mathrm{max}-V_\mathrm{min})/2$. Similar to \citet{Herenz2016} and \citet{Micheva2019}, we took the median of the upper and lower fifth percentile of the velocities to obtain $V_\mathrm{min}$ and $V_\mathrm{max}$, and propagated the full width of each percentile as a conservative estimate of the $V_\mathrm{shear}$ uncertainties. We note that the systemic stellar velocity we obtain (field \fielda~in Table \ref{tab:kin}) is $V_\mathrm{sys}^A=144.3\pm0.4$ km/s, which agrees well with the literature, for instance, \citet[][$V_\mathrm{sys}=145\pm2$ km/s]{Rogstad1979} and \citet[][$V_\mathrm{sys}=144\pm0.2$ from radio data]{Westmeier2011}.

The table is visualized in Figure \ref{fig:1Dkine}, showing the velocity, velocity dispersion, the $V_\mathrm{shear}/\sigma$ ratio, and the residual between \hii~and DIG gas kinematics in panels a), b), c), and d), respectively. Figure \ref{fig:1Dkine}a shows that the \hii~and DIG velocities are similar, implying that the two components move together. This is perhaps better visible in panel d), where $\Delta V$ is fairly flat for the inner fields, with the DIG becoming marginally faster than the \hii~gas in the outer fields \fieldp{} and \fieldq{}. In contrast, Figure \ref{fig:1Dkine}b shows an increased velocity dispersion for the DIG in the central regions, with a decreasing tendency for larger galactocentric distances, while the \hii~gas apparently maintains a constants $\sigma$. This results in an increase in $\Delta\sigma$  with galactocentric distance in panel d). Figure \ref{fig:1Dkine}c shows that $V_\mathrm{shear}/\left<\sigma\right>$ has a flat behavior that remains fairly constant within the uncertainties for both \hii~and DIG. Taking the  $V_\mathrm{shear}/\left<\sigma\right>_{\rm HII}\sim1.0$ average at face value this implies that \hii~gas is rotationally supported, while the DIG is predominantly dispersion supported, with $V_\mathrm{shear}/\left<\sigma\right>_{\rm DIG}\sim0.4$. When the propagated uncertainties are considered as well, the separation between rotationally and dispersion supported kinematics of the \hii~and DIG regions is less clear, but still systematic at all distances. The exception is field \fieldj, which covers the inter-arm region and features high extinction from a prominent dust lane \citep[see Figure 1 in ][]{Roth2018}. It is possible that the inter-arm region has different DIG kinematics, which we sample only with field \fieldj{} and partially with field \fielde, which apparently covers a fraction of the spiral arm and also another inter-arm region with a visible dust lane \citep{Roth2018}.

\begin{table}
    \caption{[\ion{N}{II}]-based electron temperature limits. For field J, an actual value could be calculated, but with very large uncertainties.}
    \label{tab:temden}
    \centering
    \begin{tabular}{ccc}
    \hline\hline
    Field& $T_e$ \hii & $T_e$ DIG\\
     & K & K \\
    A & & \\
    B & & \\
    C & & \\
    D & <9040 & <10458\\
    E & <9938 & <10550\\
    I & <5950& <12300\\
    J & $9178^{+2329}_{-3223}$ & $15611^{+9555}_{-8788}$\\
    P & <20230& \\
    Q & <25870& \\
    \end{tabular}
\end{table}

In Table \ref{tab:kinematics}, the median $\sigma_{\rm HII}=21.2\pm0.2$ km s$^{-1}$, and $\sigma_{\rm DIG}\sim25.1\pm0.1$ km s$^{-1}$. The DIG velocity dispersion is higher than for \hii~gas, which is consistent with the DIG being hotter than \hii~gas in spiral galaxies, including the Milky Way \citep[e.g.,][]{Madsen2006,Haffner2009,DellaBruna2020}. The expected temperature difference in the literature is $\Delta T_\mathrm{e} \sim 2\mbox{-}4$ kK at most. We measured a difference in average velocity dispersion of $\Delta \sigma\sim3.9$ km s$^{-1}$. When we assume that all of this difference is due to the thermal component of the velocity dispersion, $\sigma_\mathrm{th}$, we can calculate the corresponding electron temperature as $T_\mathrm{e} [K] = \sigma_\mathrm{th}^2 m_\mathrm{H} / k_\mathrm{B} $, where $m_\mathrm{H}$ and $k_\mathrm{B}$ are the mass of a hydrogen atom and the Boltzmann constant, respectively.  With $\Delta \sigma\sim3.9$ km s$^{-1}$ , we obtain $T_\mathrm{e} \sim1.8$ kK, which is well within the range observed in the literature to date. The difference per field is shown in Figure \ref{fig:1Dkine}e, where we have again assumed that all of the velocity dispersion is purely thermal, and show the corresponding difference in temperature between \hii~and DIG. It is clear from the figure that at galactocentric distance of D$\gtrsim 75$ arcsec all of the observed velocity dispersion difference between \hii~and DIG could be accounted for simply by temperature differences. We note that this does not mean that turbulence is negligible in these fields because the DIG is heterogeneous \citep[e.g.,][]{Haffner2009,Madsen2006} and may be hotter than or of equal temperature as the \hii~gas. The implied temperature differences are consistent with the velocity dispersion being dominated by its thermal component. For the innermost regions A and B, however, the figure suggests that the velocity dispersion cannot be exclusively thermal because that would correspond to a $\Delta T_\mathrm{e} = 6\mbox{-}10$ kK, that is, the DIG would have to be $6\mbox{-}10$ kK hotter than the \hii~gas, which is much higher than the expected temperature differences found in the literature. We therefore conclude that in these central regions, the DIG must be significantly more turbulent than the \hii~gas and the DIG in the rest of the fields. Alternatively, similar to the spirals in \citet{denBrok2020}, the DIG emission might partially originate from a somewhat thicker layer than the \hii~gas.

In Appendix \ref{sec:starkine} we show the complementary figure for the stellar kinematics. We do not dwell on the stars, other than to note that the stellar continuum is always slower than both the \hii~and DIG regions, and always with a higher velocity dispersion. This observation is consistent with the early-to-intermediate type spirals in \citet{VegaBeltran2001}, for example. As these authors noted, these differences in kinematics between the gas and the stars can be explained by the gas being confined to the disk and supported by rotation, as implied by the measured $V_\mathrm{shear}/\left<\sigma\right>_{\rm HII}\sim1.0$, while the observed stellar continuum mostly belongs to the bulge and is dispersion supported, with $V_\mathrm{shear}/\left<\sigma\right>_{\rm stars}\sim0.3$. 

Two-dimensional velocity maps of the stellar continuum, \hii, and DIG regions are presented in Figure \ref{fig:v}, and the corresponding velocity dispersion maps are shown in Figure \ref{fig:vdisp}. The data trace the velocity field of the galaxy. Most of the pointings show a velocity redshift from the central field \fielda, with the inter-arm field \fieldj~showing a mix of red- and blueshifted regions. The line of midpoint velocity, defined on stars in field \fielda, is clearly visible for the stellar component (white in Figure \ref{fig:v}, top panel), becomes diluted for the \hii~regions (Figure \ref{fig:v}, middle panel), and appears to be offset in the DIG (Figure \ref{fig:v}, bottom panel).  In Figure \ref{fig:vdisp} the velocity dispersion shows an increase toward the outskirts (fields \fieldp, \fieldq) for stars (top panel) and a reversed trend for the DIG (bottom panel). We note that the velocity dispersion in the inter-arm field \fieldj~is lower for both stars and DIG. The \hii~regions (middle panel) show a more complex velocity dispersion behavior that is apparently uncoupled from the distance of the pointing to the galactic center in field \fielda.

\section{Stacked spectra}\protect\label{sec:stacks}
To increase the S/N in the spectra, we average-stacked all \hii-region spaxels per field and ran pPXF on the stacks to extract nebular emission lines and kinematics. To obtain a galaxy stack, we also stacked all spaxels from all fields. The same method was used to obtain a per field and galaxy stack for the DIG regions. These spectra are shown in Figure \ref{fig:stackspec} in the appendix. 

In the stacked spectra of non-AO fields, we identify the faint auroral \niiauroral~line for both \hii~and DIG stacks. However, the propagated uncertainties in the temperature-sensitive \niiauroral/\niileft~ratio are too large for meaningful results of the electron temperature and density. For example, for field \fieldj, the only field in which the combination of [\ion{N}{II}] and [\ion{S}{ii}] ratios is strictly within the theoretical limit, we obtain $(T_e, N_e) = (9178^{+2329}_{-3223} \text{K}, 1.9^{+14.8}_{-1.9} \text{cm}^{-3})$ for \hii~regions and $(T_e, N_e) = (15611^{+9555}_{-8788} \text{K}, 12.7^{+42.3}_{-12.7} \text{cm}^{-3})$ for DIG regions. This implies that $T_\mathrm{e}$ for the DIG could be 1 kK to over 10 kK hotter than for the \hii~regions. This result is hardly constraining. For the rest of the non-AO fields, we were able to obtain upper limits on $T_\mathrm{e}$ within the uncertainties at best. We list them in Table \ref{tab:temden}. Within the uncertainties, the \hii~region electron densities in field \fieldj~are indistinguishable from those of the DIG. Taken at face value, even within the uncertainties, the $N_\mathrm{e}$ for \hii~regions in field \fieldj~are at the low end of the expected densities \citep[e.g.,][$N_\mathrm{e}\in(10,1000)$ cm$^{-3}$]{Osterbrock2006}, which is consistent with our sample containing extremely faint \hii~regions. 

From these stacked spectra, we calculated the DIG/\ha~fractions per field. They are listed in Table \ref{tab:digfrac}. In order to test the robustness of these fractions, we checked how the sizes of the \hii~regions and the physical locations of the DIG regions affect the measured DIG contribution. For this purpose, we expanded the \hii~region masks by the gap annulus of five spaxels and simultaneously shifted the DIG annulus by the same amount away from the \hii~regions. This resulted in an increasing number of \hii~spaxels by a factor of $2.0, 2.0, 2.0, 2.15, 2.66, 2.48, 2.38, 1.97$, and $2.68$ for fields A, B, C, D, E, I, J, P, and Q, respectively. Simultaneously, the number of DIG spaxels increased by a factor of $1.24, 1.24,1.24, 1.16, 1.22, 1.34,1.20,1.22,$ and $1.13$, respectively. This affects the DIG/\hii~fractions by $< 10\%$, namely, by $+4.1\%, +4.1\%, +4.1\%, +9.3\%, -4.3\%, +5.9\%, +2.4\%, +5.7\%,\text{ and } +0.7\%$, respectively. We therefore conclude that our definition of \hii~and DIG regions is robust, and that the DIG fractions are not overly sensitive to the exact shape of our \hii~regions. Doubling the \hii~region size does not influence the conclusions of our analysis. 

\section{Discussion}\protect\label{sec:discuss}
In Section \ref{sec:prop} we presented a number of properties that can be extracted from our data for the \hii~and DIG regions. We obtained no convincing differences between the two types of objects in terms of metallicity, velocity, and velocity dispersion, and marginal differences in electron density and extinction. The DIG apparently has a lower density and E(B-V) values on average. The DIG fraction we measure per field appears to agree in general with expectations from the literature \citep[e.g.,][]{Oey2007, Chevance2020,Belfiore2022,DellaBruna2020}, as shown in Table \ref{tab:digfrac}. 

\begin{table}
    \caption{DIG fraction per field}
    \label{tab:digfrac}
    \centering
    \begin{tabular}{ll}
    \hline\hline
    Field& DIG fraction\\
    A & 0.52 \\
    B & 0.48 \\
    C & 0.42 \\
    D & 0.46\\
    E & 0.64\\
    I & 0.51\\
    J & 0.15\\
    P & 0.48 \\
    Q & 0.77\\
    \end{tabular}
\end{table}

   \begin{figure}
   \centering
   \includegraphics[width=8.5cm]{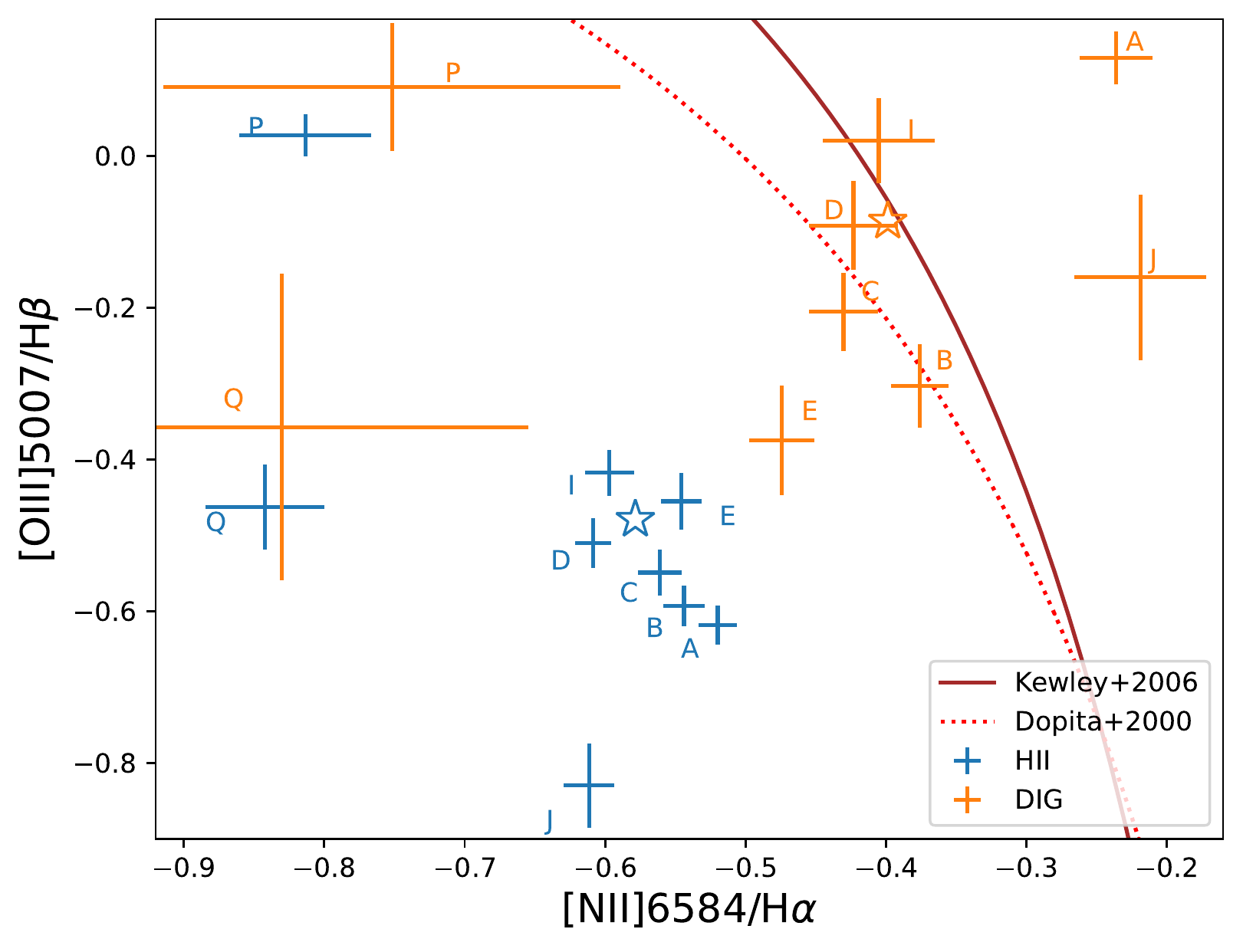}
   \includegraphics[width=8.5cm]{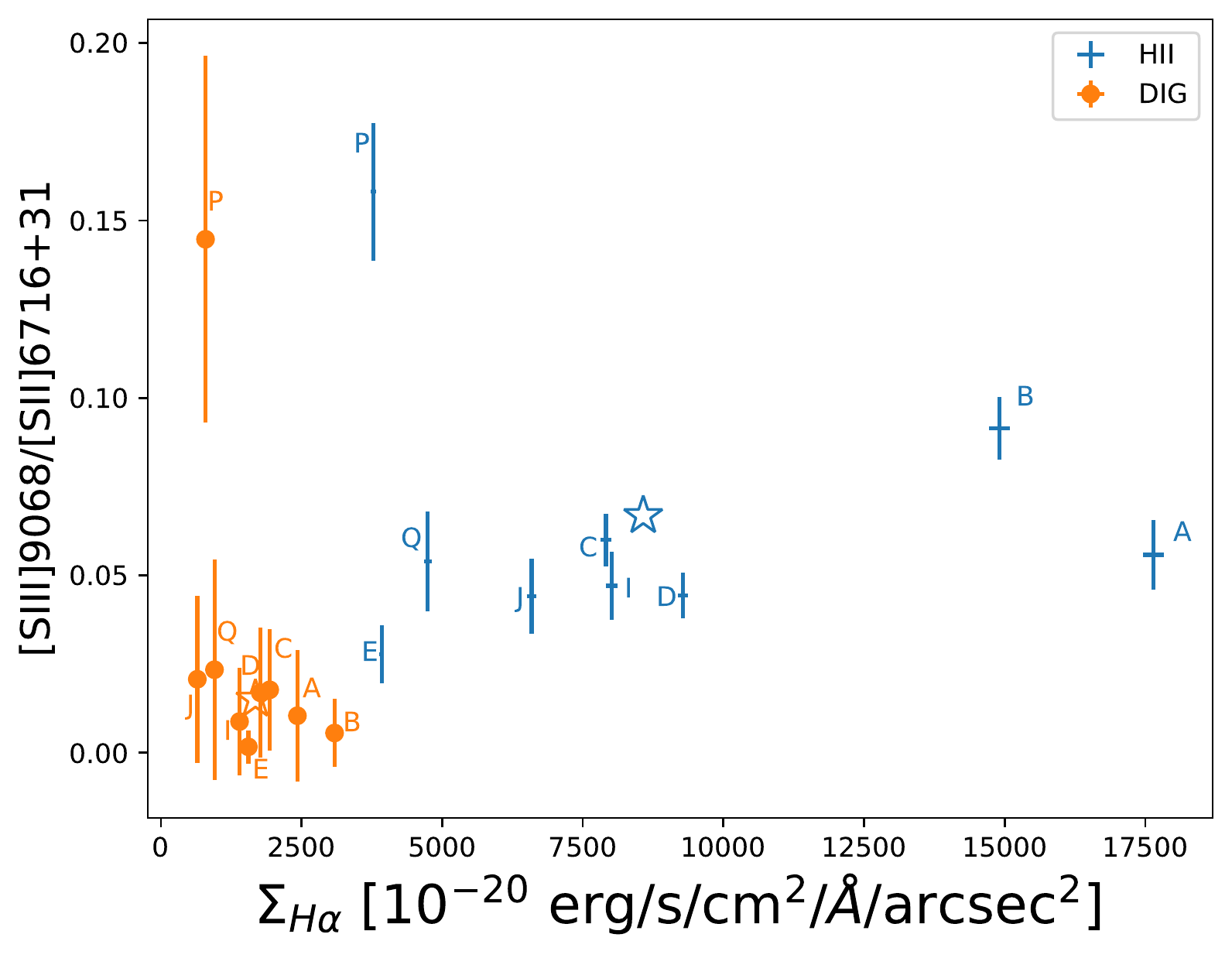}
   \includegraphics[width=8.5cm]{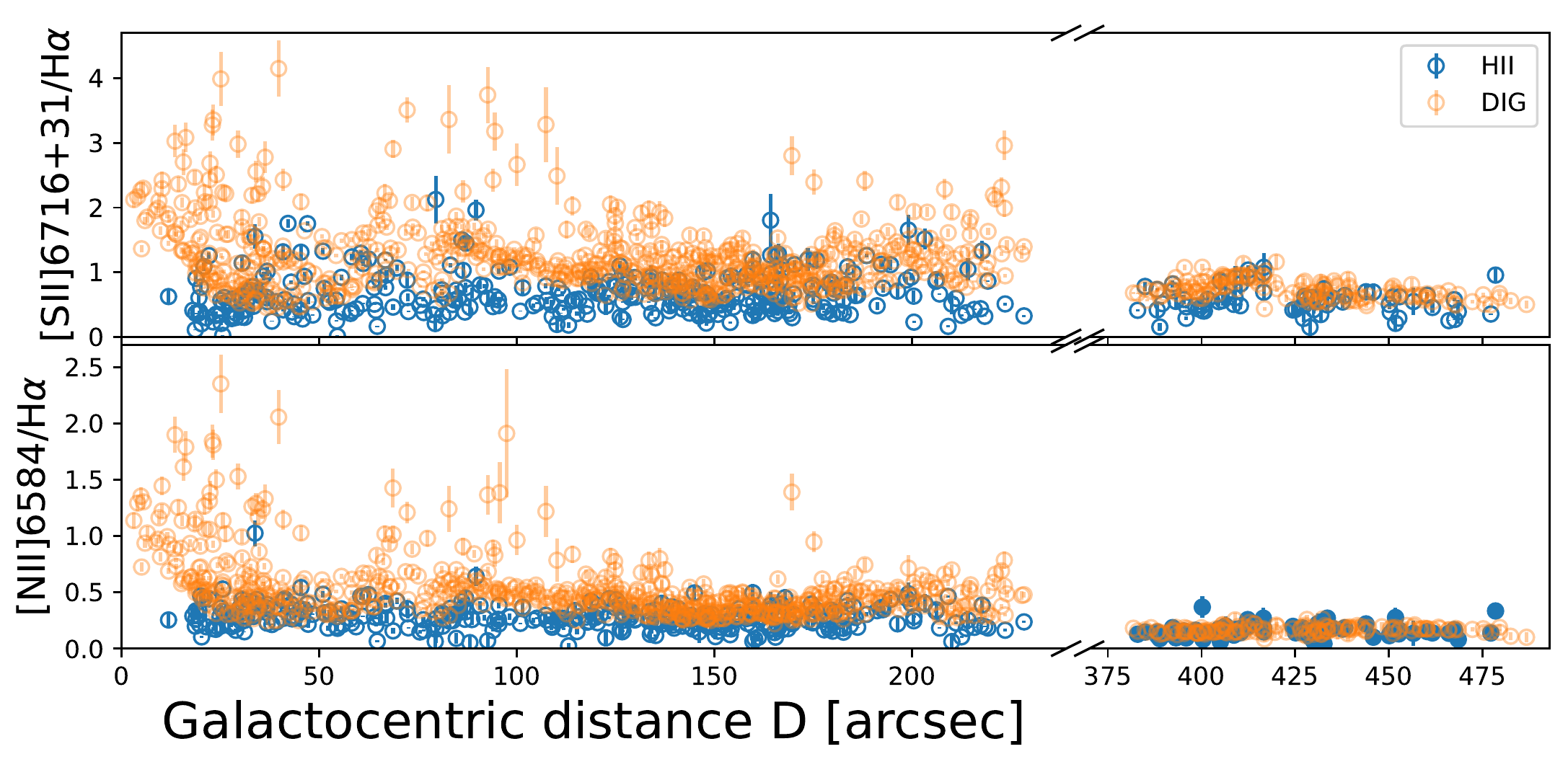}
      \caption{Evidence of HOLMES. (Top) BPT diagram for average stacked spectra per field. (Middle) High-to-low sulfur ionization ratio as a function of \ha~surface brightness for average stacked spectra. (Bottom) Low-ionization ratios of sulfur and nitrogen with galactocentric distance for individual \hii~and DIG regions. The total spectrum of stacking all spaxels from all fields is marked with a star.}
         \label{fig:dighii}
   \end{figure}
     
\citet{Belfiore2022} reported that the main differences between the DIG and \hii~regions in spiral galaxies are enhanced low-ionization line ratios toward the centers of spiral galaxies, an offset of some DIG regions from the locus of \hii~regions in the BPT diagram extending beyond the star-formation region and into the Low-ionization nuclear emission-line/Active galactic nuclei (LINER/AGN) domain, and a decreasing \oiii/\hb~ratio with the \ha~surface brightness of the DIG. Their \hii~and DIG regions, however, are brighter than ours, with $\log{L(\text{H}\alpha [\text{erg s}^{-1}])\sim36}$. Our regions are much fainter, with a median $\log{L(\text{H}\alpha [\text{erg s}^{-1}])=34.7}$, and thus fall in a brightness range that has been studied much less extensively. It is therefore prudent to verify whether these trends are still present at these extremely faint levels. In what follows, we have found no indication that the extremely faint, manually added large DIG regions described in Section \ref{sec:data} and shown in Figure \ref{fig:showcase} behave differently than normal DIG regions created by binary dilation of the \hii~region masks.

In Figure \ref{fig:dighii} we show the BPT diagram for DIG and \hii~field stacks, as well as the \citet{Kewley2006} limit for star-forming regions. In this parameter space, the \hii~regions are separated from the corresponding DIG in the field in all fields. Similar to the findings in \citet{Belfiore2022}, the DIG is found predominantly in the top right corner of the diagram, crossing into the region associated mainly with LINERs and AGN. This is likely due to the increasing contribution to DIG ionization from HOLMES with a harder spectrum than \hii~regions \citep{FloresFajardo2011,Belfiore2022}. In addition, the DIG ionization state is expected to be lower than that of the \hii~regions \citep[][]{Madsen2006}, which we also observe in the middle panel of Figure \ref{fig:dighii}, using the \siii/\siisum~ratio as a tracer of the ionization parameter. The same figure demonstrates that the DIG high-to-low ionization line ratio is flat within the uncertainties (fields \fielda-\fieldj, \fieldq) or decreasing (fields \fieldp~and, e.g., \fieldb) with increasing \ha~surface brightness. Either of these trends requires the added emission from HOLMES \citep{FloresFajardo2011, Belfiore2022}. The final argument in support of the HOLMES contribution to DIG ionization is the behavior of low-ionization line ratios such as \sii/\ha~and \nii/\ha~with galactocentric distance. For the DIG, an increase in these ratios is observed with increasing distance to the galactic plane in an edge-on spiral \citep{FloresFajardo2011} and with galactocentric distance in $19$ low-inclination spirals from the PHANGS-MUSE survey \citep{Belfiore2022}. The bottom panel of Figure \ref{fig:dighii} shows the expected systematic enhancement toward small distances in both of these line ratios in the DIG, but not in the \hii~regions.  

\begin{figure}
   \centering
   \includegraphics[width=8.5cm]{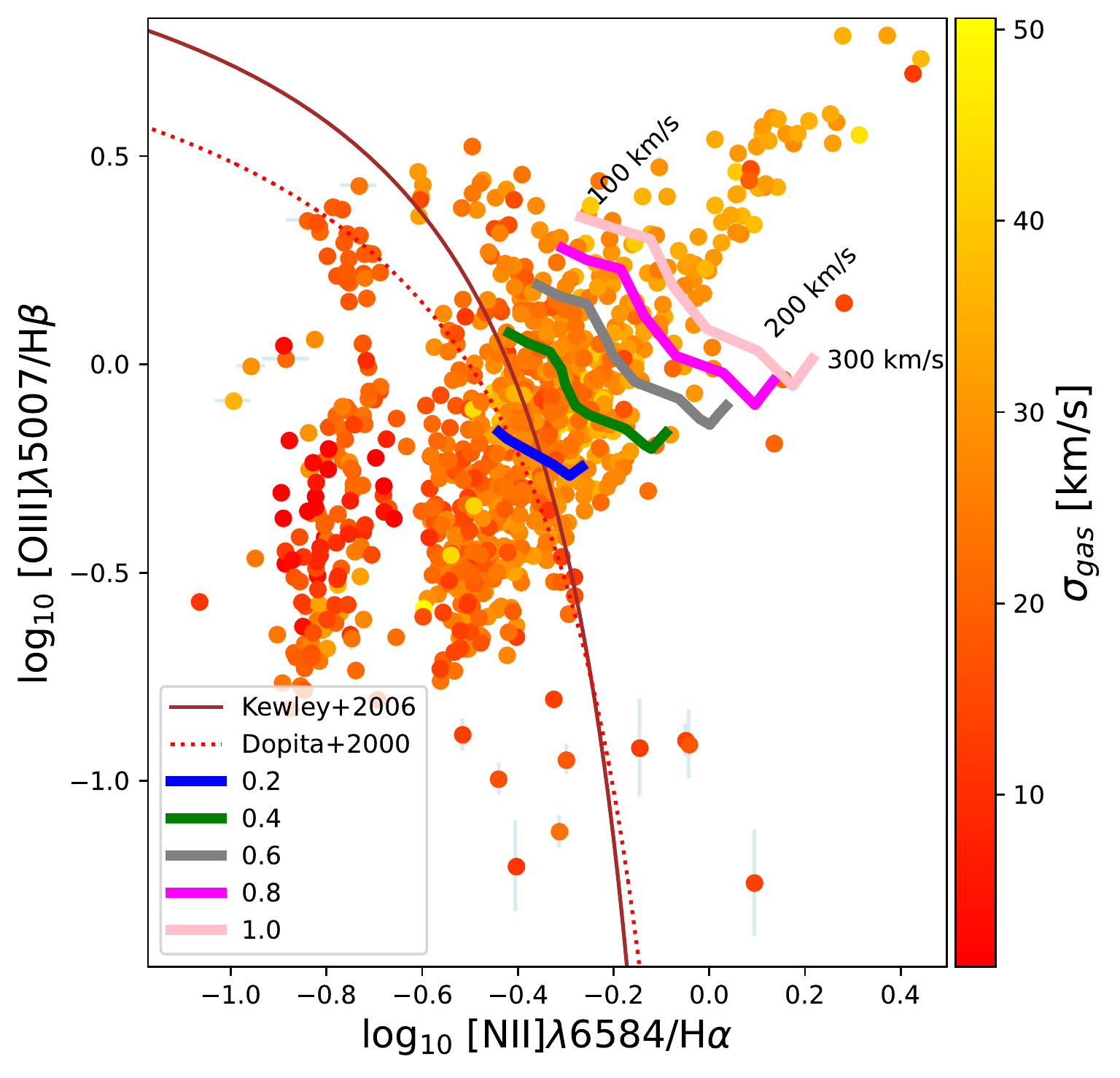}
      \caption{BPT-$\sigma$ diagram for DIG regions. The shock models of \citet{Ho2014} are overplotted for different shock velocities. The line segments are color-coded by shock fractions, as indicated in the legend.}
         \label{fig:bpt2}
   \end{figure}

\begin{figure}
   \centering
   \includegraphics[width=8.5cm]{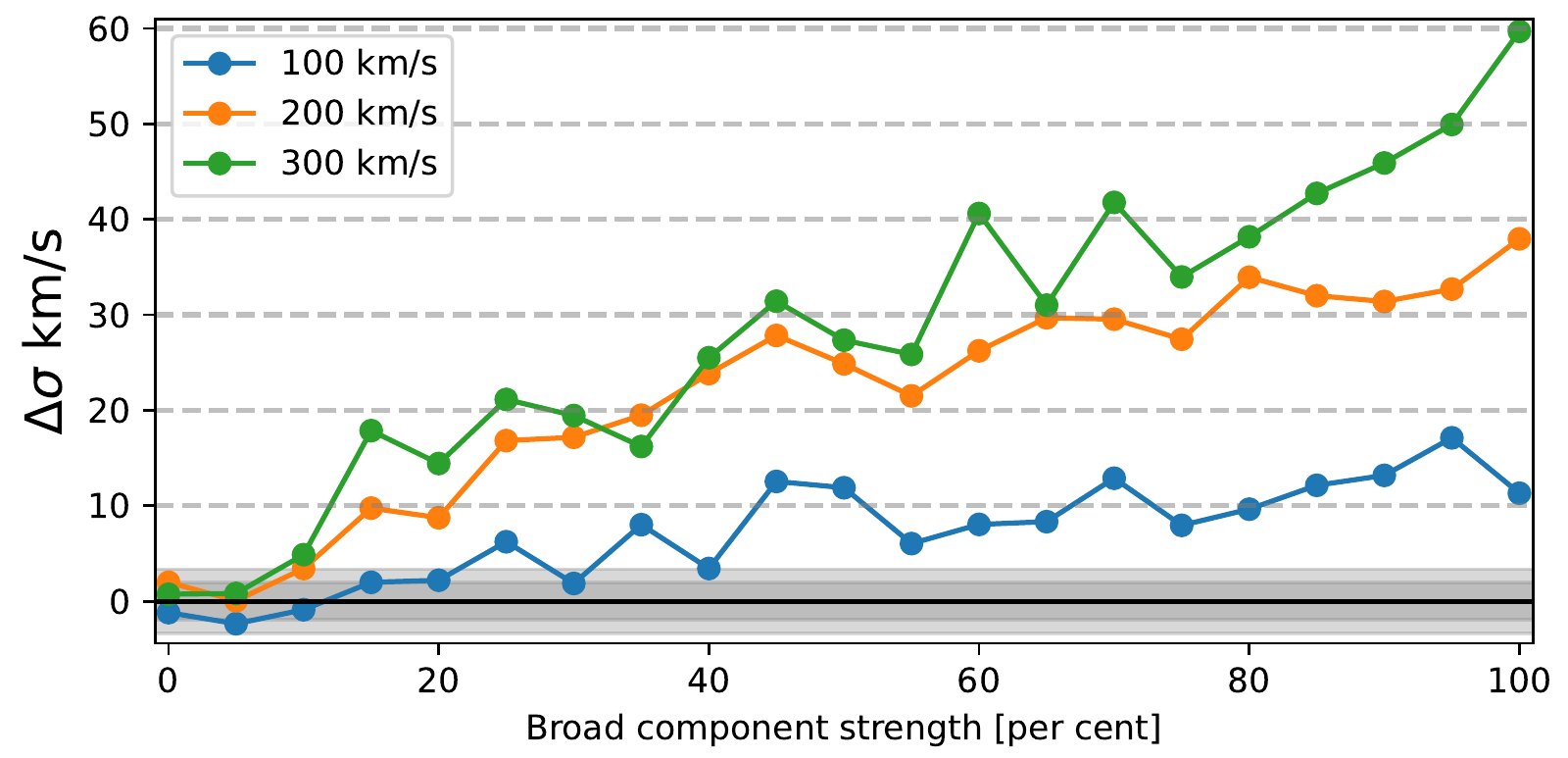}
      \caption{pPXF recovered velocity dispersion of composite lines and a single narrow line. The x-axis represents the per cent increase of the strength of the broad component relative to the narrow line. The gray-shaded horizontal area demarcates the average uncertainty in the DIG velocity dispersion measurement (single narrow line results).}
         \label{fig:testsigma}
   \end{figure}

Despite any contribution of HOLMES to the ionization of the DIG, it is likely that the DIG line ratios are further enhanced by contribution from shocks due to supernova explosions.  There is a positive correlation between the velocity dispersion and the shock velocity in a scenario in which multiple shocks propagate in random directions \citep{Ho2014}. In a BPT diagram, this correlation translates into an increased velocity dispersion with the \nii/\ha~or \sii/\ha~line ratios. Similar to \citet{Oparin2018} and \citet{DellaBruna2020}, we construct in Figure \ref{fig:bpt2}  the BPT diagram for the DIG and color-code the data points with their gas velocity dispersion. We observe an increase in velocity dispersion toward the top right corner of the diagram, in the LINER region. Overplotted are the MAPPINGS IV shock models of \citet{Ho2014} for different shock fractions ($20\mbox{-}100\%$). The line segments connect model grids of different shock velocities. All of the measured DIG velocity dispersions in any field (Table \ref{tab:kin}) are well below the shock velocities of the models ($100, 200,\text{ and } 300$ km/s), but we note that we did not account for multiple components in the line extraction with pPXF. To investigate the effect of fitting a single line to a narrow and broad line composite velocity dispersion, we performed the following test.
We created a narrow line with a velocity dispersion $\sigma=22$ km/s and three broad lines of $100, 200,\text{and } 300$ km/s each. Then we created composite lines by varying the contribution of the broad component from $5$ to $100\%$ of the strength of the narrow line. Finally, we extracted the lines with pPXF and examined the recovered velocity dispersion for each composite line as a function of the fractional contribution of the broad-line component. This is visualized in Figure \ref{fig:testsigma}, where we plot the residual of the velocity dispersion of the composite line and the narrow-line baseline ($22$ km/s). For a $100$ km/s broad component, up to $20\%$ of the contribution remains completely undetected by pPXF within the uncertainties, and it only increases the recovered velocity dispersion by $\sim10$ km/s up to $\sim90\%$ of the contribution by the broad component. This means that even if significantly strong broad components with a velocity dispersion of $100$ km/s exist, we would not detect them, and would instead measure a single line of a velocity dispersion that is marginally broadened by $\sim10$ km/s. Similar arguments can be made for the $200$ and $300$ km/s shocks, but with a much smaller contribution range of up to $\sim20\%$ of the line strength. We therefore cannot exclude the possibility that shocks of these velocities exist in our data.  

By comparing Figure \ref{fig:bpt2} to the upper panel of Figure \ref{fig:dighii}, we see that the shock models clearly predict a substantial contribution of shock ionization ($0.2\mbox{-}0.4$) for the DIG regions in fields \fieldc, \fieldb, \fieldd, and \fieldi, and a dominant contribution ($\sim0.6$) in fields \fielda~and \fieldj. Field \fielda~is the center of the galaxy and contains several morphological features reminiscent of ancient SNR arcs; this has been reported in \citet{Roth2018}. This field also contains several X-ray emission sources including a black hole X-ray binary \citep[][]{Read2001} and a bipolar microquasar jet \citep{McLeod2019}, and hence we intuitively expect field \fielda~to be kinematically active. Field \fieldj~is the inter-arm region, in which a shock can form along the trailing edges of a spiral arm \citep[e.g.,][]{Foyle2010}. Fields \fieldp~and \fieldq~are at the largest galactocentric distances in our data and are apparently consistent with a zero shock fraction contribution.  

Finally, in Section \ref{sec:ml2} in the appendix, we determine whether machine learning can detect the subtle differences between \hii~and DIG regions. However, the two object types completely overlap in the UMAP 2D projection, implying that the \hii~(DIG) regions show larger differences within themselves than when compared to the DIG (\hii) regions.

\section{Conclusions}\protect\label{sec:conclude}
We presented MUSE data of nine pointings in the flocculent spiral galaxy NGC 300. For all fields, we identified emission line regions via a dendrogram algorithm on preliminary \ha~images. We extracted nebular emission lines such as the Balmer and Paschen lines, and the strongest forbidden lines such as \sii, \siii, \oiii, and \nii~via pPXF. We proceeded to separate the detections into \hii, SNR, and PNe classes. All extracted properties are provided in a catalog of emission line objects. We also extracted DIG regions via dilated \hii-region masks. The \hii~sample is extremely faint, with $\log_{10}L_\mathrm{H\alpha} [\text{erg}/\text{s}]\in(33.2, 37.3)$, and a median of $34.7$. This makes our sample one of the faintest \hii~samples to date. The DIG is similarly faint, with $\log_{10}L_\mathrm{H\alpha} [\text{erg}/\text{s}]\in(33.6, 36.5)$, and a median of $34.9$. We examined the properties of the \hii~and DIG regions in terms of kinematics, abundances, density, and extinction. Our conclusions from this analysis are listed below.  
\begin{enumerate}
      \item The distribution of \siirat~ratios for the \hii~and DIG regions peak at the extreme low-density limit. For a handful of objects, we calculated a range of possible electron densities and found averages of $100$ and $\sim23$ cm$^{-3}$ for the \hii~and DIG regions, respectively.  
      \item Most of the DIG is consistent with no extinction, $E(B-V)=0$. The \hii~average is $E(B-V)=0.1$. For the handful of DIG objects with nonzero extinction, we find $\left<E(B-V)\right>=0.05$ on average. 
      \item The O3N2-calibration of the strong line method gives oxygen abundances $12+\log{O/H}$ consistent with those of the direct-$T_e$ method, while the S-calibration on average underestimates the oxygen metallicity. In the inner fields, the average metallicity shows a completely flat profile with galactocentric distance, without evidence of a metallicity gradient or increase in the central region. The metallicity of the DIG, $8.48$ on average, is consistent with that of the \hii~regions, $8.53$ on average, at any galactocentric distance.
      \item The \hii~and DIG regions move together with similar velocities. The DIG becomes marginally faster than the \hii~gas in the outer fields.
      \item The average velocity dispersion is $21.2\pm0.2$ km/s for \hii~gas and $25.1\pm0.1$ km/s for DIG. This difference can be fully attributed to the thermal velocity dispersion component, which would correspond to the DIG being $1.8$kK hotter than the \hii~regions. 
      \item The DIG has an increased velocity dispersion in the central galactic region, consistent with models of a dominant ($\sim60\%$) contribution of shocks to the DIG ionization. 
      \item The DIG fraction per field varies between $42\mbox{-}77\%$ of \ha. The inter-arm region field \fieldj~shows a much lower DIG fraction of $15\%$. These numbers are consistent with the literature.
      \item The DIG has a lower ionization state than \hii~gas, as traced by the high-to-low ionization line ratio \siii/\siisum.
      \item Signs of a contribution to DIG ionization by hot low-mass evolved stars are detected:
      \begin{enumerate}[i)]
          \item a completely flat trend of the DIG \siii/\siisum~ratio with \ha~surface brightness, in contrast to a positive correlation for \hii~regions, 
          \item low ionization line ratios such as \sii/\ha~and \nii/\ha~ show a systematic enhancement toward small galactocentric distances, in contrast to a flat trend for \hii~regions.
      \end{enumerate}
          \item Unsupervised machine-learning algorithms such as UMAP/HDBscan are unable to distinguish between DIG and \hii~regions, implying that both the DIG and the \hii~regions are so heterogeneous that the differences within them are larger than between them. 
          \item The differences between extremely faint \hii~and DIG regions follow the same trends as their brighter counterparts.
   \end{enumerate}

On a final note, we acknowledge that we have completely omitted any comparison of the number of available ionizing photons and the necessary photons that the DIG regions imply. This is beyond the scope of this paper and will be addressed in a separate paper that is currently in preparation. 

\begin{appendix} %First appendix
\section{Attempt of classifying emission line objects via machine learning}\protect\label{sec:ml}
Machine learning has been developed to facilitate the analysis of large data, and we wish to test whether it can help classify the objects into \hii, PNe, and SNR. An unsupervised machine-learning classification algorithm is apparently most suitable for this task; t-SNE\footnote{t-distributed stochastic neighbor embedding} \citep{vanDerMaaten2008} and UMAP\footnote{uniform manifold approximation and projection for dimension reduction} \citep{McInnes2018} are used widely. They reduce the dimensionality of the data and can project the data in 2D, clustering points of significantly similar features.  We selected UMAP because it appears to have advantages over t-SNE in the selection of its cost function, which better preserves global structure in the projection. 

\begin{table}
    \caption{Features for the 2D projection algorithm runs.}
    \label{tab:mlfeatures}
    \centering
    \begin{tabular}{ll}
    \hline\hline
    $\#$& Features\\
    2 &  [\ion{O}{iii}]/\hb, [\ion{N}{ii}]/\ha \\
    3 &  [\ion{O}{iii}]/\hb, [\ion{N}{ii}]/\ha, [\ion{S}{ii}] ratio \\
    4 &  [\ion{O}{iii}]/\hb, [\ion{N}{ii}]/\ha, [\ion{S}{ii}] ratio, $N_{pix}$ \\
    5 &  [\ion{O}{iii}]/\hb, [\ion{N}{ii}]/\ha, [\ion{S}{ii}] ratio, $N_{pix}$, $\sigma_\mathrm{gas}$ \\
    6 &  [\ion{O}{iii}]/\hb, [\ion{N}{ii}]/\ha, [\ion{S}{ii}] ratio, $N_{pix}$, $\sigma_\mathrm{gas}$, D \\
    \end{tabular}
\end{table}

 The choice of features for the machine-learning algorithm is more important than the selection of the algorithm. The features have to be relevant to the desired classification outcome. We tested several sets of features and list them in Table \ref{tab:mlfeatures}. It is possible to add more features and explore subgroups of each object class (\hii, PNe, and SNR), but the more features are added, the harder the interpretation of the resulting clustering projection. Another important step is the normalization of the features by removing the mean and scaling by the variance. This is necessary in order to ensure that all features are within the same numerical range and hence have an equal probability to influence the learning algorithm. A problem we immediately ran into is that emission line fluxes are not independent features, and hence scaling along columns does not seem suitable. We have explored both options, scaling per row, that is, scaling each individual object by some feature, and scaling per column, that is, scaling one feature across all objects. 

The most important UMAP parameter is \verb+n_neighbours+, which determines the number of neighboring points that influence the clustering of dendrograms in the 2D projection. We experimented with \verb+n_neighbours+$=5,6,7,8,9,10,12,15,16,17,18,19,20,21,22,23,35,70,100$, and $300$. Increasing the value of \verb+n_neighbours+ resulted in an ever more tightly clustered projection. For display purposes, we chose \verb+n_neighbours+$=23$, but we verified that all $20$ projections give very similar results in terms of cluster identification and classification.  In addition, UMAP is a stochastic algorithm in the way it optimizes and speeds up the various approximation steps. Hence, different runs of UMAP can produce slightly different results, depending on the random state that is selected. To ensure reproducibility, we set the random seed to $42$ in this section. For all $20$ values of \verb+n_neighbours,+ we verified that changing the random state to one, for example,$\text{}$ gives very similar results and does not change the outcome. Finally, we set \verb+min_dist+ to zero, meaning that the points were allowed to cluster so tightly in the 2D projection that they completely overlapped, and \verb+spread+ to 5, which increases the distance between clusters but does not dilute the clustering itself. 

\begin{figure*}
   \centering
   \includegraphics[width=16cm]{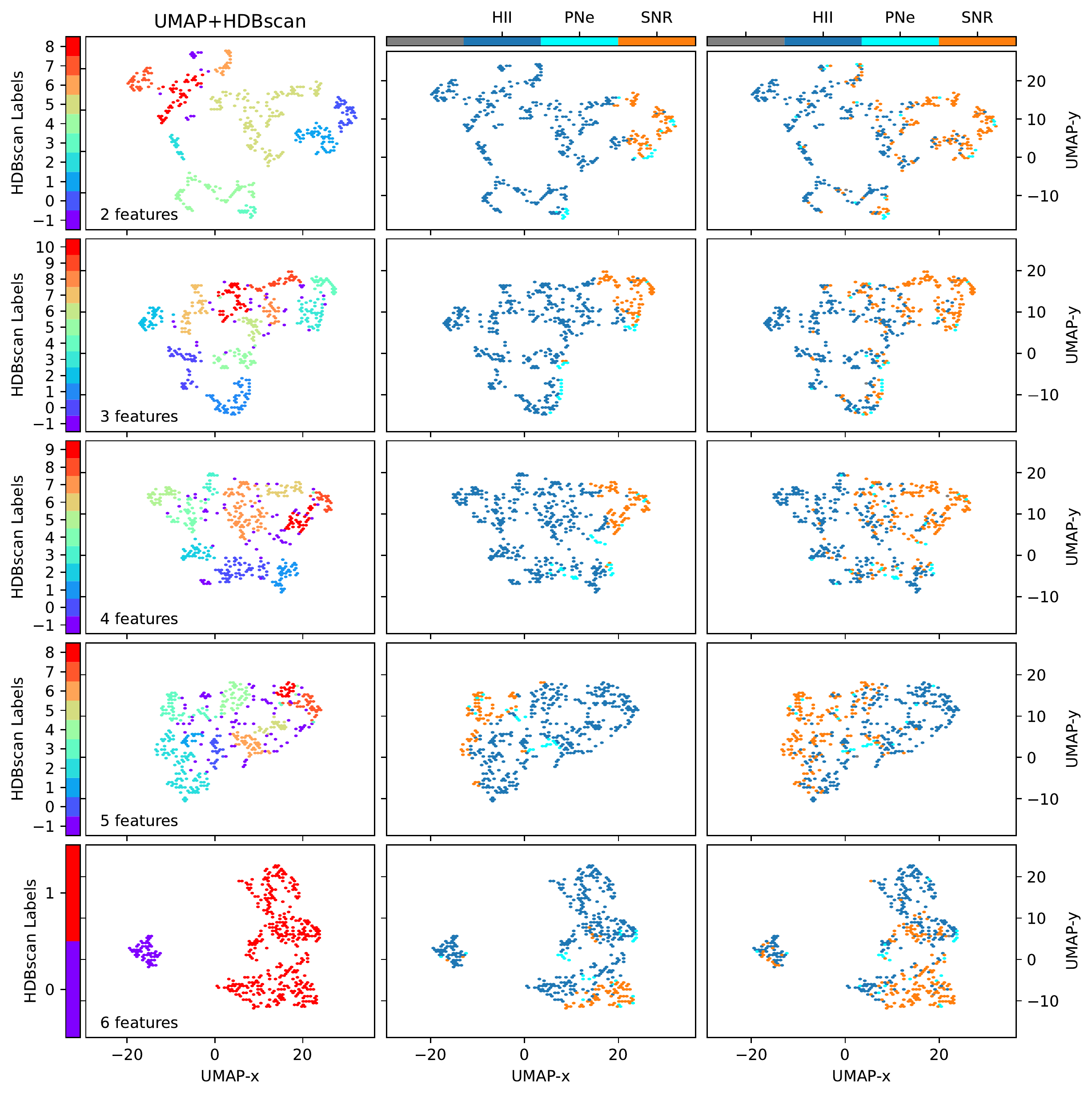}
      \caption{Sample classification via machine learning. (Left) UMAP 2D projection with HDBscan cluster labels. (Middle) UMAP 2D projection with \nii-based BPT diagram labels. (Right) UMAP 2D projection with \siisum-based BPT diagram labels. The inset text shows the number of features in each run. All displayed runs are with n\_neighbours$=20$.  }
         \label{fig:ml_res}
   \end{figure*}

The UMAP results for \verb+n_neighbours+ $=20$ and \verb+random_seed+ $=1$ are shown in Figure \ref{fig:ml_res}. No clear separation of \hii, SNR, and PNe is obtained for any feature set. At best, the run with two features pushes the SNR and PNe towards the edges of the 2D projection, as shown in the middle and right panels in the first row of Figure \ref{fig:ml_res}. This is not reflected in the clustering found by HDBscan in the first panel, however. Instead, all three classes of objects in the different subclusters have identical labels. The six-feature run shows two well-defined clusters in the HDBscan panel, which is the effect of adding the galactocentric distance as a feature in this run. The smaller cluster contains fields \fieldp~and \fieldq, and the larger cluster contains all remaining fields. We conclude that unsupervised machine-learning algorithms such as UMAP and t-SNE are not suitable for classifying emission line objects into \hii, SNR, and PNe. 

%
%                                                One column figure
%----------------------------------------------------------------- 
%----------------------------------------------------------------- 

\section{Justification of the \siirat~cutoff}\protect\label{sec:S2S2}
The theoretical limits of \siirat~are $0.6806\leq$ \siirat$\leq 2.351$. 
Simply due to statistics, we would expect there to be a distribution of \siirat~measurements around some characteristic (typical) value. In NGC 300, the characteristic value is at $0.675\pm0.036$. This poses a problem because it is located practically at the theoretical limit, and we now have to decide how many of the data points falling to the left of the limit need to be included in the sample. 

\begin{figure}%f1
\includegraphics[width=8.8cm]{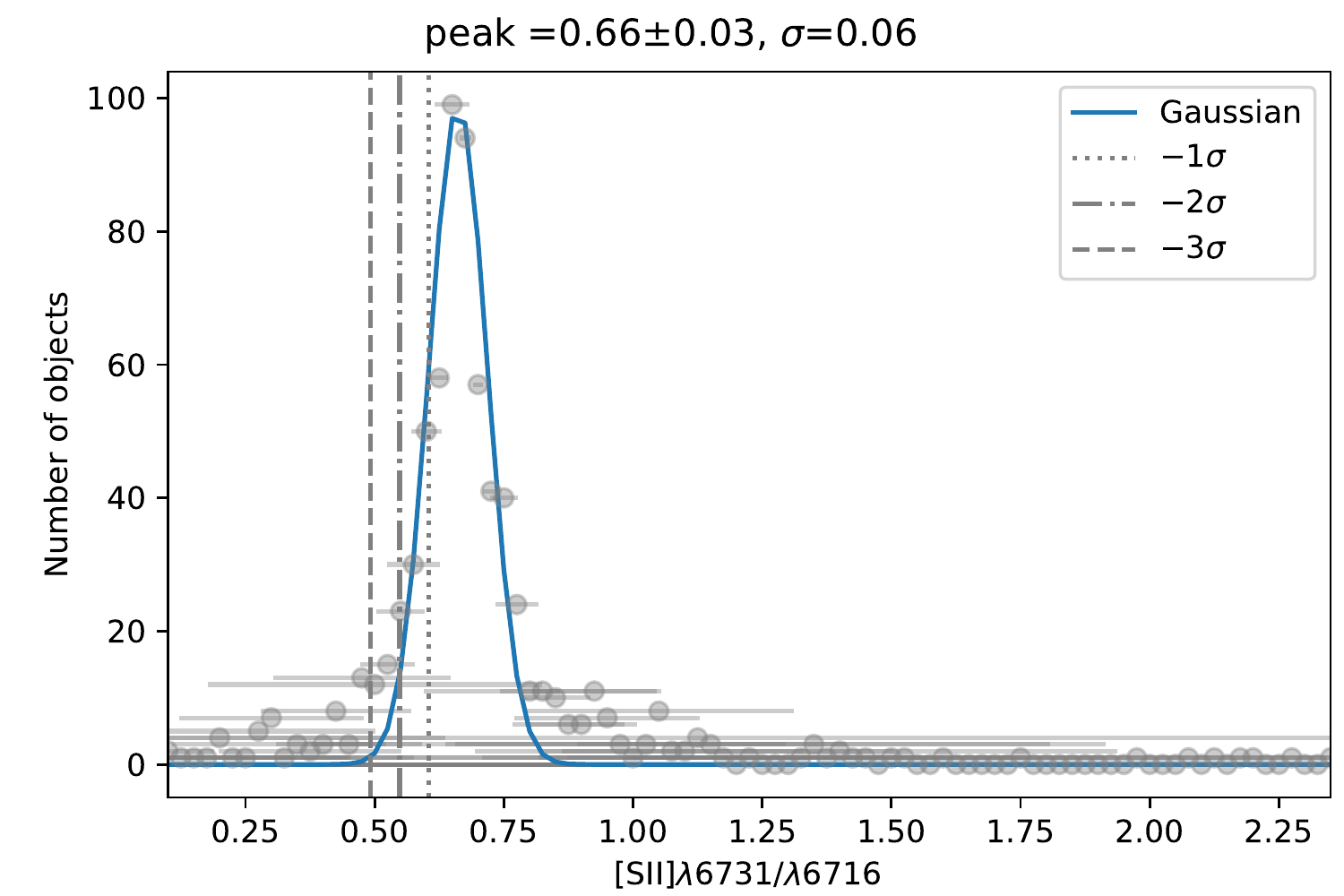}
\includegraphics[width=8.8cm]{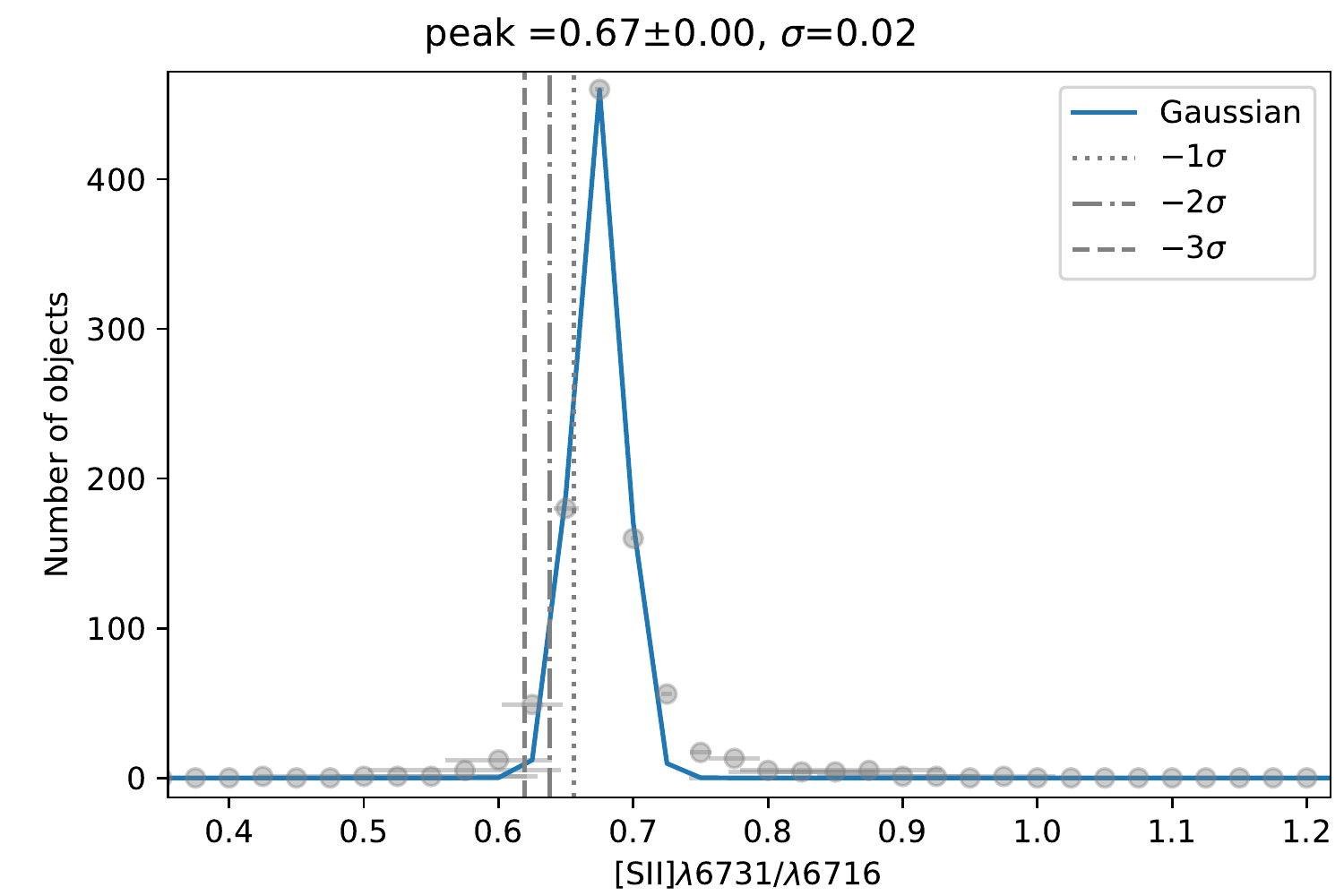}
\caption{Justification of the \siirat~cutoff. (Top) Histogram of the \siirat~for background-subtracted regions (gray markers with error bars), with the best-fit Gaussian model (solid blue line) and vertical lines indicating the location of $-1\sigma$ (dotted line), $-2\sigma$ (dash-dotted), and $-3\sigma$ (dashed) plotted for convenience. The uncertainties are propagated from the points in each bin. (Bottom) Same as top panel, but for DIG regions.}
\label{fig:s2s2hist}
\end{figure}

When the line fluxes of the \siidoublet~individually follow normal distributions, their ratio is not necessarily a Gaussian, although it can be approximated by a Gaussian if the coefficient of variation CV$=\sigma_\mathrm{gauss}/\mu_\textrm{gauss}$ is sufficiently small \citep[][CV$<0.1$]{DiazFrances2013}. The top panel of Figure \ref{fig:s2s2hist} shows a histogram of the \siirat~with propagated uncertainties of the data points falling inside each bin. We used astropy to fit a 1D Gaussian to the data points and obtained $\sigma_\mathrm{gauss}=0.06$ and a peak location $\mu_\textrm{gauss}=0.66$. This gives CV$=0.09,$ and hence we can safely continue to approximate the distribution of \siirat~data points with a Gaussian.

To the left of the Gaussian peak, up to $34.1\%$, $13.6\%$, and $2.1\%$ of the data points are expected to be found in the three intervals ($-1\sigma$, peak), ($-2\sigma$, $-1\sigma$), and ($-3\sigma$, $-2\sigma$), respectively. For the background-subtracted spectra, we calculated the fractions of data points in these intervals as $14.8\%$, $7.3\%$, and $3.7\%$, respectively. The first two intervals are consistent with the data being normally distributed, and hence we kept these points in the final sample. In contrast, we expect only $2.1\%$ of the data to fall between $-3\sigma$ and $-2\sigma$, but instead, we measure $3.7\%$, implying that in addition to statistical scatter there is also some noisy data with low S/N. In summary, we discarded all background-subtracted spectra that were associated with \siirat~more than $-2\sigma$ away from the peak. This corresponds to a limit of \siirat$\leq 0.548$. In a similar fashion, the bottom panel of Figure \ref{fig:s2s2hist} shows the histogram and fit for the DIG data. We discarded all DIG spectra with \siirat$=\leq 0.638$.

We note that we cannot apply a cutoff on the right side of the Gaussian peak because in addition to statistics, the physical structures present in the NGC\,300 fields affect the distribution. 

\section{Stellar kinematics}
\begin{figure}%f1
\includegraphics[width=8.5cm]{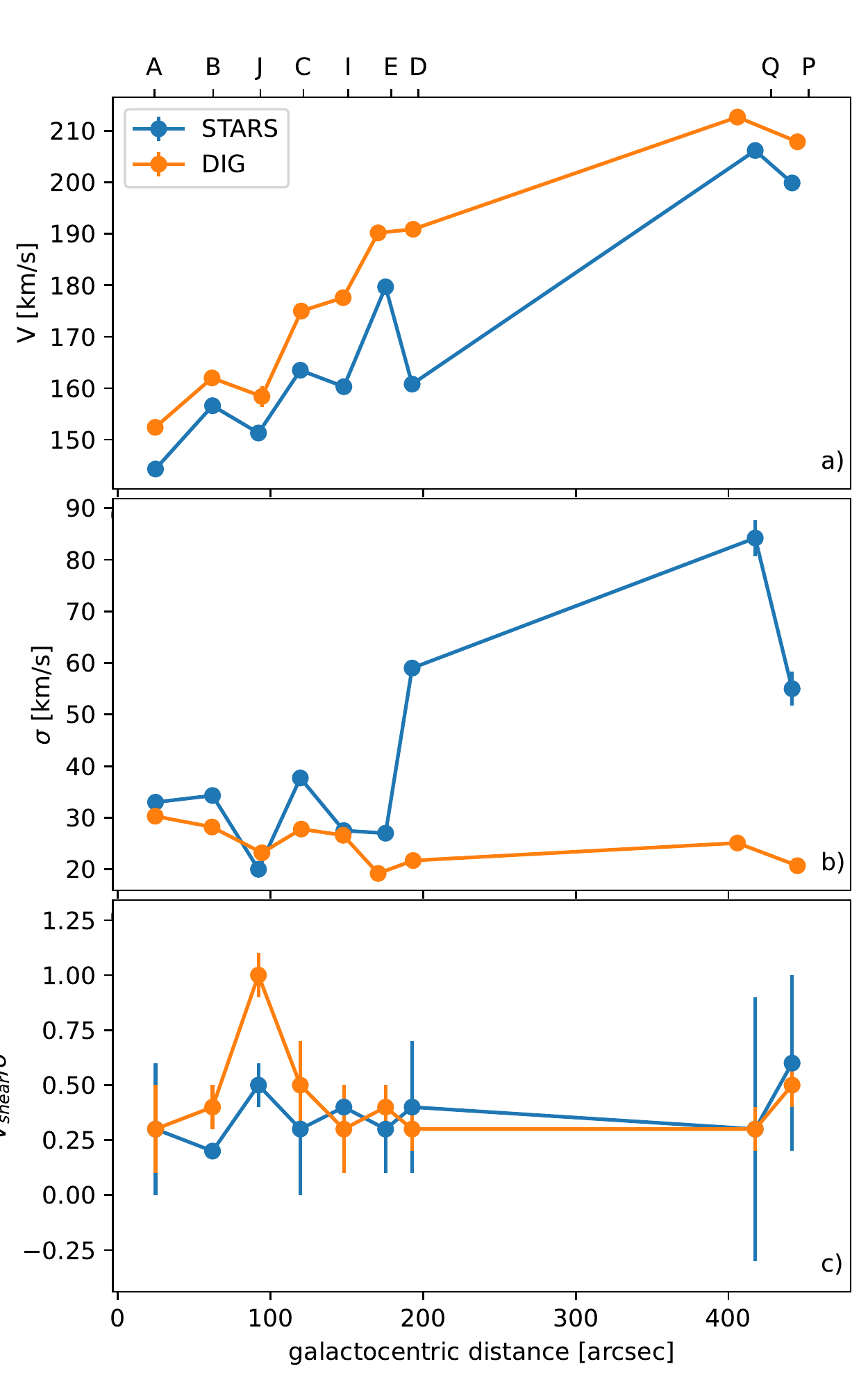}
\caption{Sample stellar kinematics. (a) Velocity. (b) Velocity dispersion. (c) $V_{\rm shear}/\sigma$ for the stellar continuum and the DIG. Panels b) and c) have the same legend as panel a). All panels share the galactocentric distance in arcsec as x-axis. The twin axis shows the corresponding field labels. }
\label{fig:starkine}
\end{figure}

\protect\label{sec:starkine}
For completeness, we present the stellar kinematics in Figure \ref{fig:starkine}.

\section{Stacked spectra}
\begin{figure*}
\includegraphics[width=17.5cm]{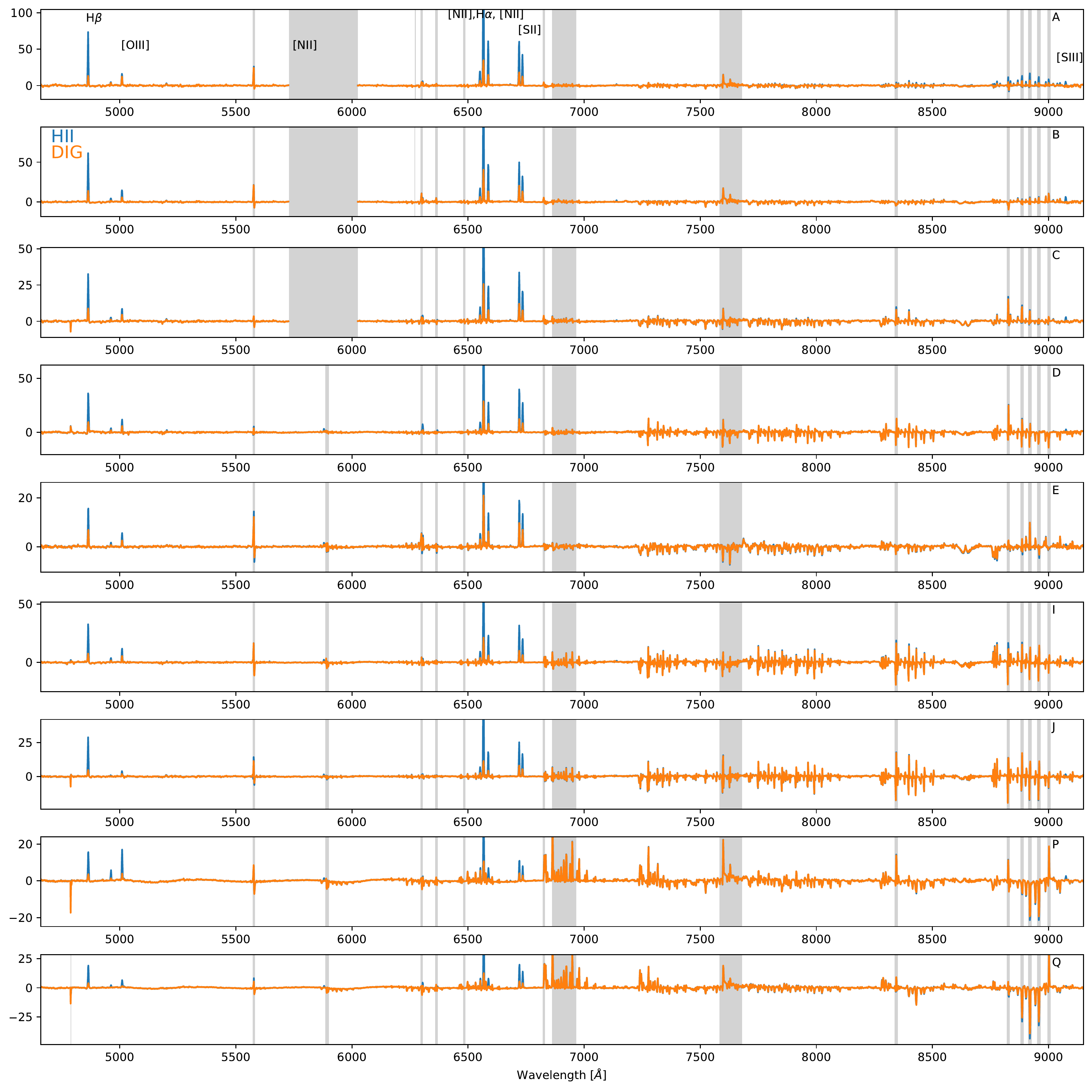}
\caption{Residual average stacked spectra for \hii~(blue) and DIG (orange) for each field. The field is indicated in the top right corner of each panel. The main nebular emission lines that are of interest in this work are annotated in the top panel. Bad pixels that were omitted in the pPXF fit are grayed out. The y-axis has units of $10^{-20}$erg/s/cm$^2$/$\AA$ and is truncated at half the \ha~flux.}
\label{fig:stackspec}
\end{figure*}

\protect\label{sec:stackspec}
The average stacked spectra per field are shown in Figure \ref{fig:stackspec}. For both \hii~and DIG spectra, only the residuals of the original spectra minus the pPXF best-fit stellar continuum are shown. For all fields, the figures are capped at half of the maximum \ha~peak value in order to provide more detail for the fainter lines. We note that the DIG contribution and the initial stellar background are removed from the \hii~regions already before the pPXF fit, as explained in Section \ref{sec:stellarbkg}.  

\section{Attempt to separate DIG and \hii~regions via machine learning}
\begin{figure*}
\includegraphics[width=17.5cm]{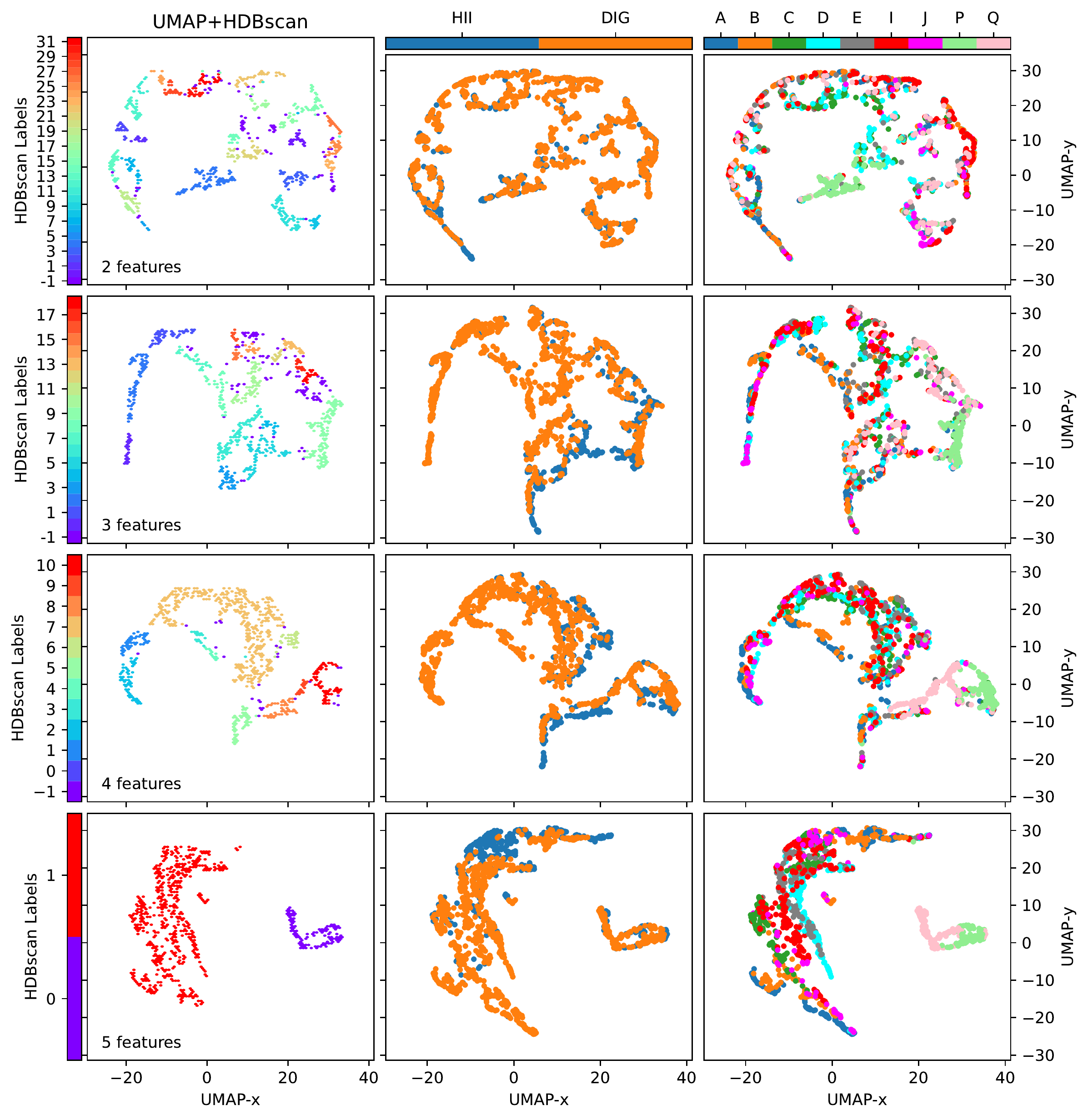}
\caption{Sample classification in to \hii~and DIG via machine learning. (Left) UMAP 2D projection with HDBscan cluster labels. (Middle) UMAP 2D projection with DIG/\hii~labels. (Right) UMAP 2D projection with field labels. The inset text shows the number of features in each run. All displayed runs are with n\_neighbours$=20$. }
\label{fig:ml2}
\end{figure*}

\protect\label{sec:ml2}
With the same method as in Section \ref{sec:ml}, but for the feature sets in Table \ref{tab:mlfeatures2}, we determined whether the UMAP projection separates the DIG and \hii~regions. The result is shown in Figure \ref{fig:ml2}. The two types of objects overlap in all runs, and moreover, the labeling of the HDBscan algorithm implies larger differences within different subgroups in the DIG or the \hii~regions than between the DIG and \hii~regions as a whole.  

\begin{table}
    \caption{UMAP features for DIG/\hii~separation.}
    \label{tab:mlfeatures2}
    \centering
    \begin{tabular}{ll}
    \hline\hline
    $\#$& Features\\
    2 &  [\ion{S}{iii}]/[\ion{S}{ii}], $\Sigma_{H\alpha}$ \\
    3 &  [\ion{S}{iii}]/[\ion{S}{ii}], $\Sigma_{H\alpha}$, [\ion{S}{ii}]/\ha \\
    4 &  [\ion{S}{iii}]/[\ion{S}{ii}], $\Sigma_{H\alpha}$, [\ion{S}{ii}]/\ha, [\ion{N}{ii}]/\ha \\
    5 &  [\ion{S}{iii}]/[\ion{S}{ii}], $\Sigma_{H\alpha}$, [\ion{S}{ii}]/\ha, [\ion{N}{ii}]/\ha, D \\
    \end{tabular}
\end{table}

\end{appendix}
\begin{acknowledgements}
      PMW received support from BMBF Verbundforschung (ErUM, project VLT-BlueMUSE, grant 05A20BAB).
      NC gratefully acknowledges funding from the Deutsche Forschungsgemeinschaft (DFG) – CA 2551/1-1. 
      CM acknowledges support from grant UNAM / PAPIIT - IN101220. 
      %Part of this work was supported by the German
      %\emph{Deut\-sche For\-schungs\-ge\-mein\-schaft, DFG\/} project
      %number ????.
\end{acknowledgements}

\bibliographystyle{aa}
\bibliography{ngc300}

\begin{thebibliography}{98}
\expandafter\ifx\csname natexlab\endcsname\relax\def\natexlab#1{#1}\fi

\bibitem[{{Alloin} {et~al.}(1979){Alloin}, {Collin-Souffrin}, {Joly}, \&
  {Vigroux}}]{Alloin1979}
{Alloin}, D., {Collin-Souffrin}, S., {Joly}, M., \& {Vigroux}, L. 1979, \aap,
  78, 200

\bibitem[{{Amayo} {et~al.}(2021){Amayo}, {Delgado-Inglada}, \&
  {Stasi{\'n}ska}}]{Amayo2021}
{Amayo}, A., {Delgado-Inglada}, G., \& {Stasi{\'n}ska}, G. 2021, \mnras, 505,
  2361

\bibitem[{{Bacon} {et~al.}(2010){Bacon}, {Accardo}, {Adjali}, {Anwand},
  {Bauer}, {Biswas}, {Blaizot}, {Boudon}, {Brau-Nogue}, {Brinchmann},
  {Caillier}, {Capoani}, {Carollo}, {Contini}, {Couderc}, {Daguis{\'e}},
  {Deiries}, {Delabre}, {Dreizler}, {Dubois}, {Dupieux}, {Dupuy}, {Emsellem},
  {Fechner}, {Fleischmann}, {Fran{\c c}ois}, {Gallou}, {Gharsa}, {Glindemann},
  {Gojak}, {Guiderdoni}, {Hansali}, {Hahn}, {Jarno}, {Kelz}, {Koehler},
  {Kosmalski}, {Laurent}, {Le Floch}, {Lilly}, {Lizon}, {Loupias}, {Manescau},
  {Monstein}, {Nicklas}, {Olaya}, {Pares}, {Pasquini}, {P{\'e}contal-Rousset},
  {Pell{\'o}}, {Petit}, {Popow}, {Reiss}, {Remillieux}, {Renault}, {Roth},
  {Rupprecht}, {Serre}, {Schaye}, {Soucail}, {Steinmetz}, {Streicher}, {Stuik},
  {Valentin}, {Vernet}, {Weilbacher}, {Wisotzki}, \& {Yerle}}]{Bacon2010}
{Bacon}, R., {Accardo}, M., {Adjali}, L., {et~al.} 2010, in Proc.~{SPIE}, Vol.
  7735, {Ground-based and Airborne Instrumentation for Astronomy III}

\bibitem[{{Baldwin} {et~al.}(1981){Baldwin}, {Phillips}, \&
  {Terlevich}}]{Baldwin1981}
{Baldwin}, J.~A., {Phillips}, M.~M., \& {Terlevich}, R. 1981, \pasp, 93, 5

\bibitem[{{Belfiore} {et~al.}(2022){Belfiore}, {Santoro}, {Groves},
  {Schinnerer}, {Kreckel}, {Glover}, {Klessen}, {Emsellem}, {Blanc}, {Congiu},
  {Barnes}, {Boquien}, {Chevance}, {Dale}, {Diederik Kruijssen}, {Leroy},
  {Pan}, {Pessa}, {Schruba}, \& {Williams}}]{Belfiore2022}
{Belfiore}, F., {Santoro}, F., {Groves}, B., {et~al.} 2022, \aap, 659, A26

\bibitem[{Biemont {et~al.}(1999)Biemont, Fremat, \& Quinet}]{BIEMONT1999117}
Biemont, E., Fremat, Y., \& Quinet, P. 1999, Atomic Data and Nuclear Data
  Tables, 71, 117

\bibitem[{{Binette} {et~al.}(2009){Binette}, {Flores-Fajardo}, {Raga},
  {Drissen}, \& {Morisset}}]{Binette2009}
{Binette}, L., {Flores-Fajardo}, N., {Raga}, A.~C., {Drissen}, L., \&
  {Morisset}, C. 2009, \apj, 695, 552

\bibitem[{{Bradley} {et~al.}(2006){Bradley}, {Knapen}, {Beckman}, \&
  {Folkes}}]{Bradley2006}
{Bradley}, T.~R., {Knapen}, J.~H., {Beckman}, J.~E., \& {Folkes}, S.~L. 2006,
  \aap, 459, L13

\bibitem[{{Bresolin} {et~al.}(2009){Bresolin}, {Gieren}, {Kudritzki},
  {Pietrzy{\'n}ski}, {Urbaneja}, \& {Carraro}}]{Bresolin2009}
{Bresolin}, F., {Gieren}, W., {Kudritzki}, R.-P., {et~al.} 2009, \apj, 700, 309

\bibitem[{{Bresolin} {et~al.}(2005){Bresolin}, {Pietrzy{\'n}ski}, {Gieren}, \&
  {Kudritzki}}]{Bresolin2005}
{Bresolin}, F., {Pietrzy{\'n}ski}, G., {Gieren}, W., \& {Kudritzki}, R.-P.
  2005, \apj, 634, 1020

\bibitem[{{Cappellari}(2017)}]{Cappellari2017}
{Cappellari}, M. 2017, MNRAS, 466, 798

\bibitem[{{Cappellari} \& {Emsellem}(2004)}]{Cappellari2004}
{Cappellari}, M. \& {Emsellem}, E. 2004, PASP, 116, 138

\bibitem[{{Cardelli} {et~al.}(1989){Cardelli}, {Clayton}, \&
  {Mathis}}]{Cardelli1989}
{Cardelli}, J.~A., {Clayton}, G.~C., \& {Mathis}, J.~S. 1989, \apj, 345, 245

\bibitem[{{Chevance} {et~al.}(2020){Chevance}, {Kruijssen}, {Hygate},
  {Schruba}, {Longmore}, {Groves}, {Henshaw}, {Herrera}, {Hughes}, {Jeffreson},
  {Lang}, {Leroy}, {Meidt}, {Pety}, {Razza}, {Rosolowsky}, {Schinnerer},
  {Bigiel}, {Blanc}, {Emsellem}, {Faesi}, {Glover}, {Haydon}, {Ho}, {Kreckel},
  {Lee}, {Liu}, {Querejeta}, {Saito}, {Sun}, {Usero}, \&
  {Utomo}}]{Chevance2020}
{Chevance}, M., {Kruijssen}, J.~M.~D., {Hygate}, A. P.~S., {et~al.} 2020,
  \mnras, 493, 2872

\bibitem[{{Collins} \& {Rand}(2001)}]{Collins2001}
{Collins}, J.~A. \& {Rand}, R.~J. 2001, \apj, 551, 57

\bibitem[{{Deharveng} {et~al.}(1988){Deharveng}, {Caplan}, {Lequeux},
  {Azzopardi}, {Breysacher}, {Tarenghi}, \& {Westerlund}}]{Deharveng1988}
{Deharveng}, L., {Caplan}, J., {Lequeux}, J., {et~al.} 1988, \aaps, 73, 407

\bibitem[{{Della Bruna} {et~al.}(2020){Della Bruna}, {Adamo}, {Bik},
  {Fumagalli}, {Walterbos}, {{\"O}stlin}, {Bruzual}, {Calzetti}, {Charlot},
  {Grasha}, {Smith}, {Thilker}, \& {Wofford}}]{DellaBruna2020}
{Della Bruna}, L., {Adamo}, A., {Bik}, A., {et~al.} 2020, \aap, 635, A134

\bibitem[{{den Brok} {et~al.}(2020){den Brok}, {Carollo}, {Erroz-Ferrer},
  {Fagioli}, {Brinchmann}, {Emsellem}, {Krajnovi{\'c}}, {Marino}, {Onodera},
  {Tacchella}, {Weilbacher}, \& {Woo}}]{denBrok2020}
{den Brok}, M., {Carollo}, C.~M., {Erroz-Ferrer}, S., {et~al.} 2020, \mnras,
  491, 4089

\bibitem[{D{\'i}az-Franc{\'e}s \& Rubio(2013)}]{DiazFrances2013}
D{\'i}az-Franc{\'e}s, E. \& Rubio, F.~J. 2013, Statistical Papers, 54, 309

\bibitem[{{Dopita} {et~al.}(2000){Dopita}, {Kewley}, {Heisler}, \&
  {Sutherland}}]{Dopita2000}
{Dopita}, M.~A., {Kewley}, L.~J., {Heisler}, C.~A., \& {Sutherland}, R.~S.
  2000, \apj, 542, 224

\bibitem[{{Faesi} {et~al.}(2016){Faesi}, {Lada}, \& {Forbrich}}]{Faesi2016}
{Faesi}, C.~M., {Lada}, C.~J., \& {Forbrich}, J. 2016, \apj, 821, 125

\bibitem[{{Faesi} {et~al.}(2014){Faesi}, {Lada}, {Forbrich}, {Menten}, \&
  {Bouy}}]{Faesi2014}
{Faesi}, C.~M., {Lada}, C.~J., {Forbrich}, J., {Menten}, K.~M., \& {Bouy}, H.
  2014, \apj, 789, 81

\bibitem[{{Ferguson} {et~al.}(1996){Ferguson}, {Wyse}, {Gallagher}, \&
  {Hunter}}]{Ferguson1996}
{Ferguson}, A. M.~N., {Wyse}, R. F.~G., {Gallagher}, J.~S., I., \& {Hunter},
  D.~A. 1996, \aj, 111, 2265

\bibitem[{{Ferland} {et~al.}(2017){Ferland}, {Chatzikos}, {Guzm{\'a}n},
  {Lykins}, {van Hoof}, {Williams}, {Abel}, {Badnell}, {Keenan}, {Porter}, \&
  {Stancil}}]{Ferland2017}
{Ferland}, G.~J., {Chatzikos}, M., {Guzm{\'a}n}, F., {et~al.} 2017, \rmxaa, 53,
  385

\bibitem[{{Fesen} {et~al.}(1985){Fesen}, {Blair}, \& {Kirshner}}]{Fesen1985}
{Fesen}, R.~A., {Blair}, W.~P., \& {Kirshner}, R.~P. 1985, \apj, 292, 29

\bibitem[{{Flores-Fajardo} {et~al.}(2011){Flores-Fajardo}, {Morisset},
  {Stasi{\'n}ska}, \& {Binette}}]{FloresFajardo2011}
{Flores-Fajardo}, N., {Morisset}, C., {Stasi{\'n}ska}, G., \& {Binette}, L.
  2011, \mnras, 415, 2182

\bibitem[{Foyle {et~al.}(2010)Foyle, Rix, Walter, \& Leroy}]{Foyle2010}
Foyle, K., Rix, H.-W., Walter, F., \& Leroy, A.~K. 2010, The Astrophysical
  Journal, 725, 534

\bibitem[{{Frew} \& {Parker}(2010)}]{Frew2010}
{Frew}, D.~J. \& {Parker}, Q.~A. 2010, \pasa, 27, 129

\bibitem[{{Galvin} {et~al.}(2012){Galvin}, {Filipovi{\'c}}, {Crawford}, {Wong},
  {Payne}, {De Horta}, {White}, {Tothill}, {Dra{\v{s}}kovi{\'c}}, {Pannuti},
  {Grimes}, {Cahall}, {Millar}, \& {Laine}}]{Galvin2012}
{Galvin}, T.~J., {Filipovi{\'c}}, M.~D., {Crawford}, E.~J., {et~al.} 2012,
  \apss, 340, 133

\bibitem[{{Gazak} {et~al.}(2015){Gazak}, {Kudritzki}, {Evans}, {Patrick},
  {Davies}, {Bergemann}, {Plez}, {Bresolin}, {Bender}, {Wegner}, {Bonanos}, \&
  {Williams}}]{Gazak2015}
{Gazak}, J.~Z., {Kudritzki}, R., {Evans}, C., {et~al.} 2015, \apj, 805, 182

\bibitem[{{Gieren} {et~al.}(2005){Gieren}, {Pietrzy{\'n}ski}, {Soszy{\'n}ski},
  {Bresolin}, {Kudritzki}, {Minniti}, \& {Storm}}]{Gieren2005}
{Gieren}, W., {Pietrzy{\'n}ski}, G., {Soszy{\'n}ski}, I., {et~al.} 2005, \apj,
  628, 695

\bibitem[{{Gross} {et~al.}(2019){Gross}, {Williams}, {Pannuti}, {Binder},
  {Garofali}, \& {Hanvey}}]{Gross2019}
{Gross}, J., {Williams}, B.~F., {Pannuti}, T.~G., {et~al.} 2019, \apj, 877, 15

\bibitem[{{Haffner} {et~al.}(2009){Haffner}, {Dettmar}, {Beckman}, {Wood},
  {Slavin}, {Giammanco}, {Madsen}, {Zurita}, \& {Reynolds}}]{Haffner2009}
{Haffner}, L.~M., {Dettmar}, R.~J., {Beckman}, J.~E., {et~al.} 2009, Reviews of
  Modern Physics, 81, 969

\bibitem[{{Hakobyan} {et~al.}(2007){Hakobyan}, {Petrosian}, {Yeghazaryan}, \&
  {Boulesteix}}]{Hakobyan2007}
{Hakobyan}, A.~A., {Petrosian}, A.~R., {Yeghazaryan}, A.~A., \& {Boulesteix},
  J. 2007, Astrophysics, 50, 426

\bibitem[{{Herenz} {et~al.}(2016){Herenz}, {Gruyters}, {Orlitova}, {Hayes},
  {{\"O}stlin}, {Cannon}, {Roth}, {Bik}, {Pardy}, {Ot{\'\i}-Floranes},
  {Mas-Hesse}, {Adamo}, {Atek}, {Duval}, {Guaita}, {Kunth}, {Laursen},
  {Melinder}, {Puschnig}, {Rivera-Thorsen}, {Schaerer}, \&
  {Verhamme}}]{Herenz2016}
{Herenz}, E.~C., {Gruyters}, P., {Orlitova}, I., {et~al.} 2016, \aap, 587, A78

\bibitem[{{Herenz} {et~al.}(2020){Herenz}, {Hayes}, \& {Scarlata}}]{Herenz2020}
{Herenz}, E.~C., {Hayes}, M., \& {Scarlata}, C. 2020, A\&A, 642, A55

\bibitem[{{Hillis} {et~al.}(2016){Hillis}, {Williams}, {Dolphin}, {Dalcanton},
  \& {Skillman}}]{Hillis2016}
{Hillis}, T.~J., {Williams}, B.~F., {Dolphin}, A.~E., {Dalcanton}, J.~J., \&
  {Skillman}, E.~D. 2016, \apj, 831, 191

\bibitem[{{Ho} {et~al.}(2014){Ho}, {Kewley}, {Dopita}, {Medling}, {Allen},
  {Bland-Hawthorn}, {Bloom}, {Bryant}, {Croom}, {Fogarty}, {Goodwin}, {Green},
  {Konstantopoulos}, {Lawrence}, {L{\'o}pez-S{\'a}nchez}, {Owers}, {Richards},
  \& {Sharp}}]{Ho2014}
{Ho}, I.~T., {Kewley}, L.~J., {Dopita}, M.~A., {et~al.} 2014, \mnras, 444, 3894

\bibitem[{{Holwerda} {et~al.}(2005){Holwerda}, {Gonzalez}, {Allen}, \& {van der
  Kruit}}]{Holwerda2005}
{Holwerda}, B.~W., {Gonzalez}, R.~A., {Allen}, R.~J., \& {van der Kruit}, P.~C.
  2005, \aj, 129, 1396

\bibitem[{{Howard} {et~al.}(2018){Howard}, {Pudritz}, {Harris}, \&
  {Klessen}}]{Howard2018}
{Howard}, C.~S., {Pudritz}, R.~E., {Harris}, W.~E., \& {Klessen}, R.~S. 2018,
  \mnras, 475, 3121

\bibitem[{{Kang} {et~al.}(2016){Kang}, {Zhang}, {Chang}, {Wang}, \&
  {Cheng}}]{Kang2016}
{Kang}, X., {Zhang}, F., {Chang}, R., {Wang}, L., \& {Cheng}, L. 2016, \aap,
  585, A20

\bibitem[{{Kauffmann} {et~al.}(2003){Kauffmann}, {Heckman}, {Tremonti},
  {Brinchmann}, {Charlot}, {White}, {Ridgway}, {Brinkmann}, {Fukugita}, {Hall},
  {Ivezi{\'c}}, {Richards}, \& {Schneider}}]{Kauffmann2003}
{Kauffmann}, G., {Heckman}, T.~M., {Tremonti}, C., {et~al.} 2003, \mnras, 346,
  1055

\bibitem[{{Kewley} {et~al.}(2006){Kewley}, {Groves}, {Kauffmann}, \&
  {Heckman}}]{Kewley2006}
{Kewley}, L.~J., {Groves}, B., {Kauffmann}, G., \& {Heckman}, T. 2006, \mnras,
  372, 961

\bibitem[{{Kotulla} {et~al.}(2009){Kotulla}, {Fritze}, {Weilbacher}, \&
  {Anders}}]{Kotulla2009}
{Kotulla}, R., {Fritze}, U., {Weilbacher}, P., \& {Anders}, P. 2009, MNRAS,
  396, 462

\bibitem[{Kramida {et~al.}(2021)Kramida, {Yu.~Ralchenko}, Reader, \& {and NIST
  ASD Team}}]{NIST_ASD}
Kramida, A., {Yu.~Ralchenko}, Reader, J., \& {and NIST ASD Team}. 2021, {NIST
  Atomic Spectra Database (ver. 5.9), [Online]. Available:
  {\tt{https://physics.nist.gov/asd}} [2021, November 25]. National Institute
  of Standards and Technology, Gaithersburg, MD.}

\bibitem[{Kumari {et~al.}(2019)Kumari, Maiolino, Belfiore, \&
  Curti}]{Kumari2019}
Kumari, N., Maiolino, R., Belfiore, F., \& Curti, M. 2019, Monthly Notices of
  the Royal Astronomical Society, 485, 367

\bibitem[{{Lindegren} {et~al.}(2018){Lindegren}, {Hern{\'a}ndez}, {Bombrun},
  {Klioner}, {Bastian}, {Ramos-Lerate}, {de Torres}, {Steidelm{\"u}ller},
  {Stephenson}, {Hobbs}, {Lammers}, {Biermann}, {Geyer}, {Hilger}, {Michalik},
  {Stampa}, {McMillan}, {Casta{\~n}eda}, {Clotet}, {Comoretto}, {Davidson},
  {Fabricius}, {Gracia}, {Hambly}, {Hutton}, {Mora}, {Portell}, {van Leeuwen},
  {Abbas}, {Abreu}, {Altmann}, {Andrei}, {Anglada}, {Balaguer-N{\'u}{\~n}ez},
  {Barache}, {Becciani}, {Bertone}, {Bianchi}, {Bouquillon}, {Bourda},
  {Br{\"u}semeister}, {Bucciarelli}, {Busonero}, {Buzzi}, {Cancelliere},
  {Carlucci}, {Charlot}, {Cheek}, {Crosta}, {Crowley}, {de Bruijne}, {de
  Felice}, {Drimmel}, {Esquej}, {Fienga}, {Fraile}, {Gai}, {Garralda},
  {Gonz{\'a}lez-Vidal}, {Guerra}, {Hauser}, {Hofmann}, {Holl}, {Jordan},
  {Lattanzi}, {Lenhardt}, {Liao}, {Licata}, {Lister}, {L{\"o}ffler},
  {Marchant}, {Martin-Fleitas}, {Messineo}, {Mignard}, {Morbidelli}, {Poggio},
  {Riva}, {Rowell}, {Salguero}, {Sarasso}, {Sciacca}, {Siddiqui}, {Smart},
  {Spagna}, {Steele}, {Taris}, {Torra}, {van Elteren}, {van Reeven}, \&
  {Vecchiato}}]{Lindegren2018}
{Lindegren}, L., {Hern{\'a}ndez}, J., {Bombrun}, A., {et~al.} 2018, A\&A, 616,
  A2

\bibitem[{{Luridiana} {et~al.}(2015){Luridiana}, {Morisset}, \&
  {Shaw}}]{Luridiana2015}
{Luridiana}, V., {Morisset}, C., \& {Shaw}, R.~A. 2015, \aap, 573, A42

\bibitem[{{Madsen} {et~al.}(2006){Madsen}, {Reynolds}, \&
  {Haffner}}]{Madsen2006}
{Madsen}, G.~J., {Reynolds}, R.~J., \& {Haffner}, L.~M. 2006, \apj, 652, 401

\bibitem[{{Marino} {et~al.}(2013){Marino}, {Rosales-Ortega}, {S{\'a}nchez},
  {Gil de Paz}, {V{\'\i}lchez}, {Miralles-Caballero}, {Kehrig},
  {P{\'e}rez-Montero}, {Stanishev}, {Iglesias-P{\'a}ramo}, {D{\'\i}az},
  {Castillo-Morales}, {Kennicutt}, {L{\'o}pez-S{\'a}nchez}, {Galbany},
  {Garc{\'\i}a-Benito}, {Mast}, {Mendez-Abreu}, {Monreal-Ibero}, {Husemann},
  {Walcher}, {Garc{\'\i}a-Lorenzo}, {Masegosa}, {Del Olmo Orozco},
  {Mour{\~a}o}, {Ziegler}, {Moll{\'a}}, {Papaderos},
  {S{\'a}nchez-Bl{\'a}zquez}, {Gonz{\'a}lez Delgado}, {Falc{\'o}n-Barroso},
  {Roth}, {van de Ven}, \& {Califa Team}}]{Marino2013}
{Marino}, R.~A., {Rosales-Ortega}, F.~F., {S{\'a}nchez}, S.~F., {et~al.} 2013,
  \aap, 559, A114

\bibitem[{Martin {et~al.}(1993)Martin, Kaufman, \& Musgrove}]{Martin1993}
Martin, W.~C., Kaufman, V., \& Musgrove, A. 1993, Journal of Physical and
  Chemical Reference Data, 22, 1179

\bibitem[{Martin {et~al.}(1990)Martin, Zalubas, \& Musgrove}]{Martin1990}
Martin, W.~C., Zalubas, R., \& Musgrove, A. 1990, Journal of Physical and
  Chemical Reference Data, 19, 821

\bibitem[{{Mathis}(2000)}]{Mathis2000}
{Mathis}, J.~S. 2000, \apj, 544, 347

\bibitem[{{McInnes} {et~al.}(2018){McInnes}, {Healy}, \&
  {Melville}}]{McInnes2018}
{McInnes}, L., {Healy}, J., \& {Melville}, J. 2018, arXiv e-prints,
  arXiv:1802.03426

\bibitem[{{McLeod} {et~al.}(2019){McLeod}, {Scaringi}, {Soria}, {Pakull},
  {Urquhart}, {Maccarone}, {Knigge}, {Miller-Jones}, {Plotkin}, {Motch},
  {Kruijssen}, \& {Schruba}}]{McLeod2019}
{McLeod}, A.~F., {Scaringi}, S., {Soria}, R., {et~al.} 2019, \mnras, 485, 3476

\bibitem[{{Micheva} {et~al.}(2019){Micheva}, {Herenz}, {Roth}, {{\"O}stlin}, \&
  {Girichidis}}]{Micheva2019}
{Micheva}, G., {Herenz}, E.~C., {Roth}, M.~M., {{\"O}stlin}, G., \&
  {Girichidis}, P. 2019, \aap, 623, A145

\bibitem[{{Mondal} {et~al.}(2019){Mondal}, {Subramaniam}, \&
  {George}}]{Mondal2019}
{Mondal}, C., {Subramaniam}, A., \& {George}, K. 2019, Journal of Astrophysics
  and Astronomy, 40, 35

\bibitem[{{Morisset} {et~al.}(2015){Morisset}, {Delgado-Inglada}, \&
  {Flores-Fajardo}}]{Morisset2015}
{Morisset}, C., {Delgado-Inglada}, G., \& {Flores-Fajardo}, N. 2015, \rmxaa,
  51, 103

\bibitem[{{Morisset} {et~al.}(2020){Morisset}, {Luridiana},
  {Garc{\'\i}a-Rojas}, {G{\'o}mez-Llanos}, {Bautista}, {Mendoza}, \&
  {Claudio}}]{Morisset2020}
{Morisset}, C., {Luridiana}, V., {Garc{\'\i}a-Rojas}, J., {et~al.} 2020, Atoms,
  8, 66

\bibitem[{{Munari} {et~al.}(2005){Munari}, {Sordo}, {Castelli}, \&
  {Zwitter}}]{Munari2005}
{Munari}, U., {Sordo}, R., {Castelli}, F., \& {Zwitter}, T. 2005, A\&A, 442,
  1127

\bibitem[{{Niederhofer} {et~al.}(2016){Niederhofer}, {Hilker}, {Bastian}, \&
  {Ercolano}}]{Niederhofer2016}
{Niederhofer}, F., {Hilker}, M., {Bastian}, N., \& {Ercolano}, B. 2016, \aap,
  592, A47

\bibitem[{{Oey} {et~al.}(2007){Oey}, {Meurer}, {Yelda}, {Furst},
  {Caballero-Nieves}, {Hanish}, {Levesque}, {Thilker}, {Walth},
  {Bland-Hawthorn}, {Dopita}, {Ferguson}, {Heckman}, {Doyle}, {Drinkwater},
  {Freeman}, {Kennicutt}, {Kilborn}, {Knezek}, {Koribalski}, {Meyer}, {Putman},
  {Ryan-Weber}, {Smith}, {Staveley-Smith}, {Webster}, {Werk}, \&
  {Zwaan}}]{Oey2007}
{Oey}, M.~S., {Meurer}, G.~R., {Yelda}, S., {et~al.} 2007, \apj, 661, 801

\bibitem[{{Oparin} \& {Moiseev}(2018)}]{Oparin2018}
{Oparin}, D.~V. \& {Moiseev}, A.~V. 2018, Astrophysical Bulletin, 73, 298

\bibitem[{{Osterbrock} \& {Ferland}(2006)}]{Osterbrock2006}
{Osterbrock}, D.~E. \& {Ferland}, G.~J. 2006, {Astrophysics of gaseous nebulae
  and active galactic nuclei}

\bibitem[{{Pettini} \& {Pagel}(2004)}]{Pettini2004}
{Pettini}, M. \& {Pagel}, B. E.~J. 2004, \mnras, 348, L59

\bibitem[{{Pilyugin} \& {Grebel}(2016)}]{Pilyugin2016}
{Pilyugin}, L.~S. \& {Grebel}, E.~K. 2016, \mnras, 457, 3678

\bibitem[{{Pilyugin} {et~al.}(2019){Pilyugin}, {Grebel}, {Zinchenko},
  {Nefedyev}, \& {V{\'\i}lchez}}]{Pilyugin2019}
{Pilyugin}, L.~S., {Grebel}, E.~K., {Zinchenko}, I.~A., {Nefedyev}, Y.~A., \&
  {V{\'\i}lchez}, J.~M. 2019, \aap, 623, A122

\bibitem[{Podobedova {et~al.}(2009)Podobedova, Kelleher, \&
  Wiese}]{Podobedova2009}
Podobedova, L.~I., Kelleher, D.~E., \& Wiese, W.~L. 2009, Journal of Physical
  and Chemical Reference Data, 38, 171

\bibitem[{{Read} \& {Pietsch}(2001)}]{Read2001}
{Read}, A.~M. \& {Pietsch}, W. 2001, \aap, 373, 473

\bibitem[{{Reynolds}(1984)}]{Reynolds1984}
{Reynolds}, R.~J. 1984, \apj, 282, 191

\bibitem[{{Riener} {et~al.}(2018){Riener}, {Faesi}, {Forbrich}, \&
  {Lada}}]{Riener2018}
{Riener}, M., {Faesi}, C.~M., {Forbrich}, J., \& {Lada}, C.~J. 2018, \aap, 612,
  A81

\bibitem[{{Robitaille} {et~al.}(2019){Robitaille}, {Rice}, {Beaumont},
  {Ginsburg}, {MacDonald}, \& {Rosolowsky}}]{Robitaille2019}
{Robitaille}, T., {Rice}, T., {Beaumont}, C., {et~al.} 2019, {astrodendro:
  Astronomical data dendrogram creator}, Astrophysics Source Code Library,
  record ascl:1907.016

\bibitem[{{Rodr{\'\i}guez} {et~al.}(2016){Rodr{\'\i}guez}, {Baume}, \&
  {Feinstein}}]{Rodriguez2016}
{Rodr{\'\i}guez}, M.~J., {Baume}, G., \& {Feinstein}, C. 2016, \aap, 594, A34

\bibitem[{{Rogstad} {et~al.}(1979){Rogstad}, {Crutcher}, \&
  {Chu}}]{Rogstad1979}
{Rogstad}, D.~H., {Crutcher}, R.~M., \& {Chu}, K. 1979, \apj, 229, 509

\bibitem[{{Roth} {et~al.}(2021){Roth}, {Jacoby}, {Ciardullo}, {Davis}, {Chase},
  \& {Weilbacher}}]{Roth2021}
{Roth}, M.~M., {Jacoby}, G.~H., {Ciardullo}, R., {et~al.} 2021, \apj, 916, 21

\bibitem[{{Roth} {et~al.}(2018){Roth}, {Sandin}, {Kamann}, {Husser},
  {Weilbacher}, {Monreal-Ibero}, {Bacon}, {den Brok}, {Dreizler}, {Kelz},
  {Marino}, \& {Steinmetz}}]{Roth2018}
{Roth}, M.~M., {Sandin}, C., {Kamann}, S., {et~al.} 2018, \aap, 618, A3

\bibitem[{{Roussel} {et~al.}(2005){Roussel}, {Gil de Paz}, {Seibert}, {Helou},
  {Madore}, \& {Martin}}]{Roussel2005}
{Roussel}, H., {Gil de Paz}, A., {Seibert}, M., {et~al.} 2005, \apj, 632, 227

\bibitem[{{Sabin} {et~al.}(2013){Sabin}, {Parker}, {Contreras}, {Olgu{\'\i}n},
  {Frew}, {Stupar}, {V{\'a}zquez}, {Wright}, {Corradi}, \&
  {Morris}}]{Sabin2013}
{Sabin}, L., {Parker}, Q.~A., {Contreras}, M.~E., {et~al.} 2013, \mnras, 431,
  279

\bibitem[{{S{\'e}rsic}(1966)}]{Sersic1966}
{S{\'e}rsic}, J.~L. 1966, \zap, 64, 212

\bibitem[{{Shields}(1990)}]{Shields1990}
{Shields}, G.~A. 1990, \araa, 28, 525

\bibitem[{{Stasi{\'n}ska} {et~al.}(2013){Stasi{\'n}ska}, {Pe{\~n}a},
  {Bresolin}, \& {Tsamis}}]{Stasinska2013}
{Stasi{\'n}ska}, G., {Pe{\~n}a}, M., {Bresolin}, F., \& {Tsamis}, Y.~G. 2013,
  \aap, 552, A12

\bibitem[{{Stetson}(1987)}]{Stetson1987}
{Stetson}, P.~B. 1987, \pasp, 99, 191

\bibitem[{{Str{\"o}bele} {et~al.}(2012){Str{\"o}bele}, {La Penna}, {Arsenault},
  {Conzelmann}, {Delabre}, {Duchateau}, {Dorn}, {Fedrigo}, {Hubin}, {Quentin},
  {Jolley}, {Kiekebusch}, {Kirchbauer}, {Klein}, {Kolb}, {Kuntschner}, {Le
  Louarn}, {Lizon}, {Madec}, {Pettazzi}, {Soenke}, {Tordo}, {Vernet}, \&
  {Muradore}}]{Stroebele2012}
{Str{\"o}bele}, S., {La Penna}, P., {Arsenault}, R., {et~al.} 2012, in
  Proc.~{SPIE}, Vol. 8447, {Adaptive Optics Systems III}

\bibitem[{{Tayal} \& {Zatsarinny}(2010)}]{Tayal2010}
{Tayal}, S.~S. \& {Zatsarinny}, O. 2010, \apjs, 188, 32

\bibitem[{{Tomi{\v{c}}i{\'c}} {et~al.}(2017){Tomi{\v{c}}i{\'c}}, {Kreckel},
  {Groves}, {Schinnerer}, {Sandstrom}, {Kapala}, {Blanc}, \&
  {Leroy}}]{Tomicic2017}
{Tomi{\v{c}}i{\'c}}, N., {Kreckel}, K., {Groves}, B., {et~al.} 2017, \apj, 844,
  155

\bibitem[{{Tomi{\v{c}}i{\'c}} {et~al.}(2021){Tomi{\v{c}}i{\'c}}, {Vulcani},
  {Poggianti}, {Werle}, {M{\"u}ller}, {Mingozzi}, {Gullieuszik}, {Wolter},
  {Radovich}, {Moretti}, {Franchetto}, {Bellhouse}, \& {Fritz}}]{Tomicic2021}
{Tomi{\v{c}}i{\'c}}, N., {Vulcani}, B., {Poggianti}, B.~M., {et~al.} 2021,
  \apj, 922, 131

\bibitem[{{Toribio San Cipriano} {et~al.}(2016{\natexlab{a}}){Toribio San
  Cipriano}, {Garc{\'\i}a-Rojas}, {Esteban}, {Bresolin}, \&
  {Peimbert}}]{Toribio2016}
{Toribio San Cipriano}, L., {Garc{\'\i}a-Rojas}, J., {Esteban}, C., {Bresolin},
  F., \& {Peimbert}, M. 2016{\natexlab{a}}, \mnras, 458, 1866

\bibitem[{{Toribio San Cipriano} {et~al.}(2016{\natexlab{b}}){Toribio San
  Cipriano}, {Garc{\'\i}a-Rojas}, {Esteban}, {Bresolin}, \&
  {Peimbert}}]{ToribioSanCipriano2016}
{Toribio San Cipriano}, L., {Garc{\'\i}a-Rojas}, J., {Esteban}, C., {Bresolin},
  F., \& {Peimbert}, M. 2016{\natexlab{b}}, \mnras, 458, 1866

\bibitem[{van~der Maaten \& Hinton(2008)}]{vanDerMaaten2008}
van~der Maaten, L. \& Hinton, G. 2008, Journal of Machine Learning Research, 9,
  2579

\bibitem[{{Vega Beltr{\'a}n} {et~al.}(2001){Vega Beltr{\'a}n}, {Pizzella},
  {Corsini}, {Funes}, {Zeilinger}, {Beckman}, \& {Bertola}}]{VegaBeltran2001}
{Vega Beltr{\'a}n}, J.~C., {Pizzella}, A., {Corsini}, E.~M., {et~al.} 2001,
  \aap, 374, 394

\bibitem[{{Vila-Costas} \& {Edmunds}(1992)}]{VilaCostas1992}
{Vila-Costas}, M.~B. \& {Edmunds}, M.~G. 1992, \mnras, 259, 121

\bibitem[{{Vriend}(2015)}]{Vriend2015}
{Vriend}, W.-J. 2015, in Science Operations 2015: Science Data Management - An
  ESO/ESA Workshop, 1

\bibitem[{{Weilbacher} {et~al.}(2018){Weilbacher}, {Monreal-Ibero}, {Verhamme},
  {Sandin}, {Steinmetz}, {Kollatschny}, {Krajnovi{\'c}}, {Kamann}, {Roth},
  {Erroz-Ferrer}, {Marino}, {Maseda}, {Wendt}, {Bacon}, {Dreizler}, {Richard},
  \& {Wisotzki}}]{Weilbacher2018}
{Weilbacher}, P.~M., {Monreal-Ibero}, A., {Verhamme}, A., {et~al.} 2018, A\&A,
  611, A95, paper {I}

\bibitem[{{Weilbacher} {et~al.}(2020){Weilbacher}, {Palsa}, {Streicher},
  {Bacon}, {Urrutia}, {Wisotzki}, {Conseil}, {Husemann}, {Jarno}, {Kelz},
  {P{\'e}contal-Rousset}, {Richard}, {Roth}, {Selman}, \&
  {Vernet}}]{Weilbacher2020}
{Weilbacher}, P.~M., {Palsa}, R., {Streicher}, O., {et~al.} 2020, A\&A, 641,
  A28

\bibitem[{{Westmeier} {et~al.}(2011){Westmeier}, {Braun}, \&
  {Koribalski}}]{Westmeier2011}
{Westmeier}, T., {Braun}, R., \& {Koribalski}, B.~S. 2011, \mnras, 410, 2217

\bibitem[{{Youngblood} \& {Hunter}(1999)}]{Youngblood1999}
{Youngblood}, A.~J. \& {Hunter}, D.~A. 1999, \apj, 519, 55

\bibitem[{{Zurita} {et~al.}(2002){Zurita}, {Beckman}, {Rozas}, \&
  {Ryder}}]{Zurita2002}
{Zurita}, A., {Beckman}, J.~E., {Rozas}, M., \& {Ryder}, S. 2002, \aap, 386,
  801

\bibitem[{{Zurita} {et~al.}(2000){Zurita}, {Rozas}, \& {Beckman}}]{Zurita2000}
{Zurita}, A., {Rozas}, M., \& {Beckman}, J.~E. 2000, \aap, 363, 9

\end{thebibliography}

\end{document}